\documentclass[showpacs,showkeys,preprintnumbers,amsmath,amssymb,times,graphicx]{revtex4}

\usepackage{graphicx} 
\usepackage{dcolumn}
\usepackage{bm}

\usepackage{color}
\usepackage{amsmath}
\usepackage{amsfonts}
\usepackage{amssymb}
\usepackage{amsbsy}
\usepackage[figuresright]{rotating}
\usepackage[font=small,labelfont=bf]{caption}
\usepackage{natbib}

\def\3{\ss}

\def\q0{\phantom{1}}

\def\thetabn{\Theta_{Bn}}

\def\parp2{\frac{\partial^{2}}{\partial p^{2} }}

\def\m21{$2^{\circ}\times 1^{\circ}$}

\def\ts{\thinspace}

\def\ne8{Ne\ts{$\scriptstyle {\rm VIII}$} }

\newcommand{\ag}{{Ann.\ Geo\-phy\-si\-c\ae}}

\newcommand{\grl}{{Geo\-phys.\ Res.\ Lett.}}

\newcommand{\jgr}{{J.\ Geo\-phys.\ Res.}}

\newcommand{\ssr}{{Space Sci.\,Rev.}}

\newcommand{\pf}{{Phys. Fluids}}
\newcommand{\npg}{{Nonlin.\ Process.\ Geophys.}}

\newcommand{\pop}{{Phys.\ Plasmas}}

\def\ion[#1 #2]{#1\,{\sc #2}}
\def\lamb[#1]{#1\,{\AA}}
\def\serts89{SERTS-89}
\def\fe12{Fe\,{\sc xii}}

\def\mb[#1]{\makebox[0.15cm][l]{#1}}

\begin{document}




\title{Fundamentals of  Non-relativistic Collisionless Shock Physics: \\  II. Basic Equations and Models}

\author{R. A. Treumann$^\dag$ and C. H. Jaroschek$^{*}$}\email{treumann@issibern.ch}
\affiliation{$^\dag$ Department of Geophysics and Environmental Sciences, Munich University, D-80333 Munich, Germany  \\ 
Department of Physics and Astronomy, Dartmouth College, Hanover, 03755 NH, USA \\ 
$^{*}$Department Earth \& Planetary Science, University of Tokyo, Tokyo, Japan
}%

\begin{abstract} This paper develops the basic sets of equations which lead to the conservation laws describing collisionless plasma shock waves. We discuss the evolution of shock waves by wave steepening, derive the Rankine-Hugoniot conditions for magnetogasdynamic shocks, discuss various analytical models of shock formation, and discuss the basic instabilities which may become important in collisionless shock physics. We then present a survey of the theory of anomalous resistivity in the quasilinear limit and beyond and discuss mechanisms of shock particle reflection as far as they have been investigated in the published literature. The content of the chapter is the following: 1. Wave steepening, describing simple waves and steepening due to nonlinearity, balnced by dissipation in Burgers' shocks, by dispersive effects in the Korteweg-de Vries equation, the Sagdeev-Potential method,  2. Basic equations, presenting kinetic theory and the transition to moment equations in the fluid description, 3. Rankine-Hugoniot relations, giving the jump conditions across shocks, explicit MHD solution for perpendicular shocks and parallel shocks, and high Mach number conditions, 4. Waves and instabilities, giving the general dispersion relation, describing low-$\beta$-shocks, whistler and Alfv\'en shocks, the various shock-relevant instabilities, 5. Anomalous transport for the various electrostatic wave-particle interactions, general description of anomalous resistivity, shock particle reflection from potential and specularly, hole formation, 6. Briefing on numerical simulation techniques, giving a short idea on this important field and its methods.
\end{abstract}
\pacs{}
\keywords{}
\maketitle

\section{Wave Steepening}\index{waves!steepening}
\noindent Shocks have a certain width $\Delta$ and a certain jump in density $N$, temperature $T$, pressure $\textsf{P}$ and magnetic field ${\bf B}$ across this width from a given upstream value to a downstream value. This jump is by no means infinitesimal. At the contrary it is usually several times the upstream value in magnitude. Thus, looked at as a wave, a shock is a highly nonlinear wave structure of wavelength $\Delta$ with amplitude that cannot be neglected compared with the upstream value. Therefore, the basic equations describing a shock cannot be linearized as is usually done in considering wave phenomena. These equation must be solved in their full nonlinearities. This, however, is barely possible and can be done analytically only in very rare cases which usually are not of interest. 
On the other hand, shocks evolve inside the plasma from small disturbances. It is thus reasonable to ask for the nonlinear evolution of such a small harmonic disturbance in order to learn, how a disturbance can evolve into a very large amplitude shock ramp. 

\subsection{Simple waves: Steepening and breaking}\index{waves!simple}
\noindent The simplest way to do this is to consider the evolution of so-called simple waves \citep[see, e.g.,][]{Witham1974}. Simple waves are one-dimensional sinusoidal disturbances of the  plasma velocity of the form $V(x)=A\sin kx$ moving on the plasma background at speed $c$. The total derivative of this disturbance is given simply by
\begin{equation}\label{chap2-eq-simplewaves}
\frac{{\rm d}V}{{\rm d} t}\equiv \frac{\partial V}{\partial t}+V\frac{\partial V}{\partial x}=0
\end{equation}
and in the absence of any forces and friction is assumed zero. Hence this is the equation which describes the evolution of the disturbance as long as no friction or other force comes into play which in the initial state is a reasonable assumption. 
\begin{figure}[t!]
\hspace{0.0cm}\centerline{\includegraphics[width=0.7\textwidth,clip=]{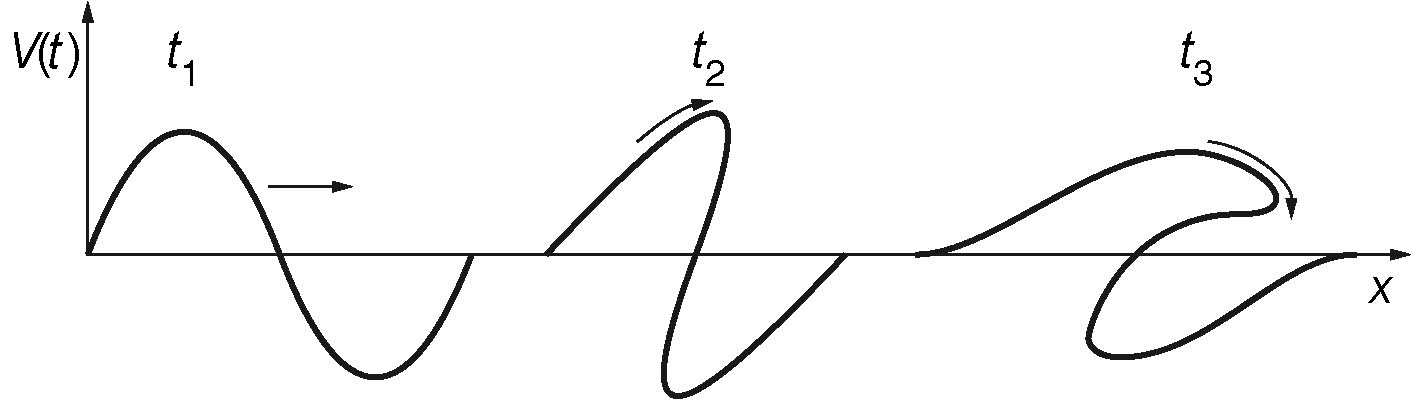} }
\caption[1]
{\footnotesize Schematic of the steepening and wave breaking phenomenon, illustrated for three successive times $t_1, t_2, t_3$. At $t_2$ the wave has steepened to maximum, and in $t_3$ it collapses in the absence of any retarding effects.}\label{chap2-fig-steep}
\end{figure}

\begin{figure}[t!]
\hspace{0.0cm}\centerline{\includegraphics[height=0.225\textheight,width=0.9\textwidth,clip=]{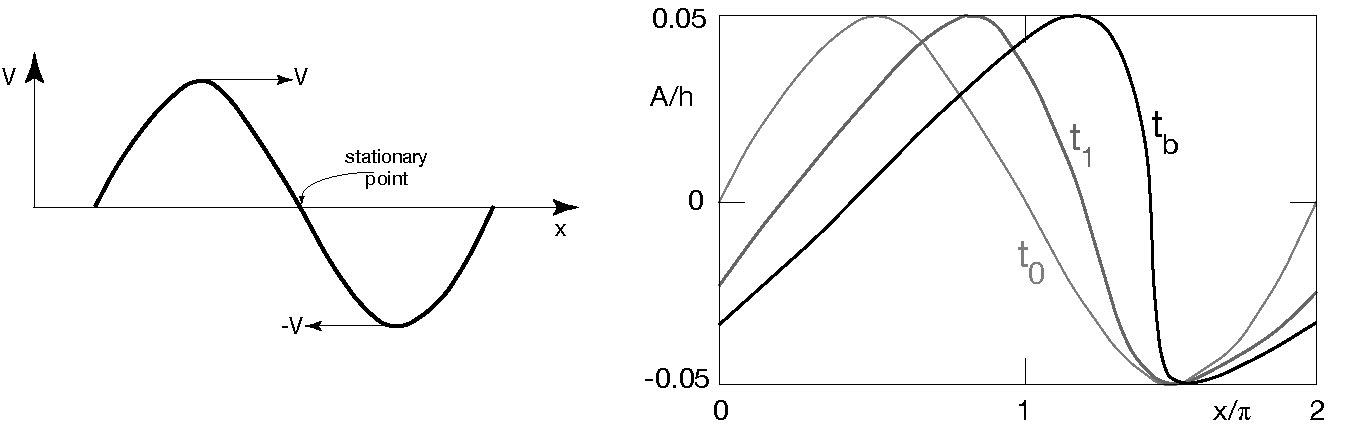} }
\caption[1]
{\footnotesize Steepening and final breaking of an initial sinusoidal (simple) wave. {\it Left}: Initial wave form in wave frame showing the nonlinear action of the wave on its own shape. {\it Right}: Calculation of wave form steepening  in a shallow fluid of depth $h$ \citep[after][]{Zahibo2007}.  The wave profile is shown at the initial time $t_0$, intermediate time $t_1$, and breaking time $t_b$ when the wave starts turning over. Steepening of the profile is well expressed.}\label{chap2-fig-breaking}
\end{figure}
Let us now investigate how such a disturbance will evolve when it propagates through the plasma. Clearly during propagation the main effect on the shape of the disturbance arises from the second nonlinear term which can be written as $Vk\cos kx$. Inserting for $V$ this becomes $\frac{1}{2}\sin 2kx$. Hence, the nonlinear term in the above equation (\ref{chap2-eq-simplewaves}) generates harmonic sidebands of half wavelength and half the amplitude of the original wave. These waves, by the same mechanism, also generate sidebands on their own now at quarter original wavelength and amplitude, and so on with increasingly shorter wavelengths. The total amplitude is the superposition of all these sideband harmonics
\begin{equation}
V_k(x)=\sum\limits_l\frac{A}{2^l}\sin 2^lkx, \qquad (l=0,1,2,\dots)
\end{equation}
All these waves and sidebands propagate at the same velocity $c$. This can be easily seen when, for propagating waves, replacing $kx\to (x-ct)$ in the above expression. They locally superpose and add to the wave amplitude. Because ever shorter wavelengths contribute, the wave steepens until the gradients become so short that other processes take over. If this does not happen, the wave will turn over and break. 

This is illustrated in Figure\,\ref{chap2-fig-steep} and \ref{chap2-fig-breaking} in the co-moving frame of a sinusoidal wave. The left part of the figure shows the nonlinear mechanism. Since the velocity is largest at the maxima it speeds up the motion of the maxima with respect to the remaining parts of the wave profile. Moreover, the action on the positive and negative maxima are oppositely directed, and the wave starts forming a ramp corresponding top a shock front. This happens at time $t_b$. For times $t>t_b$ the wave will turn over and collapse. This can be prevented only by additional processes which set on when the wavelength of the ramp becomes so short that in Eq.\,(\ref{chap2-eq-simplewaves}) terms of higher gradients in the velocity  must be taken into account.

Eq.\,(\ref{chap2-eq-simplewaves}) can in fact been understood as the lowest order equation describing the evolution of a wave packet of wave number $k$. In general in the wave frame of reference its right-hand side is a function $F(V)$ that can be expanded with respect to $V$. The first higher order term in this expansion turns out to be second order in the spatial derivative $\nabla={\hat {\bf x}}\partial_x$, where ${\hat{\bf x}}$ is the unit vector in the direction of x. The next higher order term is third order in $\nabla$, and so on. Up to third order the resulting equation then reads
\begin{equation}\label{chap2-eq-nlwave}
\frac{\partial V}{\partial t}+V\frac{\partial V}{\partial x}=\frac{\partial }{\partial x}D\frac{\partial V}{\partial x} - \beta \frac{\partial^3 V}{\partial x^3} + \cdots
\end{equation}
The first term on the right is a diffusive term \index{diffusion}\index{coefficient!diffusion}with diffusion coefficient $D(x)$. The third term with arbitrary coefficient $\beta$ is the lowest order contribution of wave dispersion to the evolution of the wave shape and amplitude. This can be most easily seen when taking the linearized equation, assuming $V$ to be a small disturbance only of the wave speed $c$ and neglecting the nonlinearity in the second term on the left by approximating it with $c\nabla V$, and subsequently Fourier analysing for a harmonic perturbation $V=A\exp i(kx-\omega t)$ with wave number $k$ and frequency $\omega$. This procedure yields (for constant $D$) the following {\it dispersion relation}
\begin{equation}
\omega -kc+k^3\beta=- ik^2D 
\end{equation}
On the left of this equation there is the relation between the frequency $\omega$ and wave number $k$, while on the right appears an imaginary term that depends on the diffusion coefficient $D$ and square of the wave number $k$. Imaginary terms in frequency imply damping. Hence, as we did express above,  the second order spatial derivative term in Eq.\,(\ref{chap2-eq-nlwave}) corresponds to diffusive dissipation of flow energy, while (for real $\beta$) the third term  in Eq.\,(\ref{chap2-eq-nlwave}) causes the wave to disperse, i.e. waves of different wave-numbers, respectively different wavelengths, propagate at different phase velocities. 

\subsection{Burgers' dissipative shock solution }\index{equation!Burgers}
\noindent Returning to the nonlinear equation (\ref{chap2-eq-nlwave}) we can thus conclude from this analysis that a wave if not breaking will steepen so long until either the diffusive or the dispersive terms on the right start competing with the nonlinearity in which case the wave may assume a stationary that is balanced either by diffusion or by dispersion. Diffusion implies balance by real dissipation with energy transformed into heat, while dispersion implies that the `dangerous' short wavelength waves which cause the steepening either run the wave out or do not catch up with the wave when the slope of the wave profile exceeds a certain steepness. \index{shocks!dissipative}
In the dispersive case the wave will either exhibit short wavelength fluctuations in front of the steepened profile or behind it, depending on whether the shorter wavelength sidebands are retarded or accelerated. However, we can also conclude that dispersion alone should be unable to generate a shock since in the simple form discussed here it does not produce irreversible dissipation and hence no heating and increase in entropy. For a shock profile to be created some kind of diffusive process will be necessary. 
\subsubsection{Stationary Burgers equation}
\begin{figure}[t!]
\hspace{0.0cm}\centerline{\includegraphics[height=0.225\textheight,width=0.9\textwidth,clip=]{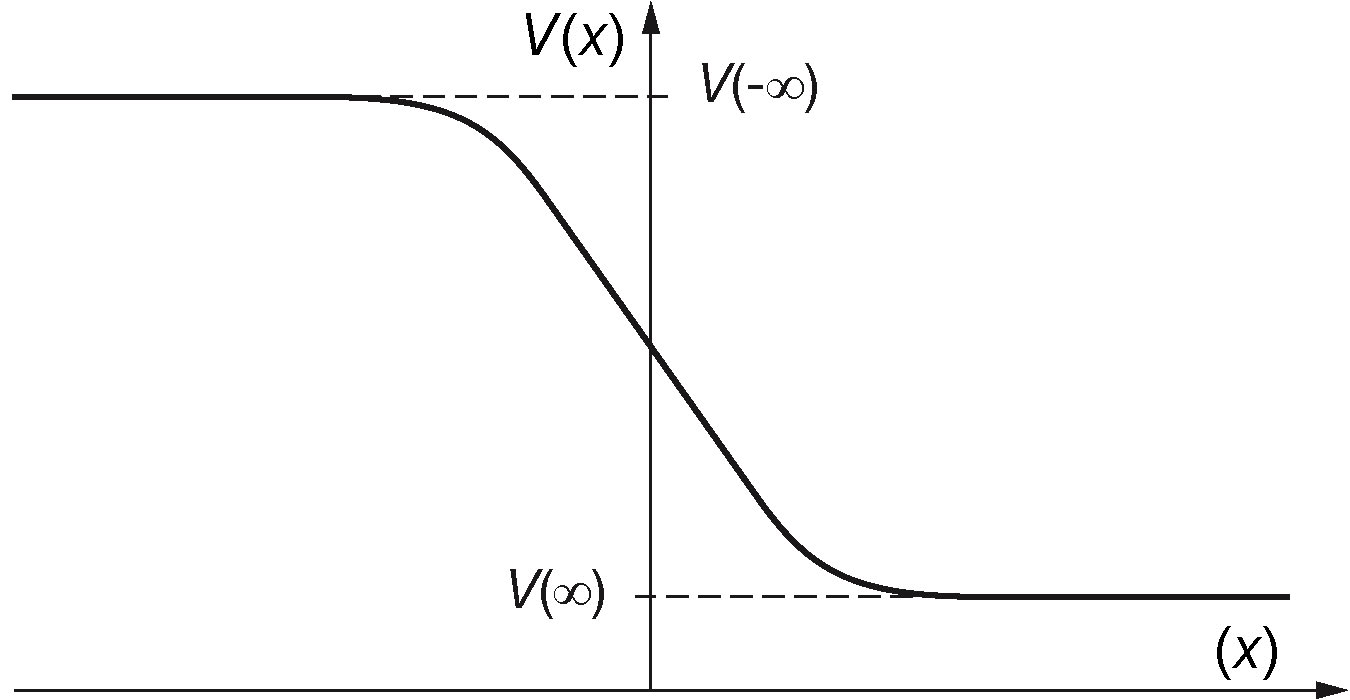} }
\caption[1]
{\footnotesize The stationary solution of the Burgers equation is a smooth shock ramp of width $\Delta=2D/c$, depending on shock velocity $c$ and diffusion coefficient $D$. }\label{chap2-fig-burgers}
\end{figure}

\noindent Eq.\,(\ref{chap2-eq-nlwave}) allows to distinguish two extreme cases. The first case is that of purely diffusive compensation of the nonlinear steepening. In this case the dispersive term can be neglected, and one obtains the Burgers equation
\begin{equation}\label{chap2-eq-burgers}
\frac{\partial V}{\partial t}+V\frac{\partial V}{\partial x}=D\frac{\partial^2 V}{\partial x^2}
\end{equation}
which is a non-linear diffusion (or heat conduction) equation. In contrast \index{diffusion!nonlinear}to the ordinary linear heat conduction \index{equation!heat conduction}equation the Burgers equation possesses stationary solutions due to the above mentioned compensation of diffusive spread by nonlinear steepening. These stationary solutions can be found when transforming to a coordinate system moving with the wave by introducing the new coordinate $y=x-ct$. Then Burgers' equation becomes
\begin{equation}
D\frac{\partial^2 V}{\partial y^2}=(V-c)\frac{\partial V}{\partial y}
\end{equation}
We are interested only in solution which are regular at infinity with vanishing derivatives. Introducing the variable $V'=V-c$ the first integral is easily obtained. Integrating a second time the solution found is then
\begin{equation}
\frac{V}{c}=1-\tanh \left(\frac{x-ct}{2D/c}\right)
\end{equation}
The form of this solution is a typical shock ramp which is displayed in Figure\,\ref{chap2-fig-burgers}. The ramp is sitting on the wave velocity $c$. Its  width is  $\Delta=2D/c$.\index{shocks!width $\Delta$} The shock solutions produced by Burgers' equation are thus propagating non-oscillatory shocks\index{shocks!non-oscillatory}; they are simple stationary constant amplitude ramps of the sort of tsunamis. We should, however, keep in mind that they are produced solely by nonlinear steepening and its compensation through diffusion. When the latter is large, the shock will be steep; in the opposite case it will be a flat ramp only, and its relative height is a function of its width.
\subsubsection{Time-dependence}
\begin{figure}[t!]
\hspace{0.0cm}\centerline{\includegraphics[width=0.8\textwidth,clip=]{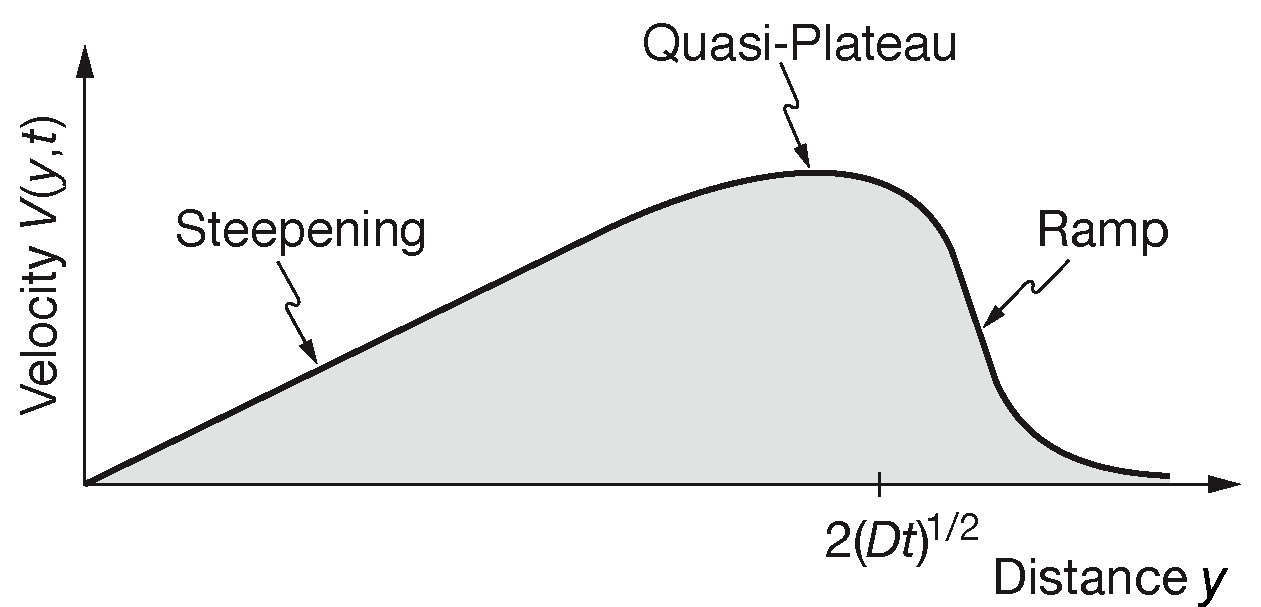} }
\caption[1]
{\footnotesize The time-asymptotic Burgers' shock solution Eq.\,(\ref{chap2-eq-burgers2}) which evolves from the initial disturbance $V?0$ through steepening and dissipative ramp formation after a given diffusive time at a location which is determined by the time $t$ and diffusion coefficient $D$. The characteristic shape of this solution contains a smooth wavelike increase up to a flat plateau followed by the shock ramp and a smooth transition to the undisturbed state. The ramp is moving to the right in the direction of original wave propagation. This is seen from the time dependence of the crest of the ramp. Clearly, most shock transitions in space do not exhibit this smooth rise inside the downstream region of the shock, indicating that the Burgers solution has pure model character which does not really confirm with the plasma reality.}\label{chap2-fig-burgers2}
\end{figure}

\noindent Since Burgers' equation is an ordinary diffusion equation it can be solved by the usual methods of treating heat conduction evolving from some initial state. Such an investigation is necessary in order to justify that the stationary state of the shock ramp Burgers solution can indeed be reached by evolution out of the initial state $V_0$. To this end one transforms the time-dependent Burgers equation (\ref{chap2-eq-burgers}) through introduction of a new variable $\phi$ via $V=-2D[\partial\ln\phi/\partial y]$ into the common form of a diffusion equation $\partial\phi/\partial t=D\partial^2\phi/\partial y^2$, which has the commonly known solution
\begin{equation}
\phi(y,t)=\frac{1}{(4\pi Dt)^\frac{1}{2}}\int\limits_{-\infty}^\infty{\rm d}\eta\exp\left[-\frac{(y-\eta)^2}{4Dt}-\frac{1}{2D}\int\limits_0^\eta V_0(\tau){\rm d}\tau\right]
\end{equation}
The initial disturbance satisfies the condition of convergence $\int_0^y{\rm d}y'V_0(y')\leq {\rm const}\cdot y$ for $y\to\infty$, which yields the requirement that $\int_{-\infty}^\infty{\rm d}y' V_0(y')= \Theta <\infty $ as well as the time-asymptotic solution
\begin{equation}\label{chap2-eq-burgers2}
V(y,t\to\infty)\simeq -2D\frac{{\rm d}}{{\rm d}y}\ln\,G\left[\frac{y}{(4Dt)^\frac{1}{2}}\right] 
\end{equation}
The dummy function $G(x)$ is a transformed version of the function $\phi$ that is given by $\pi^\frac{1}{2} G(x)= {\rm e}^{-\Theta /4D}\int_{-\infty}^x{\rm d}\eta{\rm e}^{-\eta^2}+{\rm e}^{\Theta /4D}\int_x^\infty{\rm d}\eta{\rm e}^{-\eta^2}$. This asymptotic profile Burgers shock ramp solution is shown in Figure\,\ref{chap2-fig-burgers2}. The characteristic shape of this solution contains a smooth wavelike increase up to a flat plateau followed by the shock ramp and a smooth transition to the undisturbed state. The ramp is moving to the right in the direction of original wave propagation. This is seen from the time dependence of the crest of the ramp. Clearly, most shock transitions in space do not exhibit this smooth rise inside the downstream region of the shock, indicating that the Burgers solution has pure model character which does not really confirm with the plasma reality.

\subsection{Korteweg-de Vries dispersion effects}\index{equation!Korteweg-deVries}\index{waves!dispersion}
\noindent Balancing the nonlinearity with the help of dissipation is one possibility. The remaining possibility which in the absence of dissipation becomes the dominant, is balancing nonlinearity with dispersion. In this case we can neglect the diffusion term in Eq.\,(\ref{chap2-eq-nlwave}) to obtain
the so-called Korteweg-de Vries equation\footnote{It might be of interest to note that to find a method that solves this general time-dependent equation analytically took more than a century, and for this purpose it was necessary to develop the whole apparatus of non-relativistic quantum mechanics. this was done in a seminal paper by \cite{Gardner1967}.}
\begin{equation}\label{chap2-eq-KdV}
\frac{\partial V}{\partial t}+V\frac{\partial V}{\partial x}+\beta \frac{\partial^3 V}{\partial x^3} =0
\end{equation} 

Similar to Burgers' equation, the Korteweg-de Vries equation also allows for stationary localized solutions. Such solutions are restricted to a finite spatial interval because dispersion does not cause irreversible effects. As before we assume that the stationary solution \index{equation!stationary solution}moves a speed $c$, and we introduce the co-moving coordinate $y=x-ct$, this time measured from the centre of the localized disturbance. The Korteweg-de Vries equation then transforms into the third order ordinary differential equation
\begin{equation}\label{chap2-eq-stationaryKdV}
(V-c)\frac{\partial V}{\partial y}+\beta\frac{\partial^3 V}{\partial y^3}=0
\end{equation}
Solution of this stationary equation requires the prescription of boundary conditions at $y\to\pm\infty$ for which we choose $V=\partial V/\partial y=0$. It can be shown by substitution that the function
\begin{equation}\label{chap2-eq-KdVsoliton}
V_{KdV}(x-ct)=3c\,{\rm sech}^2\left[\sqrt{\frac{c}{\beta}}\frac{(x-ct)}{2}\right]
\end{equation}
This function is a so-called {\it soliton} solution; it describes a stationary bell-shaped solitary wave pulse propagating at velocity $c$ along $x$ without any change of form. The width of this pulse is $\Delta=2\sqrt{\beta/c}$ and depends on the velocity $c$ and the dispersion parameter $\beta$ in such a way that the faster the pulse moves, the narrower the pulse becomes. In addition there is a distinct relation between the amplitude $A$ of the pulse and its width
\begin{equation}
\Delta = 2(3\beta/A)^\frac{1}{2}
\end{equation}
from where it follows that large amplitude Korteweg-de Vries solitons \index{soliton!Korteweg-e Vries}are fast and narrow. Like in the case of the Burgers equation, the Korteweg-de Vries equation is a model equation which results from the dispersive properties of nonlinear waves. However, it is interesting that it can be derived for real problems arising in plasma wave propagation, and several variants of it have in the past been applied to the plasma. Hence, in describing the nonlinear evolution of plasmas it is a more realistic model than Burgers' equation which is simply a consequence of the strongly dispersive and practically dissipation-free properties of plasmas.
\begin{figure}[t!]
\hspace{0.0cm}\centerline{\includegraphics[width=0.8\textwidth,clip=]{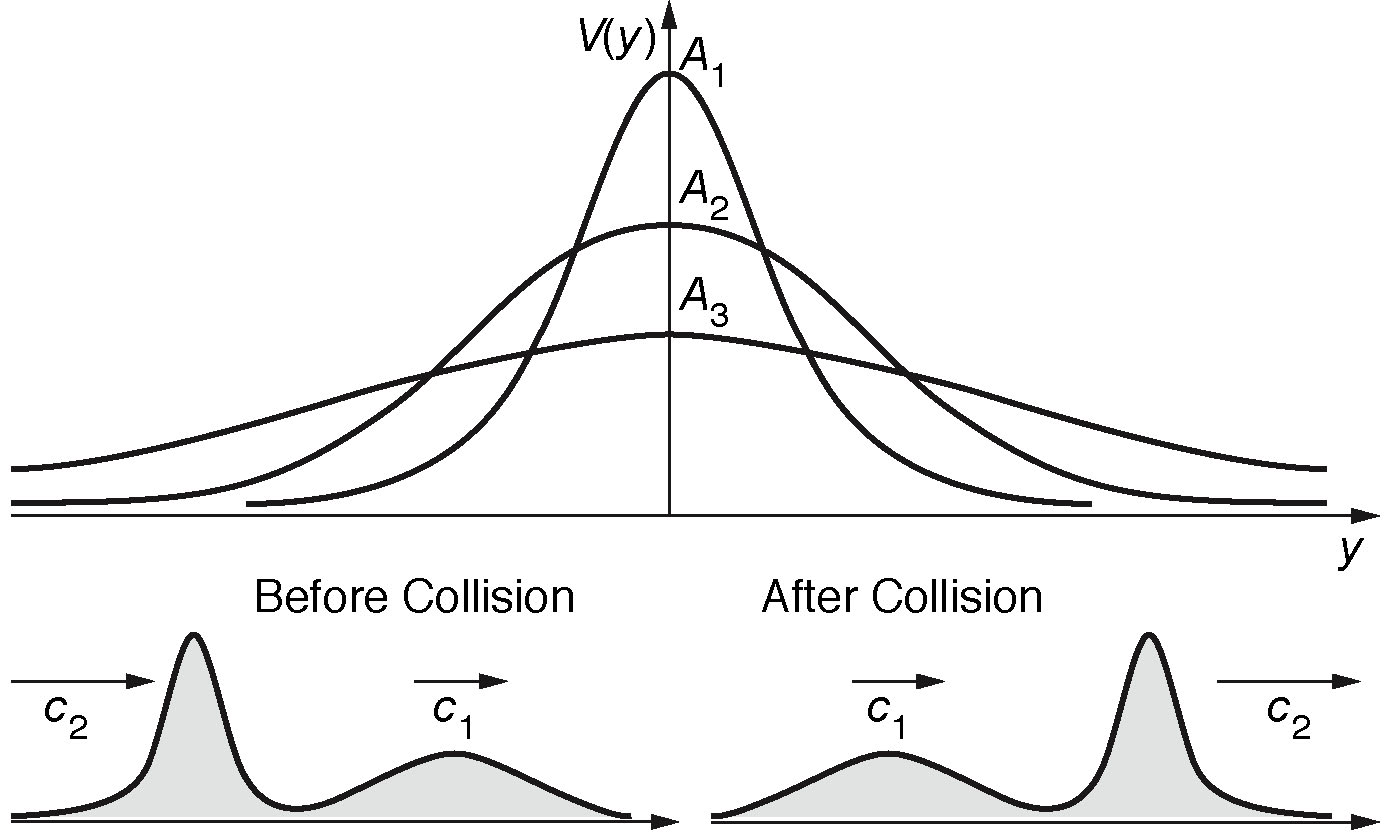} }
\caption[1]
{\footnotesize {\it Top}: Shape of Korteweg-de Vries solitons for different widths and amplitudes (labelled $A_1, A_2, A_3$). Solitons of large amplitude are narrower than those  with smaller amplitude. {\it Below}: In the interaction between two Korteweg-de Vries solitons of different speeds only the position changes while the collision has no effect on amplitude or shape. }\label{chap2-fig-kdv}
\end{figure}

The stationary Korteweg-de Vries can \index{equation!stationary solution}have a whole chain\index{soliton!chain} of such solitons as solution with the solitons having completely different amplitudes and widths. These solitons have the interesting property that they can pass through each other during mutual encounters without having any effect on their widths and amplitudes; only the phases and spatial positions of the waves from which the solitons form will change during the collision. The question of how these chains of solitons are produced is a question that can be answered only when solving the time-dependent Korteweg-de Vries problem imposing a certain initial condition similar to that imposed above on the time-dependent Burgers equation. \index{equation!time-dependent}In the case of the time-dependent Korteweg-de Vries equation the solution cannot be found in such a simple way, however. Solving it rather constitutes a major mathematical problem which requires solving an equivalent Schr\"odinger equation \citep{Gardner1967}. \index{equation!Schr\"odinger}It turns out that the soliton amplitudes in the chain which solves the Korteweg-de Vries equation are related in some manner by an infinite set of invariants of the Korteweg-de Vries equation. In addition, the time-dependent Korteweg-de Vries equation also supports wave trails which accompany the solitons forming an oscillatory (turbulent) background of spatially dependent amplitudes on which the solitons propagate. 

Clearly, these soliton chains are no shocks; they are wave pulses which after some steepening time evolve into stationarity and are completely reversible practically not leaving any effect on the plasma if one neglects the microscopic processes which take place in the plasma. But this is precisely the door where the speculation comes in about such solitons in collisionless plasma being the initial state of the formation of collisionless shocks. Because, if one can manage a soliton in the chain to move so fast that its width becomes comparable to the intrinsic plasma scales, then the wave field of the soliton should distort the microscopic particle motion causing some kind of dissipation which necessarily will turn the soliton into a shock wave by generating entropy and producing a difference between the states upstream and downstream of the soliton. The soliton in this case borrows from Burgers shock solutions, and mathematically the Korteweg-de Vries equation regains the lost dissipative Burgers second-order term becoming a Korteweg-de Vries-Burgers equation. This was, actually, the point \cite{Sagdeev1966} made intrinsically in his famous theory of shock formation\index{shocks!formation} in collisionless plasma.

\subsection{Sagdeev's Pseudo-Potential}\index{Sagdeev pseudo-potential}
\noindent The Korteweg-de Vries equation is the ideal candidate for introducing one particular notion that has become immensely important in soliton and shock research, the so-called {\it Sagdeev potential}. The Sagdeev potential is a pseudo-potential introduced in order to solve a certain class of nonlinear partial differential equations and to distinguish between solitary wave and shock solutions of these equations. This method takes advantage of the similarity of the first integral of the particular class of equations to the equation of motion of a hypothetical particle in classical mechanics. Knowledge of the Sagdeev potential then reduces the problem of solution to the mere discussion of the behaviour of a particle in the pseudo-potential well. 
\begin{figure}[t!]
\hspace{0.0cm}\centerline{\includegraphics[width=0.8\textwidth,clip=]{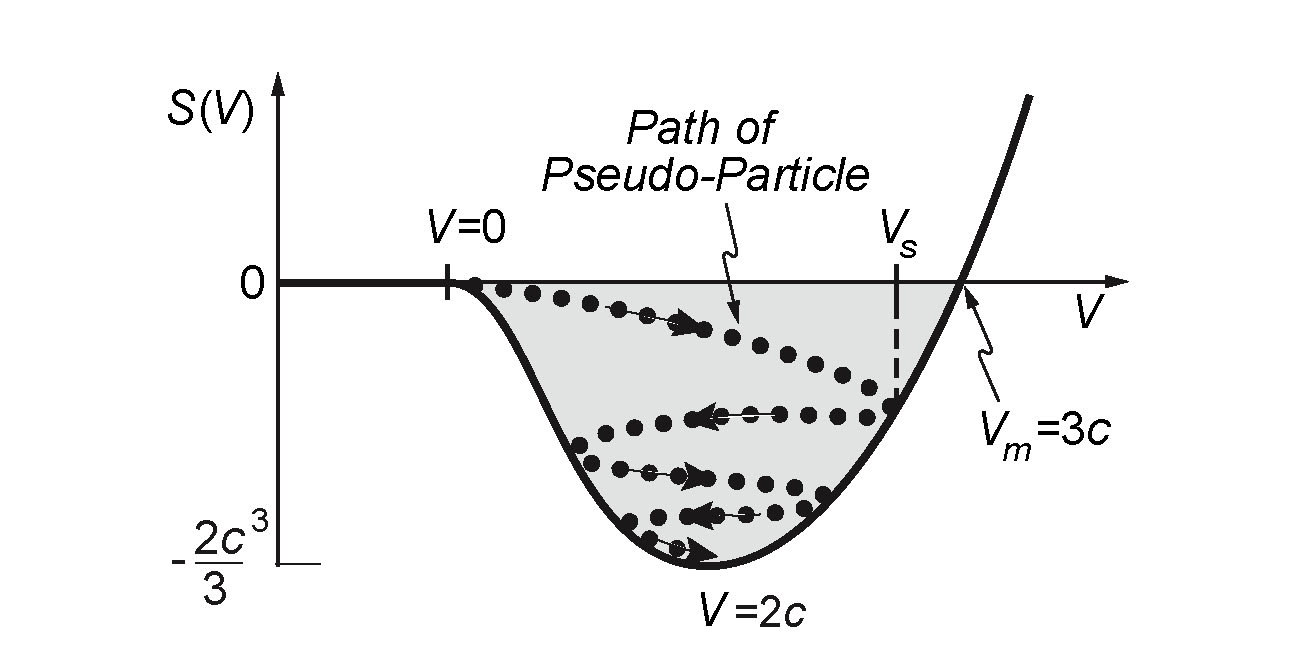} }
\caption[1]
{\footnotesize A sketch of the Korteweg-de Vries Sagdeev pseudo-potential. Solutions exist only in the region of $S(V)\leq 0$. The maximum soliton amplitude is just $V=3c$. The minimum potential is at $V=2c$. The dotted line shows the path of a pseudo-particle for shock formation in presence of dissipation when it ``steps down" the potential well to approach the ``minimum energy" state.}\label{chap2-fig-Sagdpot-2}
\end{figure}

The stationary Korteweg-de Vries equation (\ref{chap2-eq-stationaryKdV}) can be directly integrated once. Applying the boundary conditions at infinity, the integration constant in the first integration becomes zero yielding the nonlinear second-order differential equation
\begin{equation}
\beta\frac{\partial^2 V}{\partial y^2}=V\left(c-\frac{1}{2}V\right)
\end{equation}
The similarity of this equation to Newton's equation of motion of a pseudo-particle of mass $\beta$ in the force field given on the right-hand side is quite obvious. Here $V$ is the pseudo-spatial coordinate, and $y$ is the pseudo-time. This equation can be solved by multiplying it with the pseudo-velocity $\partial V/\partial y$, after which the right-hand side can be represented as the derivative of the Sagdeev pseudo-potential, $S(V)$, which in this case is a function of the (real) velocity $V$. It becomes explicit in the pseudo-energy conservation law
\begin{equation}
\frac{\beta}{2}\left(\frac{\partial V}{\partial y}\right)^2= \frac{V^2}{2}\left(c-\frac{1}{3}V\right) \equiv - S(V)
\end{equation}
after having integrated a second time and again applied the vanishing boundary conditions at infinity. Because the left-hand side of this expression is a positive quantity, solutions exist only under the condition that the Sagdeev pseudo-potential is attractive, i.e. is negative
\begin{equation}
S(V)=\frac{V^2}{2}\left(\frac{V}{3}-c\right)<0
\end{equation}
and, as we already know, solutions can exist only in the region of velocity space where $V<3c$. 
Figure\,\ref{chap2-fig-Sagdpot-2} shows a sketch of the Korteweg-de Vries Sagdeev potential. It vanishes at $V=0$ and $V=3c$ and has its minimum of $S=-3c^3/2$ at $V=2c$. In terms of energy states a pseudo-particle (soliton) can assume any of the energy levels inside the negative portion of the Sagdeev potential. The soliton with maximum amplitude $V=3c$ is at the ``highest" level $S=0$. But there can be many solitons at this level with amplitudes between 0 and $3c$. The ``most stable" soliton at the ``ground state" has minimum Sagdeev potential and amplitude $V=2c$, and there is only one soliton with such an amplitude. In the absence of dissipation all these solitons are stable. 

The actual solutions (\ref{chap2-eq-KdVsoliton}) of the Korteweg-de Vries equation can then be found from $\partial V/\partial y=\sqrt{-2S(V)/\beta}$ by simple quadrature, solving the integral
\begin{equation}
y-y_0=\int\limits_0^V\frac{{\rm d}V}{[-2S(V)/\beta]^\frac{1}{2}}
\end{equation}
We do not further discuss the stationary solution of the Korteweg-de Vries equation; it is but one example which can be solved by the Sagdeev potential method. In the literature it has been demonstrated that a very large number of other nonlinear problems in plasma related to solitary waves can be treated in the same way, sometimes under much more complicated conditions and leading to different types of solitary solutions. 

All these solutions are, however, dissipation free and do not directly lead to shock wave solutions. As \cite{Sagdeev1966} pointed out, they will turn into shock solutions whenever anomalous processes at short wavelengths cause the appearance of some kind of anomalous dissipation under the ideal conditions of non-collisionality. 

This claim is a most important insight that can, however, be based only on the kinetic theory of the microscopic interaction between waves and particles and waves and waves in plasma far from thermal equilibrium, the so-called {\it collective processes} which dominate the behaviour of high temperature plasmas in which shock waves are at home. The dotted arrowed line in Figure\,\ref{chap2-fig-Sagdpot-2} shows the presumable ``path" of such a dissipative soliton in the Sagdeev potential ``energy" space. The soliton pseudo-particle will in this case step down the potential, possibly in an oscillatory way. In the case when the system is open and energy is continuously supplied it might reach a stationary shock state with shock amplitude $V_s$ or settle at the ultimate minimum of the Sagdeev potential. The dynamics of this depends on the microphysics.

\section{Basic Equations}
\noindent Before discussing these processes and their relevance for shock wave formation, we need to briefly introduce the equations which lie at the fundament of all these processes and to discuss their macroscopic consequences. We will in the present chapter distinguish between two approaches to the description of shocks, the theoretical and the numerical approaches, respectively. The former deals with the average properties of collective plasma behaviour and the investigation of wave growth from an infinitesimal perturbation up to a large amplitude shock, the latter refers to the dynamics of macro-particles (as has been described in Chapter 1) and is independent of the average equations as it simply solves Newton's equations of motion of the many macro-particles that constitute the plasma in their self-consistent fields, where the fields are obtained from Maxwell's equations of electrodynamics. any final shock theory must combine both approaches because the fundamental basic equations cannot be solved analytically, while the numerical approach provides data which cannot be understood without a follow-up theoretical investigation tailored to serve the effects found in the numerical simulation experiments.
 
\subsection{Kinetic plasma equations}\index{equation!kinetic plasma}
\noindent Collisionless shock waves represent the final result of collective interactions in which very many particles and in addition the full electromagnetic fields are involved. It is thus quite reasonable that they cannot be described by test particle theory which considers the motion of non-interacting particles separate from other particles and fields. Test particle theory can however be applied if one is not interested in the formation of shocks but instead in its effect on small numbers of particles. This is used in the shock theory of charged particle acceleration which will be the subject of Chapter 6.

\subsubsection{Maxwell-Vlasov equations}
\noindent The basic equations on which shock physics is founded are the kinetic equations of a plasma \citep[cf., e.g.,][and others]{Montgomery1964,Tidman1971} or, at the best, some of its simplifications, in addition to the full set of the equations of electrodynamics. 

Since collisions can be neglected, and thus the Boltzmann collision term in the kinetic equations is suppressed, these equations reduce to the (non-relativistic) Vlasov-Maxwell set of equations
\begin{equation}\label{chap2-eq-vlasov}
\frac{\partial F^\pm}{\partial t} +{\bf v}\cdot\nabla F^\pm +\frac{e_\pm}{m_{\pm}}({\bf E}+{\bf v}\times{\bf B})\cdot\nabla_{\bf v}F^\pm =0
\end{equation}
where $F^\pm({\bf v},{\bf x},t)$ are the electron and ion phase space distributions, distinguished by the respective $+$ and $-$ signs, which depend on the six-dimensional phase space velocity, ${\bf v}$, and real space, ${\bf x}$, coordinates. $m_+\equiv m_i$ and $m_-\equiv m_e$ are the ion and electron masses, respectively; $e_+=e$ is the ion charge, $e_-=-e$ electron charge, $e$ the elementary charge, and ${\bf E}({\bf x},t), {\bf B}({\bf x},t)$ are the electromagnetic fields which are independent on velocity while being functions of space and time. Finally, $\nabla_{\bf v}\equiv \partial/\partial {\bf v}$ is the velocity gradient operator acting on the phase space distributions. These two\index{equation!Vlasov} Vlasov equations (\ref{chap2-eq-vlasov}) are coupled mutually and to the electromagnetic fields through Maxwell's equations
\begin{equation}\label{chap2-eq-maxwell}\left.
\begin{array}{rcccl}
\nabla\times{\bf B}\!\!&\!\!=\!\!&\!\! \mu_0\epsilon_0(\partial {\bf  E}/\partial t) +\mu_0\sum\limits_\pm e_\pm\int{\rm d}v^3 F^\pm{\bf v},  \quad\nabla\cdot{\bf B}\!\!&\!\!=\!\!&\!\! 0 \\
\nabla\times{\bf E}\!\!&\!\!=\!\!&\!\!-(\partial {\bf  B}/\partial t), \qquad\qquad\qquad\qquad\qquad \nabla\cdot{\bf E}\!\!\!\!&\!\!=\!\!&\!\!\epsilon_0^{-1} \sum\limits_\pm e_\pm\int{\rm d}v^3 F^\pm
\end{array}\right\}
\end{equation}
The second term on the right in the first of these equations\index{equation!Maxwell's} is the electric current density; the term on the right in the last of these equations is the electric space charge density (divided by the dielectric constant of vacuum, $\epsilon_0$). These equations already account for the coupling of the field to the particles through the definition of the electric current and particle densities as zeroeth and first moments of the one-particle phase space distributions. 

Shocks evolve from infinitesimal wave disturbances\index{disturbance!infinitesimal}; one hence considers two different states of the plasma with the physics of both of them contained in the above equations. These two states are first the final average slowly evolving state of the fully developed shock, and second the strongly time-dependent evolution of the infinitesimal disturbance from the thermal level where it starts up to the formation of the shock. In the first state the shock possesses a distinct shock profile while in the second state one deals with initially infinitesimal fluctuations. When the fluctuation amplitude approaches the shock strength the two different ways of looking at the shock should ideally lead to the same result. According to this distinction one divides all field and plasma quantities, $A$, into their slowly varying averages, $\langle A\rangle$, and fast fluctuations, $\delta A$, superimposed on the averages according to the prescription
\begin{displaymath}
A=\langle A\rangle +\delta A, \qquad\qquad\qquad  \langle\delta A\rangle =0
\end{displaymath}
The second part of this prescription breaks down when the fluctuations become very large or non-symmetric, but it makes life easier to deal with zero averages of small fluctuations as long as the fluctuation amplitude remains to be small. Frequently it is the only way of extracting a solution from the above nonlinear and complex set of equations (\ref{chap2-eq-vlasov}-\ref{chap2-eq-maxwell}). 

In what follows we will apply different simplifications to all these basic equations referring to the last conditions of simplification. Only the last section of the present chapter will, finally, deal with the numerical simulation technique which, in fact, will become the most important tool in the investigation of shocks in the remaining three chapters of this first part of the book.

\subsubsection{Equations for averages and fluctuations}\index{fluctuations}\index{equation!average}
\noindent Let us for simplicity temporarily indicate the averages $\langle\cdots\rangle$ of the distribution functions and fields by the subscript 0 on the unbraced quantities, and the fluctuations by small letters $f, {\bf e, b}$. Then,  on applying the above prescription of averaging to the Vlasov equation, we obtain the kinetic equation for the average distribution functions $F_0^\pm({\bf v, x}, t)$ in the form
\begin{equation}\label{chap2-eq-vlasovaverage}
\frac{\partial F^\pm_0}{\partial t} +{\bf v}\cdot\nabla F^\pm_0+\frac{e_\pm}{m_\pm}({\bf E}_0+{\bf v}\times{\bf B}_0)\cdot\nabla_{\bf v} F_0^\pm=-\frac{e_\pm}{m_\pm}\langle ({\bf e}+{\bf v}\times{\bf b})\cdot\nabla_{\bf v}f^\pm\rangle
\end{equation}
Here the average quantities are assumed to vary on much longer spatial and temporal scales\index{scales!temporal}\index{scales!spatial} than the fluctuation scales such that the condition of averaging $\langle f, {\bf e, b}\rangle=0$ remains valid. This average Vlasov equation contains a non-vanishing pseudo-collision term \index{collisions!pseudo-collision term}on its right which accounts for the effect of the correlations between the fluctuations and particles on the average distribution.  In contrast to the Vlasov equation, the Maxwell equations (\ref{chap2-eq-maxwell}) retain their form with the sole difference that the full distribution functions $F^\pm$ appearing in the expression for the electric current density in Amp\`ere's law and in the space charge term in Poisson's equation\index{equation!Poisson's} are to be replaced by their average counterparts $F^\pm_0$, yielding
\begin{equation}\label{chap2-eq-maxaverage}\left.
\hspace{-0.25cm}\begin{array}{rclcl}
\nabla\times{\bf B}_0\!\!\!\!&\!\!\!\!=\!\!\!\!&\!\!\!\! \mu_0\epsilon_0(\partial {\bf  E}_0/\partial t) +\mu_0\sum\limits_\pm e_\pm\int{\rm d}v^3 F^\pm_0{\bf v},  \quad\nabla\cdot{\bf B}_0\!\!\!\!&\!\!\!\!=\!\!\!\!& 0 \\
\nabla\times{\bf E}_0\!\!\!\!&\!\!\!\!=\!\!\!\!&\!\!\!\!-(\partial {\bf  B}_0/\partial t), \qquad\qquad\qquad\qquad\qquad \nabla\cdot{\bf E}_0&=&\epsilon_0^{-1} \sum\limits_\pm e_\pm\int{\rm d}v^3 F^\pm_0
\end{array}\right\}
\end{equation}
In order to obtain equations for the fluctuations one subtracts the set of averaged equations from the full set of equations and orders the terms for the fluctuation quantities $f=F-F_0, {\bf e}={\bf E}-{\bf E}_0, {\bf b}={\bf B}-{\bf B_0}$. This procedure leaves the Maxwell equations unchanged when all quantities appearing in them are replaced by the fluctuating quantities, and the fluctuating Vlasov equation becomes
\begin{eqnarray}\label{chap2-eq-fluctuationvlasov}
\frac{\partial f^\pm}{\partial t} \!\!\! &\!\!\!+\!\!\! &\!\!\!{\bf v}\cdot\nabla f^\pm +\frac{e_\pm}{m_{\pm}}({\bf E}_0+{\bf v}\times{\bf B}_0)\cdot\nabla_{\bf v}f^\pm \nonumber
=-\frac{e_\pm}{m_{\pm}}({\bf e}+{\bf v}\times{\bf b})\cdot\nabla_{\bf v}F_0^\pm \\ 
\!\!\! &\!\!\!-\!\!\! &\!\!\!\frac{e_\pm}{m_{\pm}}({\bf e}+{\bf v}\times{\bf b})\cdot\nabla_{\bf v}f^\pm+\frac{e_\pm}{m_\pm}{\langle} ({\bf e}+{\bf v}\times{\bf b})\cdot\nabla_{\bf v}f^\pm{\rangle}
\end{eqnarray}
Up to this stage the fluctuations are allowed to have arbitrarily large amplitudes; it is only their scales which must be much shorter than the scales of the average field quantities. This means for instance that the width of the shock transition regions should be much larger than the wavelengths of the fluctuations.

The last equation is in fact the equation that describes the evolution of fluctuations. however the coupling to the average quantities is still so strong in this equation that it can be solved only together with the average equation. In particular the average ``collision term'' \index{collisions!average term}appearing on its right provides the greatest complications. It will therefore be simplified considerably in treating real problems.

On the other hand, the ``collision term" in the average equation is the term that is responsible for anomalous dissipation and is thus the most interesting term in any theory that deals with the evolution of shock waves. For a spectrum of properly chosen fluctuations this term prevents large amplitude waves from indefinite steepening and breaking and provides the required dissipation of kinetic energy, entropy generation, and shock stabilisation. In its general version given above it should contain the whole physics of the shock including the complete collective processes which occur before real particle collisions come into play. 

However, the complexity of these equations is still too large for solving them. So one needs further simplifications in order to infer about the behaviour of shocks. The simplest and at the same time very effective simplification is to ask for the macroscopic conservation laws and the conditions of change of the plasma quantities across the shock transition layer which are in accord with the above fundamental kinetic equations. These are the magnetogasdynamic equations and the Rankine-Hugoniot jump relations. 

\subsubsection{Conservation laws}\index{conservation laws}
\noindent Following the philosophy of simplification we will first, before asking for the internal processes taking place in the shock transition, the generation of dissipation, particle reflection, entropy production etc., look into the global-- i.e. large-scale -- structure of a shock. In order to do this we need consider only the global plasma and field quantities, density $\langle N\rangle= N_0$, flow velocity $\langle {\bf V}\rangle = {\bf V}_0$ respectively momentum density $\langle N{\bf V}\rangle = N_0{\bf V}_0 $, pressure $\langle {\textsf P}\rangle = {\textsf P}_0$, magnetic field $\langle{\bf B}\rangle={\bf B}_0$, electric field $\langle{\bf E}\rangle={\bf E}_0$, current density $\langle{\bf j}\rangle ={\bf j}_0$, entropy ${\cal S}$ and so on. 

These quantities are all averages or result from average moments over the global distribution function $\langle F\rangle =F_0$. Since we will be dealing in the following only with average moments we suppress both the angular brackets and index 0. The prescription of taking moment of order $i$  is
\begin{equation}
M^{i}= \int {\rm d}v^3 {\bf v}^i F
\end{equation}
where ${\bf v}^i={\bf v}\dots{\bf v}$ is understood as the $i$-fold dyadic product. The first three moments are $N=\int {\rm d} v^3F$, $N{\bf V}=\int {\rm d} v^3{\bf v}F$, ${\textsf P}=m\int {\rm d} v^3({\bf v-V})({\bf v-V})  F$. Clearly, the diagonal of the pressure tensor ${\textsf P}$ gives the average energy density and also defines the local temperatures $T_\|, T_\perp$ parallel and perpendicular to the average magnetic field. Operating in the usual way with these definitions on the average Vlasov equation (\ref{chap2-eq-vlasovaverage}) produces the well-known full -- i.e. infinite -- set of magnetogasdynamic equations for the infinite chain of moments of $F^\pm$ for each particle species $\pm=e,i$. The first two of them are
\begin{eqnarray}\label{chap2-eq-momentequ}
\frac{\partial N^\pm}{\partial t}  +\nabla\cdot(N{\bf V})^\pm\!\!&\!\!=\!\!&\!\!0 \\
\frac{\partial (N{\bf V})^\pm}{\partial t}+\nabla\cdot(N{\bf V}{\bf V})^\pm+\frac{1}{m_\pm}\nabla\cdot{\textsf P}^\pm\!\!&\!\!=\!\!&\!\!\frac{e_\pm N^\pm}{m_\pm}({\bf E}+{\bf V}^\pm\times{\bf B})+\int{\rm d}v^3 {\bf v} {\cal C}^\pm
\end{eqnarray}
where by ${\cal C}$ the pseudo-collision term on the right of Eq.\,(\ref{chap2-eq-vlasovaverage}) is meant. Because this term conserves particle number (or mass) the zeroeth moment of it vanishes identically and does not contribute to the first (zero-order) of the above moment equations. Wave particle interaction neither changes particle number nor mass density. These are strictly conserved as is shown by the above  particle umber conservation equation.

In the first order moment equation it produces a wave friction term that has the explicit form
\begin{equation}\label{chap2-eq-qlcollisionterm}
-\frac{1}{m_\pm}\left\{\frac{1}{\mu_0}\frac{\partial}{\partial t}\langle{\bf e\times b}\rangle+\nabla\cdot\left[\left(\frac{\epsilon_0}{2}\langle{\bf e}^2\rangle+\frac{1}{2\mu_0}\langle{\bf b}^2\rangle\right)\textsf{I}-\left(\epsilon_0\langle{\bf ee}\rangle+\frac{1}{\mu_0}\langle{\bf bb}\rangle\right)\right]\right\}
\end{equation}
All these terms are in fact of nature ponderomotive force-density terms\index{force!ponderomotive} contributed by the average wave pressure gradients; the first term results from the wave Poynting moment, the second is the  gradient of a pure isotropic wave pressure, the third is related to wave pressure anisotropy. The inverse proportionality of this entire expression to the mass shows that the main contribution is due to the electron momentum density equation. The effect on the ions can be neglected as they are (in the non-relativistic case considered here) insensitive to ponderomotive effects. We note in passing, that it is this term which while affecting the motion of the electron gas will be responsible for the appearance of anomalous collisions, anomalous resistivity and viscosity, which we will discuss at a later occasion.

The two above equations do not form a complete system of equations. The first contains number density flux, the main constituent of the second equation which, as a new entity, contains the pressure. For ${\textsf P}$ one, in principle, can derive an energy conservation (heat conduction) equation which would contain the new quantity of heat flux, the next higher moment. On the other hand, one can replace the pressure equation that follows from the energy conservation law, by equations of state, ${\textsf P}(N,\gamma, T_\|, T_\perp)$, which express the pressure tensor components through density, temperature, adiabatic coefficient $\gamma$ etc. This is the usual procedure applied when investigating shock solutions. One should, however, be aware of the fact that equations of state in non-equilibrium are merely  approximations which hold under certain conditions of either isothermality -- which does not apply to shocks as they are not in thermal equilibrium -- or adiabaticity. The latter condition is quite reasonable in dealing with the fast processes taking part in the shock environment when the flow passes across the shock front in a time so short that thermalization becomes impossible.

The idea is to apply the momentum equations to an extended shock that represents a ramp in real space. In the spirit of our discussion in Chapter 1, the first step to do this is assuming that the shock is a thin planar discontinuity that moves at a certain shock velocity $U$ in the shock normal direction ${\bf n}$ across the plasma. \index{shocks!shock normal}If we confine all the micro-processes to the interior of the shock plane, i.e. if we go far enough away from the shock plane upstream and downstream, then we can apply the above dissipation-free average conservation laws to the shock and ask only for the differences in the plasma and field parameters between downstream and upstream of the shock, trying to express the downstream values in terms of the undisturbed upstream flow and field values.  In doing this, we completely neglect the ``pseudo-collision" terms on the right of these equations, since all physics that is going on will be confined to the transition region as wide as it can be. For a plane rigid stationary shock surface this assumption is good enough. However, when doing so, with the above separate conservation laws for electrons and ions, we immediately run into severe problems even in the simplest completely interaction-free case. The reason is that electrons and ions because of their different mass behave completely differently while at the same time cannot be treated separately as they are coupled through charge conservation and electrical neutrality and through their unequal contributions to the electric current density and therefore to the fields, a difficulty that has been discussed by \cite{Woods1971}.
\begin{figure}[t!]
\hspace{0.0cm}\centerline{\includegraphics[width=0.7\textwidth,clip=]{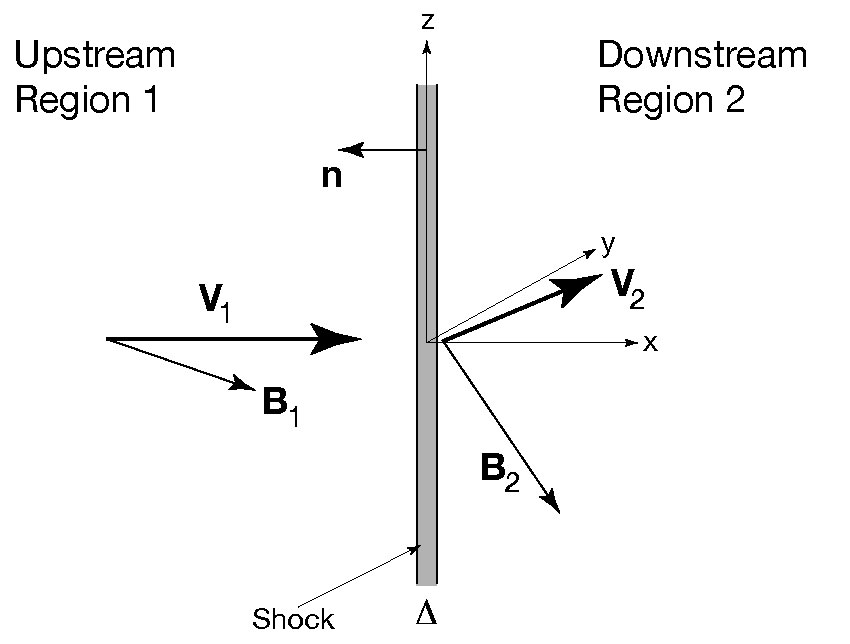} }
\caption[1]
{\footnotesize A sketch of the planar thin (width $\Delta$) shock geometry with ${\bf n}$ shock normal, upstream and downstream regions, bilk flow velocities and magnetic field vectors.}\label{chap2-fig-geometry}
\end{figure}

\section{Rankine-Hugoniot relations}\index{Rankine-Hugoniot relations}
\noindent In order to overcome this difficulty one is forced to further simplification of the conservation equations by adding up the electron and ion equations  \citep[cf., e.g.,][as for one of the many accounts available in the literature]{Baumjohann1996}. To this purpose one must define new centre-of-mass variables 
\begin{equation}
m=\sum\limits_\pm m_\pm = m_i\left(1+\frac{m_e}{m_i}\right), \quad N=\frac{\sum_\pm m_\pm N^\pm}{\sum_\pm m_\pm}, \quad {\bf V}=\frac{\sum_\pm m_\pm (N{\bf V})^\pm}{\sum_\pm m_\pm N^\pm}
\end{equation}
This leads to the magnetogasdynamic or MHD equations for a single-fluid plasma. Since the equation of continuity remains unchanged from Eq.\,(\ref{chap2-eq-momentequ}) it suffices to write down the momentum conservation equation
\begin{equation}\label{chap2-eq-onefluidmomentum}
\frac{\partial (mN{\bf V})}{\partial t} +\nabla\cdot(mN{\bf VV})=-\nabla\cdot {\textsf P} +\rho{\bf E}+{\bf j\times B}
\end{equation}
where ${\textsf P}={\textsf P_e}+{\textsf P_i}$ is the total pressure tensor, and $\rho$ is the electric charge density $\rho=e(N_i-N_e)$ which in quasi-neutral plasmas outside the shock is assumed to be zero such that the second term on the right containing the average electric field vanishes outside the shock ramp. The last term in this equation is the Lorentz force written in terms of the average current from Amp\`ere's law
\begin{equation}\label{chap2-eq-ampere}
\mu_0{\bf j}=\nabla\times {\bf B}
\end{equation}
The displacement current can be safely neglected because these equations hold only for very slow variations with frequency smaller than the ion cyclotron frequency $\omega\ll \omega_{ci}=eB/m_i$, scales much larger than the ion gyro-radius $L\gg r_{ci}= V_{i\perp}/\omega_{ci}$, and wave speeds much less than the speed of light. Note that this equation is completely collisionless. If we would have retained the pseudo-collision term on the right in the electron equation this would simply have added an electron ponderomotive force term on the right.

In fact, together with Maxwell's equations these equations are not yet complete in a double sense. They have to be completed with appropriate equations of state for the pressure components, as has been mentioned above, and they have to be completed by a relation between the current and the average electric field that appears in Maxwell's equations, i.e. with an appropriate Ohm's law. This is found by subtracting the electron and ion momentum conservation equations and turns out to be quite complicated \citep[cf., e.g.,][]{Krall1973}. In slightly simplified form Ohm's law reads
\begin{equation}\label{chap2-eq-ohmslaw}\index{Ohm's law}
{\bf E+ V\times B} =\frac{1}{eN}{\bf j\times B}-\frac{1}{eN}\nabla\cdot{\textsf P}_e+\frac{m_e}{e^2N}\frac{\partial{\bf  j}}{\partial t}
\end{equation}
Note that on the right only electron terms are contained in this expression. Also, an electron ponderomotive term -- responsible for anomalous transport effects -- would appear on the right if we would retain the pseudo-collision term.

However, even in this form even though the system is non-collisional Ohm's law is still too complex for  treating the conservation laws at a shock transition. The reason is that the right-hand side introduces second order spatial derivatives into Faraday's law through the pressure gradient and current expressions. One therefore argues that for sufficiently flat shock transitions the terms on the right can be neglected. This argument implies that one must go far enough away from the shock into a region where shock excited turbulence has decayed away in order to apply global conservation laws to the shock transition. this can be done when only the left-hand side in Ohm's law is retained and the ideal MHD frozen-in condition holds:
\begin{equation}\label{chap2-eq-idealohmslaw}\index{Ohm's law!ideal}
{\bf E = - V\times B}
\end{equation}
Assuming that the shock is plane and narrow as shown in Figure\,\ref{chap2-fig-geometry} such that any variations along the shock can be ignored and the sole variation is along the shock normal, Eqs. (\ref{chap2-eq-onefluidmomentum}-\ref{chap2-eq-idealohmslaw}), the continuity equation and Maxwell's equations become all one-dimensional and can be integrated along ${\bf n}$ over the shock transition (with regular boundary conditions at $x=\pm\infty$). Applying the definition of the shock normal ({\ref{chap1-eq-shocknormal}) in Chapter 1 and\index{shocks!shock normal} the prescription for the $\nabla$-operator in Eq. (\ref{chap1-eq-nabla}) transforms these equations into a nonlinear algebraic system of equations for the jumps $[\dots]$ of the field quantities\index{Rankine-Hugoniot relations!jump conditions}
\begin{eqnarray}\label{rankine-hugoniot1}
{\bf n}\cdot[N{\bf V}]&=& 0 \nonumber \\
{\bf n}\cdot[mN{\bf VV}]+{\bf n}\left[P+\frac{B^2}{2\mu_0}\right]-\frac{1}{\mu_0}{\bf n}\cdot[{\bf BB}]&=&0\nonumber  \\
 \left[{\bf n}\times {\bf V} \times {\bf B}\right] &=&0 \\
{\bf n\times[ B ]} &=& 0 \nonumber \\
\left[mN{\bf n\cdot V}\left\{\frac{V^2}{2}+w+\frac{1}{mN}\left(P+\frac{B^2}{\mu_0}\right)\right\}-\frac{1}{\mu_0}{\bf (V\cdot B) n\cdot B}\right] &=& 0 \nonumber
\end{eqnarray}
Here, for simplicity, the pressure has been assumed isotropic. The last equation is the rewritten energy conservation equation, where $w=c_vP/k_BN$ is the ideal gas enthalpy density, $c_v$ the specific heat, and $k_B$ Boltzmann's constant. This system of equations is the implicit form of the {\it Rankine-Hugoniot} conservation equations in ideal magnetogasdynamics (ideal MHD). In this version it contains all ideal MHD discontinuities of which shock waves are a subclass, the class of solutions with a finite flow across the discontinuity, compressions (in density), and increases in temperature $T$, pressure $P$, and entropy ${\cal S}$ across the discontinuity in the transition from upstream to downstream.

\subsection{Explicit MHD shock solutions}
\noindent We are not interested in the full set of solutions of the above system of jump conditions (\ref{rankine-hugoniot1}). We rather look for genuine shock conditions. This requires finite mass flux ${\cal F}=NV_n$ across the shock in the normal direction. The first of the Rankine-Hugoniot relations (\ref{rankine-hugoniot1}) tells that the jump $[{\cal F}]=0$. Hence ${\cal F}=$\,const, and we must sort for solutions with ${\cal F}\neq 0$, or $N_1V_{n1}=N_2V_{n2}$, in order to be dealing with a shock.

Introducing the specific volume ${\cal V}= (mN)^{-1}$, the whole system of jump conditions can be factorized \citep{Baumjohann1996} and can be written in the form
\begin{equation}\label{chap2-eq-RHfactor}
{\cal F}\left({\cal F}^2-\frac{B_n^2}{\mu_0\langle{\cal V}\rangle}\right)\left\{{\cal F}^4+{\cal F}^2\left(\frac{[P]}{[{\cal V}]} -\frac{\langle{\bf B}\rangle^2}{\mu_0\langle{\cal V}\rangle}\right)-\frac{B_n^2}{\mu_0\langle{\cal V}\rangle}\frac{[P]}{[{\cal V}]}\right\}=0
\end{equation}
In the one-fluid approximation magnetogasdynamic shock waves with ${\cal F}\neq 0$ are contained in the expression in curly braces which still depends on the jumps in pressure $[P]$ and specific volume $[{\cal V}]$ and thus on the energy conservation equation respectively the equation of state. We will not discuss this equation further as in the following more insight can be gained from explicit consideration of a few particular cases.

Under the special condition that the flow in the upstream Region 1 is along $x$ (anti-parallel to ${\bf n}$) and the upstream magnetic field ${\bf B}_1=(B_{1x}, 0,B_{1z})$ is in the $xz$-plane, and assuming ${\cal F}\neq 0$, the jump conditions Eqs.\,(\ref{rankine-hugoniot1}) simplify. Since $V_{1}, B_{1x}, B_{1z}, P_1$ are known quantities, it is convenient to introduce normalised variables for the corresponding downstream values
\begin{displaymath}
\frac{N_2}{N_1}\to N_2, \quad \frac{{\bf V}_2}{V_1}\to {\bf V}_2, \quad \frac{T_{1,2}}{mV^2_1/2}\to T_{1,2}, \quad \frac{{\bf B}_{1,2}}{\sqrt{\mu_0mN_1V_1^2}}\to {\bf B}_{1,2}
\end{displaymath}
where the temperature is taken in energy units. Instead of it we may also use the corresponding thermal speeds $v_{1,2}$ which by the above normalisation are normalised to $V_1$. This then yields the following normalised Rankine-Hugoniot relations, in which $B_n=$\,const as a consequence of the vanishing divergence of the magnetic field,
\begin{eqnarray}\label{chap2.eq-rhnormalized}
N_2V_{n2} &=& 1\nonumber \\
V_{n2}B_{z2}-V_{z2}B_{n} &=& B_{z1} \nonumber \\
B_{z2}B_n-V_{z2} &=& B_{z1}B_n \\
2N_2(v_2^2+V_{n2}^2)+B_{z2}^2 &=& 2(1+v_1^2) + B_{z1}^2\nonumber \\
V_{n2}^2+V_{z2}^2+2B_{z2}B_{z1} +5v_2^2 &=& 1+2B_{z1}^2+5v_1^2 \nonumber
\end{eqnarray}
The energy conservation equation yields the last in these expressions. There the enthalpy is taken into account giving a factor 5 in front of the thermal velocities. These five equations can be combined into a third-order equation for one of the downstream unknown quantities, for instance $V_{n2}$, expressed in terms of the upstream values
\begin{equation}\label{chap2-eq-cubic}
a_3V_{n2}^3+a_2V_{n2}^2 +a_1V_{n2}+ a_0=0
\end{equation}
where $a_0=-B_n^2[b_{z1}^2+B_n^2(1+5v_1^2)],a_1=2B_n^2(1+2{\bf B}_1^2+5v_1^2)-\frac{1}{2}B_{z1}^2, -a_2=1+5v_1^2+8B_n^2+\frac{5}{2}B_{z1}^2, a_3=4$. Below we discuss a few simple illustrative solutions of this equation.

\subsection{Perpendicular shocks} \noindent \index{shocks!perpendicular}For strictly perpendicular shocks we have $B_n=0, B_{z1}=B_1, a_0=0, a_1=-\frac{1}{2}B_1^2, -a_2=1+5v_1^2+\frac{5}{2}B_{z1}^2, a_3=4$. Equation\,(\ref{chap2-eq-cubic}) turns into a quadratic equation yielding the solution\index{shocks!perpendicular}
\begin{equation}
V_{n2}=\frac{1}{8}\left\{1+\left(1+\frac{5}{2}\beta_1\right)B_1^2+\left[\left(1+\left(1+\frac{5}{2}\beta_1\right)B_1^2\right)^2+2B_1^2\right]^\frac{1}{2}\right\} =\frac{1}{N_2}=\frac{1}{B_{2}}
\end{equation}
Since the condition for a shock to exist is that the normal velocity $V_{n2}<1$ in Region 2 we immediately conclude that  in a perpendicular shock the density and tangential magnetic field components in Region 2 increase by the same fraction as the normal velocity drops, and this fraction is determined by the plasma-$\beta$ ratio $\beta_1=2\mu_0N_1T_1/B_1^2$ in Region 1, where $T_1=T_{e1}+T_{i1}$ is the total temperature. The condition on $V_{n2}$ implies that the Mach number takes the form (now in physical units)
\begin{equation}
1< {\cal M}=\frac{1}{1+5\beta_1/6}\frac{V_1}{V_{A1}}=\frac{1}{1+5\beta_1/6}{\cal M}_A
\end{equation}
Here ${\cal M}_A$ is the Alfv\'en-Mach number which is the flow to Alfv\'en velocity ratio. In cold plasmas or plasmas containing strong magnetic fields $\beta\ll 1$, and the Mach  number is simply the Alfv\'en-Mach number. Conversely, in hot plasmas the Mach number becomes about the ordinary gasdynamic Mach number. 

For the increase in normalised temperature one finds accordingly
\begin{equation}
\frac{T_2}{T_1}=1+\frac{2}{5T_1}\left[1-N_{2}^{-2}+2{\cal M}_A^{-2}\left(1-N_{2}\right)\right] >1
\end{equation}
This is always larger than one. Perpendicular shocks cause plasma heating during shock transition time and thus cause also increase in entropy
\begin{equation}
\Delta{\cal S}\propto \ln \left[\left(\frac{T_2}{T_1}\right)^\frac{1}{\gamma-1}\frac{1}{N_2}\right]
\end{equation}
which holds under the ideal gas assumption. 

\subsection{Parallel shocks} \noindent \index{shocks!parallel}This case is not well treated in magnetogasdynamics conditions as we have explained earlier. Since the magnetic field is normal to the shock it is theoretically unaffected by the presence of the shock which therefore should become purely gasdynamic. In the above perpendicular shock jump conditions one can for this case simply delete the magnetic terms. However, this does not cover the real physics involved into parallel shocks which must be treated on the basis of kinetic theory and with the simulation tool at hand.\index{shocks!parallel}

\subsection{High Mach numbers} \noindent This limit applies when the ram pressure of the flow is very high and exceeds the thermal pressure. Then all terms including $v_{1,2}$ can be neglected. Moreover, one usually also neglects the magnetic field in this case, and the shocks become then purely flow determined with $V_{n2}=N_2^{-1}\sim B_2^{-1}\simeq \frac{1}{4}$, suggesting that both, the magnetic field and density should not increase by more than a factor of 4. \index{shocks!high-Mach number}

In fact, the observations in interplanetary space indeed confirm that all shock that have been observed there are weaken than this. However, again, this reasoning does not really apply in plasmas because at very high Mach numbers other effects come into play which are connected with the kinetic nature of a plasma, electrodynamic effects, and the differences in electron and ion motion. Ultimately relativistic effects must be taken into account. These become susceptible first for electrons, increasing their mass but at the same time distinguishing them even stronger from the inert ions, because the electron dynamics changes completely in the relativistic domain. \index{shocks!relativistic}

In addition, high Mach number shocks are supercritical and even though it seems that one could treat them in the simple way as has been done here, the kinetic effects involved into their physics inhibit such a simplistic interpretation of high-Mach number shocks. High-Mach number shocks readily become turbulent, exciting various kinds of waves which grow to large amplitudes and completely modify the environment of the shock which cannot be treated any more as quiet. In such a turbulent environment shocks assume intermittent character losing stationarity or even identity as a single ramp which the flow has to surpass when going from Region 1 to Region 2. Occasionally the distinction between two regions only may become obsolete. There might be more than one transition regions, subshocks form, the ramp will evolve its own structures. And these structures come and go, are temporarily created and damp away to make space for the evolution of other new structures. Probably, high Mach number shocks exist only temporarily at one and the same spatial location. They are highly dynamical, changing their nature, structure, shape, steepness and intensity along the surface of the shock such that they strongly deviate from one-dimensionality and even from two-dimensionality. They are time-dependent, reforming themselves continuously in different regions of space and thus cannot be described by a simple plane shock geometry of the kind we have assumed. Later in this book at the appropriate place we will consider moderately high Mach number shocks when dealing with the extended class of supercritical shocks.\index{shocks!subcritical}\index{shocks!supercritical}
\vspace{-0.1cm}
\subsubsection{Oblique shocks} \noindent Real shocks do not belong neither to the very particular classes of parallel nor perpendicular shocks. Real shocks are oblique in the sense that the upstream magnetic fields ${\bf B}_1$ are inclined with respect to the shock normal ${\bf n}$. As mentioned earlier, one distinguishes between {\it quasi-parallel} and {\it quasi-perpendicular} shocks depending on the shock normal angle $\thetabn$ being closer to $0^\circ$ or $90^\circ$. Since we will treat the properties of these shocks separately in some following chapters, we are not going to discuss them at this place.\index{shocks!oblique}

Shocks around finite size obstacles will never be really plane. The best approximations to plane shocks are interplanetary shocks. Bow shocks in front of magnetised planets, comets or other bodies are always curved. They assume all kinds of shock properties along their surfaces reaching from perpendicular to oblique and parallel. Curving their surface implies that the shock normal changes its angle with respect to the direction of the upstream flow ${\bf V}_1$. 

Since shocks evolve for Mach numbers ${\cal M}_{ms}=V_1/c_{ms} >1$ they occupy  a finite volume of space only.  For shock formation the Mach number based on $V_{1n}$ normal to the shock is relevant. Defining the flow normal angle  $\cos\Theta_{\,Vn} =|{\bf V}_1{\bf\cdot n}|$, formation of a bow shock in front of a finite size object like a magnetosphere is restricted to flow normal angles 
\begin{equation}
\Theta_{\,Vn}<\cos^{-1}({\cal M}_{ms}^{-1})
\end{equation}
For instance, at a nominal Mach number ${\cal M}_{ms}\sim 8$ shocks exist for angles $\Theta_{\,Vn}<82.8^\circ$.

\section{Waves and Instabilities}\index{instability}
\noindent It has been mentioned several times that shocks evolve from waves mainly through nonlinear wave steepening and the onset of dissipation and dispersion. Moreover, it is the various modes of waves that are responsible for the generation of anomalous dissipation, shock ramp broadening, generation of turbulence in the shock environment and shock ramp itself, as well as for particle acceleration, shock particle reflection and the successive effects. The idea is that in a plasma that consists of electrodynamically active particles the excitation of the various plasma wave modes in the electromagnetic field as collective effects is the easiest way of energy distribution and transport. There is very little momentum needed in order to accelerate a wave, even though many particles are involved in the excitation and propagation of the wave, much less momentum than accelerating a substantial number of particles to medium energy. Therefore any more profound understanding of shock processes cannot avoid bothering with waves, instabilities, wave excitation and wave particle interaction. 

\subsection{Dispersion relation}\index{dispersion relation}
\noindent Waves are a very general phenomenon of most media. However, they do not fall from sky. Instead, they evolve from small thermal fluctuations in the medium. Such fluctuations are unavoidable. In order for a wave to propagate in the medium a number of conditions need to be satisfied, however. The first is that the medium allows for a particular range of frequencies $\omega$ and wave-vectors ${\bf k}$ to exist in the medium; i.e. it allows for eigen-oscillations or eigen-modes. These ranges are specified by the dispersion relation $ D(\omega, {\bf k},\dots)=0$ which formulates the condition that the dynamical equations of the medium possess small-amplitude solutions. This dispersion relation is usually derived in the linear infinitesimally small amplitude approximation. However, nonlinear dispersion relations can sometimes also be formulated in which case ${D}$ depends on the amplitude as well. 

Plasmas are electromagnetically active media with the electromagnetic field governed by Maxwell's equations. Since there the plasma properties enter only through the material equations (i.e. current density ${\bf j}$, space charge $\rho$), the dispersion relation is most easily obtained from them. Moreover both, the current and space charge in a plasma, depend on the number densities in the plasma; i.e. the space charge variation can be included into the current variation. It is then simple matter to derive the general electromagnetic wave equation for the fluctuating fields, ${\bf e}$, on a much slower evolving background, $\langle{\bf B, E}\rangle$,
\begin{equation}\label{chap2-eq-genwaveequ}
\nabla^2{\bf e}-\nabla(\nabla\cdot{\bf e})-\epsilon_0\mu_0\frac{\partial^2{\bf e}}{\partial t^2} =\mu_0\frac{\partial {\bf j}}{\partial t}
\end{equation}
The magnetic fluctuation field ${\bf b}$ is completely determined from Maxwell's equations, and the current appearing on the right is expressed conveniently through the space-time dependence of the fluctuation-conductivity tensor ${\bf \sigma}({\bf x}, t)$ as
\begin{equation}\label{chap2-eq-gencurrent}\index{waves!wave conductivity}
{\bf j}(({\bf x}, t)=\int {\rm d}x'\int\limits_{-\infty}^t {\rm d}t'{\bf \sigma}({\bf x-x'},t-t')\cdot{\bf e}
\end{equation}
an expression that implicitly accounts for causality due to the integration over the entire past of the current up to the observation time $t$. Since the fluctuation current is a functional of the complete set of particle distribution functions (through the zeroeth, $N$, and first, $N{\bf V}$, moments) the complete evolution of the fluctuations up to a large amplitude shock is contained in these expressions. However, for practical purposes one linearizes this equation by assuming that the fluctuation-conductivity tensor is, to first order, independent of the fields ${\bf e, b}$. The above equation (\ref{chap2-eq-genwaveequ}) becomes linear under this assumption and can be Fourier analysed, with wave vector ${\bf k}$ and frequency $\omega$, yielding (with $c^2=1/\mu_0\epsilon_0$ the square of the speed of light)
\begin{equation}
\left[\left( k^2-\frac{\omega^2}{c^2}\right){\textsf I}-{\bf kk} -i\omega\mu_0{\bf \sigma}(\omega,{\bf k})\right]\cdot{\bf e}(\omega,{\bf k})=0
\end{equation}
The quantities in this expression satisfy the following symmetry relations ${\bf e}(-\omega,-{\bf k})={\bf e}^*(\omega, {\bf k}), {\bf \sigma}(-\omega, -{\bf k})={\bf \sigma}^*({\omega, {\bf k}})$. Setting the expression in brackets to zero yields the equation for the linear eigenmodes $\omega=\omega({\bf k})$. For convenience we define the dielectric tensor\index{waves!dielectric tensor}
\begin{equation}\label{chap2-eq-espilongeneral}
{\bf \epsilon}(\omega, {\bf k})\equiv{\textsf I}+\frac{i}{\omega\epsilon_0}{\bf \sigma}(\omega, {\bf k})
\end{equation}
which satisfies the same conditions as the fluctuation conductivity. Then we can finally write the general dispersion relation in the compact form as the determinant of the bracketed expression 
\begin{equation}\label{chap2-eq-generaldispersionrelation}
D(\omega,{\bf k} )\equiv {\rm Det}\left[\left( k^2-\frac{\omega^2}{c^2}\right){\textsf I}-{\bf kk} +{\bf \epsilon}(\omega,{\bf k})\right]=0
\end{equation}
The particular linear physics of the plasma is contained in the dielectric tensor through the conductivity. (We note in passing that for any classical medium the above conditions together with the dispersion relation are the equivalent of the well-known Kramers-Kronig relations of causal fluctuations in quantum mechanics \citep[see, e.g.,][]{Landau1980}.) In addition to the solution of the dispersion equation (\ref{chap2-eq-generaldispersionrelation}) the first problem consists in the determination of the linear fluctuation conductivity tensor which enters the dielectric tensor (\ref{chap2-eq-espilongeneral}). For this one needs to go to the appropriate plasma model. However, we repeat  that without this linear step one cannot obtain any susceptible information about the nature of a shock wave. This we have explained in breadth in Chapter 1 and the preceding sections. 

The linear dispersion relation Eq.\,(\ref{chap2-eq-generaldispersionrelation}) has plane wave mode solutions of the form $\propto \exp i({\bf k\cdot x}-\omega t)$, which are eigenmodes of the particular plasma model which is described by the kinetic equations (or appropriate simplifications of the kinetic equations) of the plasma. The dynamics of the plasma enters  through the wave conductivity tensor which can be determined from the Fourier transformed expression for the current density\index{dispersion relation!linear}
\begin{equation}
{\bf j}({\bf k},\omega)=\sum\limits\pm e_\pm\int {\rm d}v^3 {\bf v} f^\pm ({\bf v, k},\omega) = {\bf\sigma}({\bf k},\omega)\cdot{\bf e}({\bf k},\omega)
\end{equation}
as the first moment of the fluctuating part $f^\pm$ of the distribution function $F^\pm$. Solving for the integral and expressing $f^\pm$ through the electric wave field ${\bf e}$ yields the wanted form of the wave conductivity tensor ${\bf \sigma}$.  The problem is thus reduced to the determination of $f^\pm$ from  Eq.\,(\ref{chap2-eq-fluctuationvlasov}) where we drop the average terms on the right-hand side retaining only terms linear in the fluctuations:
\begin{equation}
\frac{\partial f^\pm}{\partial t} + {\bf v}\cdot\nabla f^\pm +\frac{e_\pm}{m_{\pm}}({\bf E}_0+{\bf v}\times{\bf B}_0)\cdot\nabla_{\bf v}f^\pm =-\frac{e_\pm}{m_{\pm}}({\bf e}+{\bf v}\times{\bf b})\cdot\nabla_{\bf v}F_0^\pm
\end{equation}
Operating with a Fourier transform on this equation then yields the following expression 
\begin{equation}
\left[1 -\frac{ie_\pm}{m_\pm({\bf k\cdot v}-\omega)}({\bf E}_0+{\bf v}\times{\bf B}_0)\cdot\nabla_{\bf v}\right]f^\pm=\frac{ie_\pm}{m_{\pm}} \frac{i({\bf e}+{\bf v}\times{\bf b})\cdot\nabla_{\bf v}F_0^\pm}{{\bf k\cdot v}-\omega}
\end{equation}
which determines $f^\pm$ in terms of the average and fluctuating field quantities, which is just what we want. One can now make assumptions about the average fields and distribution function in order to explicitly calculate $f^\pm$. Usually these assumptions are ${\bf E}_0=0, {\bf B}_0=B_0\hat{\bf z}$ with $B_0=$\,const. Then the operator $\nabla_{\bf v}=-\hat{\bf z}\partial /\partial\phi$ on the left simplifies to a mere derivative with respect to the gyration angle $\phi$ of the particles. Further assumptions on $F^\pm$ are that the average distributions are gyro-tropic, in which case the integration with respect to $\phi$ becomes trivial. 

With such assumptions it is not difficult so still tedious to solve for $f^\pm$ and finally get the conductivity tensor. The respective expressions have been given in various places and will not repeated here. Good references among others are \cite{Montgomery1964}, \cite{Gary1993}, \cite{Baumjohann1996}. One can even include weak inhomogeneities in the average distribution function and fields which is needed when considering an inhomogeneous initial state like a given soliton or shock structure and investigating its prospective stability. In this case the plasma background is not homogeneous anymore because a soliton or shock has already evolved in it and has locally modified the plasma. Any wave modes which will be excited on this modified background will then not be influenced only by sideband formation\index{waves!sideband formation}, steepening\index{waves!steepening}, nonlinearity and dispersion\index{waves!dispersion} but also by the change of the plasma properties from location to location. This implies that the waves themselves change character  and properties across a shock.

\subsubsection{Damping/growth rate}\noindent\index{waves!damping rate}\index{waves!growth rate}
The solutions of the dispersion relation are in most cases complex, and for real wave vector ${\bf k}$ can be written as $\omega({\bf k})=\omega_r({\bf k})+i\gamma(\omega_r, {\bf k}) $, where the index $r$ indicates the real part, and $\gamma$ is the imaginary part of the frequency which itself is a function of the real frequency and wave number, because each mode of given  frequency can behave differently in time, and the wave under normal conditions will be dispersive, i.e. it will not be a linear function of wave number. In most cases the amplitude of a given wave will change slowly in time, which means that the imaginary part of the frequency is small compared to the real frequency. If this is granted, then $\gamma$ can be determined by a simple procedure directly from the dispersion relation $D(\omega, {\bf k}) =D_r(\omega, {\bf k}) +iD_i(\omega, {\bf k})$, which can be written as the sum of its real $D_r$ and imaginary $D_i$ parts  because a small imaginary part $\gamma$ in the frequency changes the dispersion relation only weakly, and it can be expanded with respect to this imaginary part. Up to first order in $\gamma/\omega$ one then obtains
\begin{equation}\label{chap2-eq-growthrategeneral}
D_r(\omega_r,{\bf k})=0, \qquad\quad \gamma(\omega_r,{\bf k})=-\frac{D_i(\omega_r,{\bf k})}{\partial D_r(\omega_r,{\bf k})/\partial \omega |_{\gamma=0}}
\end{equation}
The first of these expressions determines the real frequency as function of wave number $\omega_r({\bf k})$ which can be calculated directly from the real part of the dispersion relation. The second equation is a prescription to determine the imaginary part of the frequency, i.e. the damping or growth rate of the wave.
\begin{figure}[t!]
\hspace{0.0cm}{\includegraphics[width=0.4\textwidth,clip=]{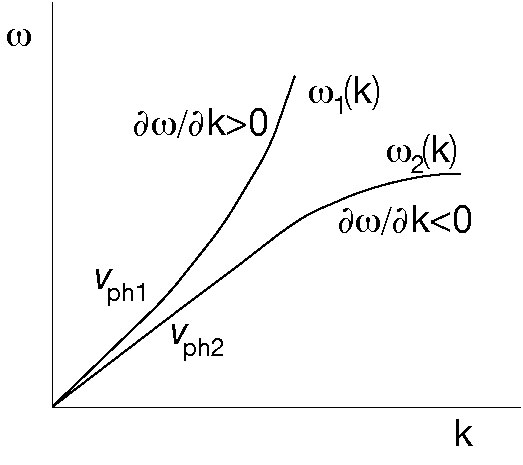} }
\hspace{0.4cm}{\includegraphics[width=0.5\textwidth,clip=]{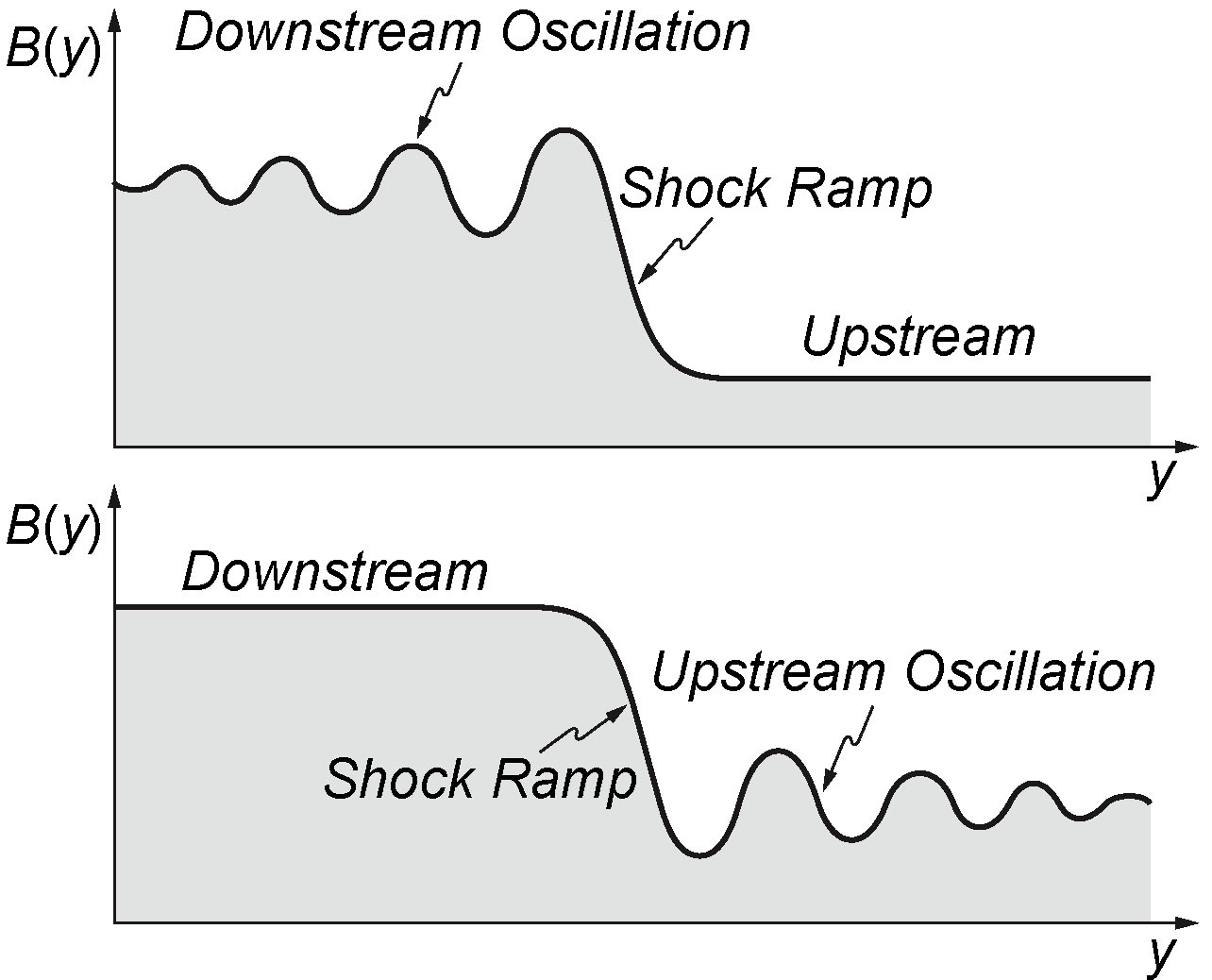} }
\caption[1]
{\footnotesize {\it Left}: Two different types of dispersions in the (real) ($\omega_r,{\bf k}$)-plane. Short waves with concave dispersion have slower group velocity than long waves and form a trail on the long wave. Short waves with convex dispersion move faster than long waves. {\it Right}: The effect of this difference in sidewave velocity on a laminar subcritical shock wave. Convex dispersions produce run-away waves which appear as spatially damped upstream oscillations (in the B-field, for instance). The trailing waves resulting from concave dispersion occur as downstream spatially damped oscillations. Maximum wave amplitudes are observed near the shock ramp in both cases.}\label{chap2-fig-dispersionsteep}
\end{figure}

\subsubsection{Remarks}\noindent\index{dispersion relation!nonlinear}
Two remarks on the dispersion relation are in place. First, the weak damping/growth rate solutions $\omega=\omega({\bf k})$ of the above general dispersion relation -- themselves called dispersion relations -- are also of use in the weakly nonlinear case. They can be understood as the lowest-order expansion term of a more general nonlinear dispersion relation $\omega=\omega({\bf k},|{\bf e, b}|^2 )$ which depends weakly on the wave amplitude or wave energy $|{\bf e, b}|^2$. When taking into account higher order expansion terms in the wave amplitude it produces other non-linear equations which govern the amplitude evolution of the wave under consideration. Such an equation is the non-linear Schr\"odinger equation we will get familiar with when discussing transport processes.

Second, from the dispersion relation $\omega_r=\omega_r({\bf k})$ one can infer in which way steepening of a wave is compensated by the dispersion of the wave. Figure\,\ref{chap2-fig-dispersionsteep} on its left shows two typical cases of (real) dispersion curves of low frequency waves in the ($\omega_r,k$)-plane from which shock waves could evolve \cite[after][]{Sagdeev1966}. Both curves have in common that they exhibit linear dispersion at long wavelengths, i.e. at small wave-numbers $k$, with slope giving the phase velocities of the waves. In this region all nonlinearly generated sidebands have same phase and group velocities causing broadening of the wave spectrum and steepening. However, at higher wave-numbers the dispersion curves start diverging from linear slope, one of the waves turning convex, the other concave. These turnovers imply a change in phase and group velocities. The convex dispersion implies that shorter wavelengths generated in the convex part\index{dispersion relation!convex}\index{dispersion relation!concave} of the dispersion curve move faster than the long waves. They will thus catch up with the long wavelength wave and run away ahead of the wave forming upstream precursors of the wave as shown for the shock in the lower part on the right. On the other hand, for the concave dispersion shorter wavelength waves fall behind the long waves. They represent a wave trail following the large amplitude long wave as is shown for the shock in the upper part on the right. Hence a simple glance at the dispersion curves already unmantles the possible properties of the expected nonlinearity and the structure of the shock. 

A word of caution is in place here, however. This reasoning does not hold for all shocks but for subcritical laminar shocks only. Supercritical higher Mach number shocks will behave in a more complicated way being much less dependent on dissipation and dispersion.

\subsection{The MHD modes - low-$\beta$ shocks}\index{shocks!low-$\beta$}
\noindent The waves from which a shock forms are the lowest-frequency plasma modes that are excited under the particular conditions of the shocked plasma. In low-$\beta$ (cold) plasma these are the tree fundamental MHD modes. Since all collisionless shocks in the heliosphere are magnetised the magnetic field has to be included, and the global shocks are not purely electrostatic even though subshocks developing in them can well behave approximately electrostatic.

We are already familiar with the three low-$\beta$ magnetogasdynamic modes, the fast and slow magnetosonic and the intermediate Alfv\'en waves. These are the lowest frequency eigenmodes of a homogeneous not necessarily isotropic plasma, i.e. when a small disturbance is present in the plasma it will propagate in one or all of these modes. Their dispersion relation follows from
\begin{equation}\label{chap2-eq-mhddisp}
D_{\rm mhd}(\omega, {\bf k})=\left(
\begin{array}{ccc}
 \omega^2-V_A^2k_\|^2-c_{ms}^2k_\perp^2 & 0  & -c_s^2k_\|k_\perp   \\
 0 &  \omega^2-V_A^2k_\|^2 &  0 \\
 -c_s^2k_\|k_\perp & 0  &   \omega^2-c_s^2k_\|^2
\end{array}
\right)=0
\end{equation}
which depends on the parallel and perpendicular components of ${\bf k}$ only in a very simple way. Moreover, it is a purely real dispersion relation lacking any imaginary part and therefore also any damping which is of course typical for a low frequency dissipation-free plasma. It is a different question of how these modes can be excited, and we will come to this at a later stage. 

The phase velocity $c_{ms}$  of these modes has been dealt with already in Eq.\,(\ref{chap1-eq-cms}). Since these waves are linear waves with no dispersion, their dispersion relation is simply $\omega=kc_{ms}(\theta)$ with $\theta$ the angle between the wave vector ${\bf k}$ and magnetic field ${\bf B}$. Figure\,\ref{chap2-fig-mhdmodes} shows the real space angular dependence of these three phase velocities for two special cases. Clearly in the direction perpendicular to the magnetic field only the fast mode propagates and, hence, strictly perpendicular  MHD shocks are fast shocks\index{shocks!perpendicular}\index{shocks!fast} as has been noted. In the  direction parallel to ${\bf B}$ all three waves can propagate.
\begin{figure}[t!]
\hspace{0.0cm}\centerline{\includegraphics[width=0.8\textwidth,clip=]{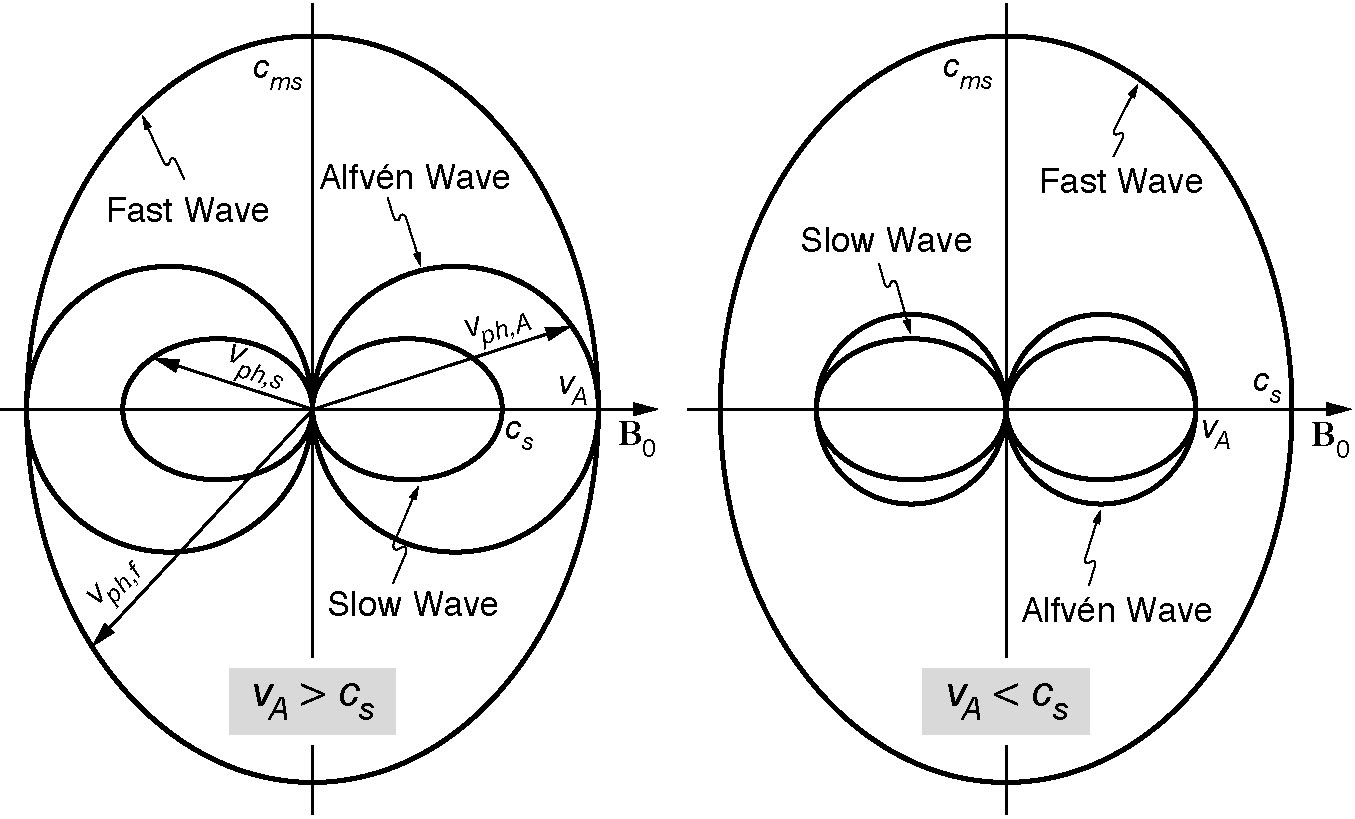} }
\caption[1]
{\footnotesize Wave vector diagram of two cases of  MHD waves in the plane of the magnetic field.}\label{chap2-fig-mhdmodes}
\end{figure}

\subsubsection{Magnetosonic solitons in cold plasma}\noindent\index{soliton!magnetosonic}
We will now show that the method of the Sagdeev pseudo-potential can be used to understand the formation of a fast mode solitary wave (or soliton) propagating strictly perpendicular to the magnetic field ${\bf B}$. This has first been shown by \cite{Davis1958}. What results from this procedure will not yet be a shock, because, as we have noted before, shock formation requires the presence of some kind of dissipation, while the equations on which the present theory is based are strictly dissipation-free (remember that we have dropped the correlation terms on the right-hand side of the kinetic equation before deriving the moment equations and that we have not yet discussed any way of how dissipation occurs when these terms are taken into account). 

In order to find the stationary solutions we are looking for, one must retain the nonlinearity in the stationary one-dimensional quasi-neutral magnetogasdynamic equations. This nonlinearity appears in the convective term $V_n\nabla_n V_n$ in the equation of motion. From constancy of normal flux $[{\cal F}]=0$ one has $NV_n=N_1V_1$ and $E_y=N_1B_1$ where the index 1 means undisturbed values far upstream. Since only electrons contribute to the current by their drift in the crossed electric and magnetic fields ${\bf E=- V\times B, B}=B\hat{\bf z}$ in the shock frame,  we must retain a small component $E_n=-BV_y$ across the shock, effectively produced by the difference in electron and ion motion and causing the shock current to flow in $y$-direction in the shock transition. It  occurs in the stationary equation of motion on the scale of the shock transition, i.e. on the scale of the ion gyro-radius, and as current ${\bf j}_y=-eNV_y\hat{\bf y}$ in Amp\`ere's law
\begin{equation}
mNV_n\frac{{\rm d}V_n}{{\rm d}x}= -eNBV_y, \qquad\qquad \frac{{\rm d}B}{{\rm d}x}=-\mu_0eNV_y
\end{equation}
Combining these equations yields the normal fluid velocity as function of the magnetic field and the initial bulk flow velocity $V_1$ at infinity
\begin{equation}\label{chap2-eq-vnormal}
V_n=V_1\left(1-\frac{B^2-B_1^2}{2\mu_0mN_1V_1^2}\right)=V_1\left(1-\frac{1}{2}\frac{V_A^2-V_{A1}^2}{V_1^2}\right)
\end{equation}
showing that the bulk velocity decreases from $V_1$ when the magnetic field $B$ increases. The second term in the first parentheses is the difference in the ratio of magnetic pressures $B^2/2\mu_0$ at the location under observance and $B_1^2/2\mu_0$ at infinity upstream to the kinetic pressure $mNV_1^2$ at infinity. This is written in terms of Alfv\'en velocities in the second parentheses on the right. 

With the help of this expression, $E_y$ and Amp\`ere's law we obtain
\begin{equation}
\mu_0eN_1V_y=\left[\frac{B^2-B_1^2}{2\mu_0mN_1V_1} -1\right]\frac{{\rm d}B}{{\rm d}x}
\end{equation}
which can be used to eliminate the velocity components and obtain an equation for the variation in the magnetic field given in the form of the energy conservation equation of a pseudo-particle of mass 1 and velocity $B$ in the Sagdeev pseudo-potential $S(B)$:
\begin{equation}\label{chap2-eq-bfield-sagdeev}
\frac{1}{2}\left(\frac{{\rm d}B}{{\rm d}x}\right)^2= -S(B), \quad S(B)=-\frac{(B-B_1)^2[(B+B_1)^2/4\mu_0mN_1V_1^2-1]}{2\lambda_e^{2}[1-(B^2-B_1^2)/2\mu_0mN_1V_1^2]^2}
\end{equation}
where $\lambda_e=c/\omega_{pe}$ is the electron inertial length (electron skin depth). The electron inertial length is the only length scale that appears in the above equation when we perform a dimensional analysis. It therefore turns out that the\index{shocks!width $\Delta$} characteristic width, $\Delta\sim \lambda_e$, of the magnetosonic solitons is of the order of the electron skin depth. This can also be seen when for small amplitude disturbances the last expression is expanded with respect to $B$. Defining $b=(B-B_1)/B_1, b_m=(B_m-B_1)/B_1, \xi=x/\Delta$ yields to first order\index{Sagdeev pseudo-potential}
\begin{equation}
\frac{{\rm d}b}{{\rm d}\xi}\simeq \pm b(b_m-b)^\frac{1}{2}\qquad \rightarrow \qquad b\simeq \frac{4b_m}{(1+\exp-|\xi|)^{2}}, \qquad \Delta= \frac{\lambda_e}{\sqrt b_m}
\end{equation}
giving the above scaling of the soliton width. In addition the inverse scaling of $\Delta$ with the maximum soliton amplitude $b_m$ is reproduced.

As discussed before, $S<0$ is required for solutions to exist. It is clear that in the absence of dissipation no shock can emerge from these stationary waves. They are solitary waves, stationary wave structures of finite spatial extensions and amplitude. 

Returning to the original variables, the maximum soliton amplitude is obtained from $S(B_m)=0$ as $B_m=B_1(2V_{1}/V_{A1}-1)$ which together with Eq.\,(\ref{chap2-eq-vnormal}) yields that $B_m<3B_1$, and consequently the Alfv\'enic Mach number for the solitons to exist is ${\cal M}_A<2$. The closer the Mach number approaches the maximum Mach number 2 the narrower the solitons become. This means that they steepen and, for Mach numbers ${\cal M}>2$, will overturn and break, because the dispersion does no longer balance the nonlinearity. Solitons cannot exist anymore at Mach numbers such high.

When a way can be found to generate dissipation in the region occupied by the soliton, then a magnetosonic soliton can evolve into a shock wave. Sagdeev's idea was that this can happen when the soliton becomes large amplitude and narrow enough such that in the steep rise of its crest sufficient dissipation could be generated by anomalous collisions and anomalous friction. These anomalous collisions would generate sufficient entropy that the states on the two sides of the soliton would differ from each other and the flow across the soliton would irreversibly change. In this case the soliton would turn into a dissipative subcritical laminar shock. Such shocks will be discussed in Chapter 3. Formally we may, of course, model the dissipative effect by simply defining some ``collision frequency $\nu$" and introducing a ``collision term"  on the right of Eq.\,(\ref{chap2-eq-bfield-sagdeev}). In order to do this we go one step back to the equation from which \,(\ref{chap2-eq-bfield-sagdeev}) has been obtained and add the collision term there:
\begin{equation}\label{chap2-eq-dampedsagdeev}
\frac{{\rm d}^2B}{{\rm d}x^2} = -\frac{\partial S(B)}{\partial B}-\nu\frac{{\rm d}B}{{\rm d}x}
\end{equation}
This equation is modelled exactly after the equation of a damped oscillator, where ${\rm d}B/{\rm d}x$ is the velocity. We should note here that this modelling has not been justified yet and in fact is not justified by any of our arguments yet. It not only requires the proof of the existence of an anomalous collision frequency $\nu$, it also requires the proof that from the kinetic equations containing the correlation terms an equation of the above structure can be derived. 

Ignoring these objections and exploiting the analogy with the damped oscillator we may conclude from the equation \,(\ref{chap2-eq-dampedsagdeev}) for the damped oscillator that the inclusion of anomalous ``collisions'' will dissipate the kinetic energy of motion of the pseudo-particle during its oscillation in the Sagdeev pseudo-potential $S(B)$ until the particle will finally come to rest at the bottom of the potential well. This is the case we have discussed earlier in connection with the Korteweg-de Vries-Burgers equation. It is drawn schematically in Figure\,\ref{chap2-fig-Sagdpot-2} when a shock wave forms from the soliton.\index{soliton!shock formation} The value of $B$ at minimum in the Sagdeev pseudo-potential where the pseudo-particle ultimately settles is the magnetic field level  
\begin{equation}
B_2 =\frac{1}{2}B_1\left[\left(8\beta_1+1\right)^\frac{1}{2}-1\right]
\end{equation}
far downstream of the shock that has formed in this dissipation process from the magnetosonic soliton. This value is determined by the upstream plasma-$\beta$ value. 

Taking this for granted, we can conclude that the damped oscillations the pseudo-particle performs on its damped downward path in the Sagdeev pseudo-potential are the spatially damped oscillations of the field $B(x)$ downstream of the shock. Moreover, the shock possesses an overshoot in $B$ at shock position $B_2<B_{\rm ov}<B_m$ which is smaller than $B_m$ but larger than $B_2$. The existence of damped downstream oscillations is in agreement with the concave shape of the dispersion relation of magnetosonic waves which, for large $k$, become dispersive and approach the lower-hybrid branch. This can be seen directly from the  dispersion relation for perpendicular ($\theta=90^\circ$) propagating magnetosonic waves
\begin{equation}
\omega_{ms}=V_A k_\perp\left(1+k^2\lambda_e^2\right)^{-\frac{1}{2}}
\end{equation}
For $k^2\lambda_e^2\ll 1$ the wave has constant phase velocity and is non-dispersive, becoming gradually dispersive with increasing $k$ when the effective phase velocity decreases. Hence, $\partial\omega/\partial k<0$ and the dispersion is concave, for very large $k\lambda_e\gg 1$ approaching the lower hybrid frequency $\omega_{lh}=\sqrt{\omega_{ce}\omega_{ci}}=\omega_{ce}\sqrt{m_e/m_i}$ where it flattens out, as shown in Figure\,\ref{chap2-fig-ms-whistler}. At oblique angles $90^\circ-\delta\gg\sqrt{m_e/m_i}\approx 1/43$ the dispersion is inverted, and $\partial\omega/\partial k>0$. Here the shorter waves run the soliton out and appear on the upstream side as spatially damped oscillations. Now their scale is  the ion inertial length $\lambda_i=c/\omega_{pi}$. However, shocks with convex dispersion where the shorter waves outrun the soliton will not exhibit a sharp shock profile. Rather they will be oscillating shocks with smoothed out ramp.

The theory presented above applies to a low-frequency plasma of velocity $V_1>>v_e,v_i$ larger than the thermal velocities of the plasma components. When the relation between $V_1$ and the thermal velocities changes, one must take into account thermal effects. These change the nature of the solitons and shock substantially. These changes will be discussed in greater depth in Chapter 3.
 
\begin{figure}[t!]
\hspace{0.0cm}\centerline{\includegraphics[width=1.0\textwidth,clip=]{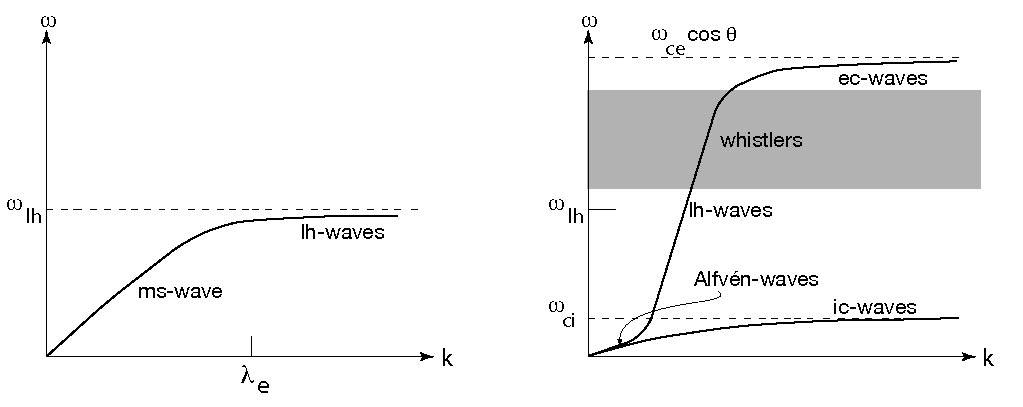} }
\caption[1]
{\footnotesize {\it Left}: Dispersion relation of perpendicular magnetosonic waves with concave dispersion at large $k$, where the wave effectively becomes a lower-hybrid wave with frequency close to the lower-hybrid frequency $\omega_{lh}$. {\it Right}: Dispersion relation in cold plasma at nearly parallel propagation. The ion branch starts from the left-handed Alfv\'en modes and goes in resonance at the ion-cyclotron frequency $\omega_{ci}$ with concave dispersion. The electron branch starts from the right-handed Alfv\'en mode, has convex dispersion, passes through the lower hybrid mode into the whistler mode $\omega_{lh}\ll\omega\ll\omega_{ce}$, assumes concave dispersion, and finally goes in resonance at the electron-cyclotron frequency $\omega_{ce}$, where it becomes the electron-cyclotron wave.}\label{chap2-fig-ms-whistler}
\end{figure}

\subsection{Whistlers and Alfv\'en shocks}\noindent\index{shocks!whistler}\index{shocks!Alfv\'en}
Many waves are capable of dispersively evolving into solitons or other similar stationary wave pulses if only their dispersion relation allows it. Among those waves we here consider only two particular cases, whistlers in cold plasma and Alfv\'en waves in low but finite $\beta$ conditions.

\subsubsection{Whistler solitons in cold plasma}\noindent
We now return to the cold plasma dispersion relation including both kinds of particles, electrons and ions. For parallel propagation ${\bf k}=k_\|\hat{\bf z}$, where ${\bf B}||\hat{\bf z}$, and the angle $\theta=0$, the dispersion relations of linear plasma waves are
\begin{equation}\label{chap2-eq-whitlerdisp}
\frac{k^2c^2}{\omega^2}=1-\frac{\omega_{pe}^2}{\omega(\omega\mp\omega_{ce})}-\frac{\omega_{pi}^2}{\omega(\omega\pm\omega_{ci})}
\end{equation}
The frequencies in the nominators are the electron and ion plasma frequencies, $\omega_{pe,pi}$, respectively. In the denominators appear the electron and ion cyclotron frequencies, $\omega_{ce,ci}$ respectively.  These relations describe right-hand and left-hand polarised waves according to the upper and lower signs.  Basically two branches coming from the resonances in the denominators are described by this relation. Figure\,\ref{chap2-fig-ms-whistler} on the right shows a plot of the two dispersion curves. For the evolution of shocks the most interesting part is the shaded whistler-mode dispersion relation.

The whistler dispersion relation\index{dispersion relation!whistler} is the (upper sign) electron part of the above dispersion relation. Neglecting the non-resonant ion contribution it reads
\begin{equation}
\frac{k^2c^2}{\omega^2}=1+\frac{\omega_{pe}^2}{\omega(\omega_{ce}-\omega)}
\end{equation}
with the second term being large because of the resonance in the denominator. Thus one can also neglect the 1 on the right finding that solutions exist only for $\omega<\omega_{ce}$ as is also seen in the above drawing. Then the dispersion relation becomes 
\begin{equation}
k^2\lambda_e^2\simeq \omega/(\omega_{ce}-\omega)\qquad\to\qquad \omega\simeq\omega_{ce}\left(1+\frac{1}{k^2\lambda_e^2}\right)^{-1}
\end{equation}
which exhibits its concave character confirming that short wavelengths whistlers will fall behind the main shock pulse. Nonlinear analysis of these waves goes back to \cite{Montgomery1959}, \cite{Sagdeev1966},  and \cite{Kakutani1966} and is based on the fluid equations  we used before. Let the plasma again be moving in $x$-direction antiparallel to the shock normal ${\bf n}$ and write the magnetic field in polar coordinates as ${\bf B}=B_\perp(0,\cos\theta,\sin\theta)$; then one again obtains the canonical Sagdeev form of the first integral of the equation of motion of a pseudo-particle at pseudo-position $B_\perp$ and pseudo-velocity d$B_\perp/$d$x$ as
\begin{equation}
\frac{1}{2}\left(\frac{{\rm d}B_\perp}{{\rm d}x}\right)^2=-S(B_\perp)
\end{equation}
The Sagdeev pseudo-potential is a complicated expression which simplifies considerably for a uniform upstream state. We introduce the normalised variables $b_\perp=B_\perp/B_{\perp m}, \xi=x/\lambda_e$ and $ \beta_\perp =B_\perp^2/\mu_0mN_1V_1^2$ writing
\begin{equation}
\frac{1}{2}\left(\frac{{\rm d}b_\perp}{{\rm d}\xi}\right)^2=-S(b_\perp)=-\frac{1}{8\beta_{\perp m}}\frac{b_\perp(b_\perp^2-1)}{(1-1/2\beta_\perp)^2}
\end{equation}
Solitons exist for $B_\perp\leq B_{\perp m}$ which is the maximum whistler soliton amplitude, and for $\beta_\perp>\frac{1}{2}$. For the maximum amplitude we have 
\begin{equation}
B_{\perp m}<\sqrt{2\mu_0mN_1V_1^2}\sim\sqrt{m_i/m_e}B_1\approx 43\, B_1
\end{equation} 
and the whistler soliton velocity $V_1\gg V_{A1}$ yielding a soliton Mach number range of
\begin{equation}
\sqrt{m_i/4m_e}<{\cal M}_A<\sqrt{m_i/2m_e} \qquad \to \qquad 22<{\cal M}_A<30
\end{equation}
which identifies the whistler solitons as being high-Mach number solitons, indeed. In case they evolve into shocks, these shocks are high-Mach number as well. This might cause other effects which have not been considered so far. Hence the present formal theory must be taken with caution in application to real problems of much lower Mach numbers. One of the neglected conditions is quasi-neutrality, which demands that $\epsilon_0E/eN\Delta x\ll 1$. This leads to the further restriction on $B_\perp$ and the pulse width $\Delta$
\begin{equation}
B_\perp/B_1\ll \left(2\pi c^2/V_{A1}^2\right)^\frac{1}{4} \sim 1.5 c/V_{A1}, \qquad \Delta\sim\Delta x \sim \lambda_i (B_1/B_\perp)
\end{equation}
where $\lambda_i=c/\omega_{pi}$ is the ion skin depth (ion inertial length). It follows that these whistler pulses should have quite large characteristic widths and, moreover, characteristic frequencies $\omega\sim\frac{1}{2}\omega_{lh}(B_\perp/B_1)$, far below the lower hybrid frequency. These properties identify the whistlers as  right-handed (rotating clockwise along $x$) high-frequency Alfv\'en wave pulses. These low frequency whistler/high frequency whistler-Alfv\'en waves can indeed been excited by a cold shock-reflected ion beam as will be shown below in the section on ion beam instabilities.

\begin{figure}[t!]
\hspace{0.0cm}\centerline{\includegraphics[width=0.75\textwidth,clip=]{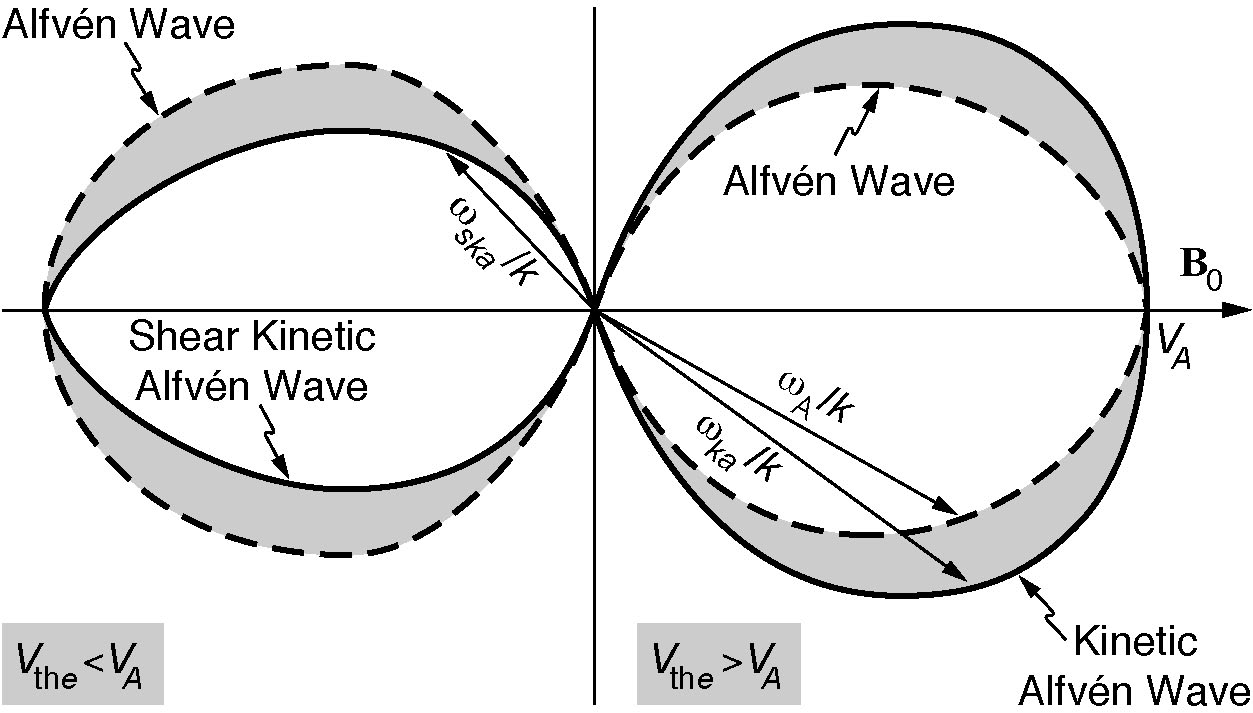} }
\caption[1]
{\footnotesize Phase velocity diagram of the three Alfv\'en wave modes in the (${\bf B, k}$)-plane of the magnetic field and wave vector for the two cases when the alfv\'en velocity is larger {\it left} or smaller {/right} than the electron thermal speed $v_{th,e}$. The ordinary Alfv\'en wave describes a circle in this plane. The two other phase velocities describe deformed curves. }\label{chap2-fig-kaw}
\end{figure}

In view of our remarks we do not give rigourous derivations of these approximate formulae. It will turn out later when discussing numerical simulations that whistlers do indeed occur at shocks and have been observed early on as well in laboratory experiments on collisionless shocks \citep[e.g.,][]{Decker1972} that, however, other fluctuations driven by ion reflection are of greater importance in structuring supercritical Mach\index{shocks!supercritical} number shocks. Still it is highly probable that whistlers are excited in shock waves as the conditions will be in favour of them when sufficient free energy is available in the shock front because of several reasons, one of them electron heating in the perpendicular direction \citep[early observations suggested their presence near shocks in space, see,][]{Rodriguez1975}. In this case whistlers become very important for producing dissipation via a short-wavelength instability, called {\it decay instability}, which had been predicted  by \cite{Galeev1963} and for which evidence has been found in the above laboratory observations by \cite{Decker1972}.

\subsubsection{Alfv\'en solitons at finite-$\beta$}
\noindent Alfv\'en waves are non-dispersive. However, when the plasma temperature increases, dispersion in the direction perpendicular to the magnetic field sets on. From the general dispersion relation in the very low-frequency limit one then obtains \cite[e.g.,][]{Baumjohann1996} for the frequency of the Alfv\'en wave
\begin{equation}\label{chap2-eq-kinalf}
\omega^2(k_\|,k_\perp) = k_\|^2V_A^2\frac{1+k_\perp^2r_{ci}^2}{1+k_\perp^2\lambda_{pe^2}}
\end{equation}
where the ion gyro-radius is slightly modified with temperature according to $r_{ci}^2\to r_{ci}^2(\frac{3}{4}+T_e/T_i)$, and $V_A^2=B^2/\mu_0mN$ is the square of the Alfv\'en speed. This dispersion relation describes two kinds of Alfv\'en waves depending on $k_\perp\sim r_{ci}^{-1}$ or $k_\perp\sim\lambda_e^{-1}$. These modes become important when the plasma-$\beta<1$. The phase velocities of the two modes together with the ordinary Alfv\'en wave are shown in Figure\,\ref{chap2-fig-kaw}. The  modes differ in their dispersive properties from ordinary Alfv\'en waves as they propagate oblique to the magnetic field, i.e. the wave energy propagates under an angle to the magnetic field for $k_\perp$ is independent of $k_\|$. 

At finite temperatures $1>\beta>m_e/m_i$ the wave is called {\it kinetic} Alfv\'en wave. Its perpendicular wavelength becomes the order of the ion gyro-radius, and the phase speed parallel to the magnetic field increases. At low temperatures $\beta<m_e/m_i$ the wave is called {\it shear} or better {\it inertial} Alfv\'en wave with perpendicular wavelength comparable to the electron skin depth, and the effective parallel phase velocity decreases. In terms of prospective Alfv\'enic shocks this means that a kinetic Alfv\'en shock in direction parallel to the external field will support oscillations upstream of the shock ramp, while an inertial Alfv\'en shock will support downstream oscillations and thus possess a sharp shock ramp.  

The dispersive properties of both kinetic Alfv\'en modes\index{waves!kinetic Alfv\'en} enable the existence of stationary wave pulses. These have been discovered first by \cite{Hasegawa1976} for the {\it kinetic} mode. For the low Alfv\'en frequencies quasi-neutrality is a good assumption. Moreover, the electrons have time enough to behave Boltzmann-like. Hence with the parallel electric potential $\phi_\|$ we have
\begin{equation}
N_e=N_0\exp(e\phi_\|/T_e)
\end{equation}
In the perpendicular direction we use the electric potential $\phi_\perp$ . Maxwell's equations, the nonlinear ion momentum conservation, and Poisson's equation then reduce to
\begin{eqnarray}\label{chap2-eq-kaweq}
\frac{\partial B_\perp}{\partial t} &=&\nabla_\perp\nabla_\|(\phi_\perp-\phi_\|) \nonumber \\
\nabla_\perp^2\nabla_\|^2(\phi_\perp-\phi_\|) &=& \mu_0\nabla_\|\frac{\partial j_\|}{\partial t} \\
\frac{\partial N_i}{\partial t} &=& \frac{1}{B_0\omega_{ci}}\nabla_\perp\left(N_i\nabla_\perp\frac{\partial\phi_\perp}{\partial t}\right)\nonumber
\end{eqnarray}
The field-aligned current $j_\|$ is carried by the hot electron component, such that its divergence is given by $\nabla_\| j_\|=e(\partial N_e/\partial t)$. introducing dimensionless variables $\xi=x\omega_{ci}/c_{ia},\zeta=z\omega_{pi}/c,\tau=\omega_{ci}t,N'=N/N_0$ and measuring the potentials in units of $e/T_e$, then transforming to a comoving coordinate system $\eta=\kappa_\perp\xi+\kappa_\|\zeta-\tau$, the whole system of equations is reduced to the nonlinear equation
\begin{equation}
\kappa_\perp^2\kappa_\|^2N'\frac{{\rm d}^2\ln N'}{{\rm d}\eta^2}=(1-N')(N'-\kappa_\|^2)
\end{equation}
which is in the form suited for the Sagdeev pseudo-potential method, yielding $({\rm d}N'|{\rm d}\eta)^2=-S(N',\kappa_\|,\kappa_\perp)$ with
\begin{equation}\label{chap2-eq-kaw-s}
S(N',\kappa_\|,\kappa_\perp)=-\frac{2N'}{\kappa_\|^2\kappa_\perp^2}\left[(1-N')(N'+\kappa_\|^2)+(1+\kappa_\|^2)N\,'\ln\,N'\right]<0
\end{equation}
for soliton solutions to exist. These solutions give the density as function of the linear coordinate $\eta$. Interestingly, there solutions which are dilutions and solutions which are compressions. The condition for existence of soliton solutions is independent of the perpendicular wavenumber. hence it is the parallel electric field that is responsible for the formation of solitons and balance of the nonlinear steepening. Solitons form only in parallel direction with the magnet field being inclined to the soliton which in the perpendicular direction is flat. When such a soliton attains dissipation and turns into a shock, it will be a quasi-parallel shock preceded by damped upstream waves\index{waves!upstream} that have outrun the shock ramp. That the shock will be quasi-parallel can be easily seen from the fact that $k_\perp\gg k_\|$, the shock front will be perpendicular to the external field, and therefore $b_\perp\ll b_\|$ as required for a quasi-parallel shock.

Inspection of Eq.\,(\ref{chap2-eq-kaw-s}) shows that the Sagdeev potential vanishes at $N'=0$, $N'=1$, and $N_m'$. Compressive (rarefaction) solitons occur at $N'>1$ ($N'<1$). Only compressive solitons are of interest in shock formation. The maximum amplitude $N'_m$ of compressive solitons follows from setting the bracket to zero. It is approximately given by the solution of\index{soliton!kinetic Alfv\'en} 
$N'+\kappa^2_\|\approx(1+\kappa_\|^2)\ln N'$ which, for $\kappa_\|^2=1$ is  $N'_m\approx 3$. The minimum of $S(N')$ for compressive solitons is found by taking the derivative of the bracket and putting it to zero. Setting $N'=1+n'$ and expanding the logarithm one finds the minimum of the Sagdeev pseudo-potential trough at 
\begin{equation}
N'_{\textsf{KAS}}= 1+n'\approx\frac{1+3\kappa^2_\|}{1+\kappa^2_\|} < 3
\end{equation}
This is the prospective maximum amplitude of a \index{shocks!kinetic Alfv\'en}kinetic Alfv\'en shock (KAS) evolving in the presence of dissipation from a compressive kinetic Alfv\'en soliton. The compression at a KAS ramp is thus limited to a factor $< 3$. Finally, the value of the Sagdeev pseudo-potential at its absolute minimum  gives the steepest gradient of the density in the shock ramp, $({\rm d}N'/{\rm d}\eta)_{\rm max} =|S(N'_{\textsf{KAS}})|$ at the turning point of the density in the ramp. This ratio provides an estimate of the width of the KAS-shock pulse 
\begin{equation}
\Delta_{\textsf{KAS}}\sim N'_{\textsf{KAS}}/|S(N'_{\textsf{KAS}})| >|3-\kappa_\|^2|^{-1}
\end{equation}
in normalised units (where in the final estimate at the right we have used the maximum shock amplitude). It depends on the given  values of $\kappa_\|$ while being independent on $\kappa_\perp$, in agreement with the fact that only values parallel to the field will affect the shock structure.

We will not investigate the case of {\it inertial} Alfv\'en wave solitons. In the heliosphere these shocks occur only in the auroral zones of strongly magnetised planets and, possibly, also deep in the solar corona only in region of very strong magnetic field and comparably small temperature. The related solitons have been discussed elsewhere \citep{Treumann1990,Berthomier1999}. Shocks forming there should always be rarefactive, containing very little plasma and having highly oscillating wakes on their downstream side containing short wavelength inertial Alfv\'en waves\index{waves!inertial Alfv\'en} which have been left behind the faster shock ramp. In fact such very diluted plasmas have indeed been inferred in the auroral zones of planets like Earth and Jupiter.\index{auroral zone}\index{planets!Earth}\index{planets!Jupiter}

\subsubsection{Remarks on the generation of dissipation}\noindent
\cite{Sagdeev1966} gave an idea of how dissipation can be generated in a shock ramp realising that shocks must contain a -- large scale -- electric potential drop (or its equivalent as, for instance, an equivalent electric field corresponding to the shock ramp density gradient) at which low energy ions will be reflected. (Note that electron will instead become accelerated by this potential across the shock.) No matter how few those ions are, they will return into the upstream medium, where they (as \cite{Sagdeev1966} had noted) become accelerated tangentially along the (perpendicular) shock surface by the upstream convection electric field until gaining enough kinetic energy to overcome the shock potential barrier, passing the shock, and escaping downstream. These ions form a current in front of the shock that carries free energy and will ultimately become unstable with respect to the {\it two-stream} instability, scattering upstream electrons and in this way cause dissipation. The physical mechanism of this process will be discussed later in this chapter in the section on anomalous transport. First we need to be informed about the instabilities and waves that are relevant with respect to shock formation. To these mechanisms we will continuously return when discussing the different types of shocks and the corresponding numerical simulations. Without them neither the existence nor the structure of shocks in collisionless plasmas can be understood. The models presented so far cannot give more than hints in which direction one has to pursue. Shock physics is too complicated for analytical theory. \index{dissipation!generation of}

\subsection{Instabilities}
\noindent An instability is the reaction of an active medium like a warm plasma to the presence of free energy. Since the available free energy keeps the plasma away from thermodynamic equilibrium, restoration of equilibrium becomes necessary. This is most easily done by exciting fluctuations to amplitudes large enough for either causing dissipation or transporting the energy away to a location where it can be dissipated by other processes. As a consequence a wave will  start growing out of the thermal fluctuation background. The selection of the frequency and wavelength range usually is ruled by resonance with the number of particles which carry the free energy. However, other ways of exciting waves are also possible. The wave that survives is that with the fastest grow, and least damping. \citep[Theoretical overviews of instabilities and dissipation related to shocks have been given by][]{Galeev1976,Sagdeev1979,Wu1984a,Papadopoulos1985,Winske1985}

Since most instabilities follow this recipe of growing out of the thermal background they start as infinitesimal disturbances which can be described by linear dispersion theory. Hence instability theory in plasma can be based on the dispersion relation Eq.\,(\ref{chap2-eq-generaldispersionrelation}) and for weakly growing waves on the expression for the growth rate (\ref{chap2-eq-growthrategeneral}). Plane linear wave modes possess a phase factor $\exp\,i({\bf k\cdot x}-\omega\,t)$. It is clear that instability sets on whenever the imaginary part of the frequency becomes positive; $\gamma(\omega_r,{\bf k}, \dots)>0$ is the growth rate of the instability of the particular wave mode that becomes unstable.

In the context of collisionless shocks the instabilities of interest can be divided in two classes. The first class contains those waves which can grow themselves to become a shock. It is clear that these waves will be of low frequency and comparably large scale because otherwise they would not evolve into a large macroscopic shock. We have already discussed a few candidates and their nonlinear evolution in the previous sections, among them magnetosonic, Alfv\'enic and whistler modes.\index{instability!shock forming} In this section we will investigate a number of waves which form secondarily after an initial seed shock ramp has grown in some way out of one of these wave modes, these are ion modes\index{waves!ion modes} which have now been identified to be responsible for structuring, shaping and reforming the shock. In fact real oblique shocks -- which are the main class of shocks in interplanetary space and probably in all space and astrophysical objects -- cannot survive without the presence of these ion waves which can therefore be considered of the wave modes that really produce shocks in a process of taking and giving between shock and waves.

The second class are waves that accompany the shock and provide \index{instability!generation of dissipation}anomalous transport coefficients like anomalous collision frequencies, friction coefficients, heat conductivity and viscosity. These waves are also important for the shock as they contribute to entropy generation and dissipation. However, they are not primary in the sense that they are not shock-forming waves. Among them there is another group that only carries away energy and information from the shock. These are high-frequency waves, mostly electrostatic in nature, produced by electrons, or when electromagnetic they are in the free-space radiation modes. In the latter case they carry the information from remote objects as radiation in various modes, radio or x-ray to Earth, informing of the existence of a shock. In interplanetary space it is only radio waves which fall into this group as the radiation measure of the heliospheric shocks is too small to map them into x-rays. These groups of waves we will briefly mention below; they will however play a more important role in Part 2 of this text when discussing measurements and observations of the various types of shock waves that are encountered in the heliosphere.  

\subsubsection{Ion-beam driven instabilities -- {\bf $\omega \lesssim \omega_{ci}$}}
\noindent The shock waves in the heliosphere are magnetised. As long as we are interested in their formation and properties we can restrict to low frequency electromagnetic waves in warm plasma. Such waves are excited by plasma streams or kinetic anisotropies in one or the other way. The simplest instability known which distorts the magnetic field by exciting Alfv\'en waves that are propagating along the magnetic field is the firehose mode. \index{instability!ion beam}

\paragraph{Firehose mode.} The firehose mode\index{instability!firehose} is the result of a pressure (or temperature) anisotropy in plasma with the parallel pressure $P_\|$ exceeding the perpendicular $P_\perp$ and magnetic $B^2/2\mu_0$ pressures. \cite{Sagdeev1966} gave a simple intuitive explanation of this instability based on the insight that the parallel thermal motion of the adiabatic magnetised ions along the magnetic field exerts a centrifugal force on the field lines. When this force exceeds the restoring forces of the magnetic pressure and perpendicular plasma pressure, the centrifugal force wins and a small excursion of the magnetic field starts growing and propagates as a wave along the magnetic flux tube like on a string. The condition for instability is 
\begin{equation}\label{chap2-eq-firehose}
P_\|-P_\perp>B^2/\mu_0
\end{equation}
Since the pressure anisotropy on the left means that there is an excess in parallel energy in the plasma, the plasma possesses free energy which by the instability is fed into the excitation of Alfv\'en waves with frequencies $\omega_A \ll \omega_{ci}$, transported away with Alfv\'en speed and ultimately dissipated in some way -- as expected. The waves excited are ordinary Alfv\'en waves, however, and not suited for shock formation. Below we will once more encounter this mode in discussing ion beam instabilities.

\paragraph{Kinetic Alfv\'en waves.} Excitation of kinetic Alfv\'en waves requires $\beta <1$ and a different process. In the solar wind the $\beta$-condition is barely satisfied except possibly in the very strong coronal magnetic fields or locally (possibly in Corotating Interaction Region boundaries when the magnetic field may become compressed without just forming a shock). Kinetic Alfv\'en waves possess a finite electric field component parallel to the magnetic field which ca accelerate electrons. However, the inverse mechanism is also possible that electrons moving along the magnetic field in the opposite direction, become retarded by this field component and feed their energy into the kinetic Alfv\'en wave. A process similar to this has been suggested by \cite{Hasegawa1979} in different context for bouncing electrons in a locally inhomogeneous magnetic field represented as $B(z)= 
 B_0(1+az^2)$. The electron beam conserves the magnetic moment when moving along the magnetic field, interacting adiabatically with the parallel wave electric field for long wavelength
\begin{equation}
E_\|=\frac{k_\perp^2r^2_{ci}}{1+k_\perp^2r^2_{ci}}\left| \frac{\omega B_\perp}{k_\perp}\right| 
\end{equation}
The condition that the electrons form a beam is that at the resonance with the wave the derivative of the electron distribution with respect to the resonant parallel electron energy of motion $\epsilon_\|=m_e\omega_b^2/2b$ is positive, $\partial F_e/\partial \epsilon_\|>0$. Here the square of the bounce frequency is $\omega_b^2=2\mu\omega_{ce}/m_e$, and $\mu=m_eV_\|^2/2B$ is the magnetic moment. The one can calculate the growth rate as
\begin{equation}
\gamma_{\,\,\textsf{KAW}}\simeq (k_\perp\lambda_e)^{-2 }(\omega^4/bk_\|^3v_e^3), \qquad\quad \omega=k_\|V_A
\end{equation}
We see that an electron beam moving in an inhomogeneous magnetic field can excite kinetic Alfv\'en waves.

Probably more important than this is, however, the interaction of ions which are reflected from a solitary pulse and move back upstream ahead of the pulse, as had been suggested by \cite{Sagdeev1966}. The reflected ions will represent a beam that is moving against the initial  plasma inflow which by itself is another ion beam neutralised by the comoving electrons. This configuration leads to a ion beam-ion beam interaction and should cause an instability because free energy is present in the two counter-streaming beams. The various instabilities this process may cause have been reviewed by \cite{Gary1993}.

\paragraph{Kinetic growth rate.} Before coming to discussing the relevant instabilities we should briefly mention the waves which can be driven by them. We already noted that in thermal plasma most waves will rest in thermal fluctuations. Once a wave which is an eigenmode of the plasma is injected it will experience thermal damping until it disappears in the background fluctuations. Hence, a wave that is assumed to grow must overcome this damping which for propagation parallel to the average magnetic field ${\bf B}_0$ in a uniform plasma is given by
\begin{equation}\label{chap2-eq-kineticgrowthrate}
\gamma(\omega, {\bf k})\simeq \sqrt{\frac{{\pi}}{2}}\sum\limits_s\frac{\omega_s^2}{2\omega v_s}\left(V_s-\frac{\omega}{k_\|}\right)\exp\left[\frac{(\omega\pm\omega_{cs}-k_\|V_s)^2}{2k_\|^2v_s^2}\right]
\end{equation}
where the index $s$ identifies the species, $v_s=\sqrt{2T_s/m_s}$ is the thermal speed of the species $s$, $V_s$ its average parallel bulk drift velocity, $\omega_{s}\equiv\omega_{ps}$ its plasma frequency, and we have dropped the index $r$ (for real) at $\omega$ which in this expression is understood as real anyway. Note that we are going to take into account several different ion species and thus need an extra index to distinguish between them all. The simplified cold dispersion relation is
\begin{equation}
\omega^2-k^2c^2-\sum\limits_s\frac{\omega_s^2(\omega-k_\|V_s)}{\omega-k_\|V_s\pm\omega_{cs}}=0
\end{equation}
which determines the approximate real frequency. However, when thermal effect are included, then there is no way to avoid the numerical solution of the full kinetic dispersion relation. In the following we will generally refer to such numerical solutions.

It is obvious from the expression for the damping rate, that for sufficiently large average drift velocities $V_s$ of species $s$ larger than the phase velocity $\omega/k_\|$ of the wave this particular species contributes a positive term to the damping rate $\gamma$ which, when large enough can dominate the entire damping rate. In this way streaming is one way to cause instability. 
In the absence of streaming $\gamma$ is independent of $V_s$ and is negative for a thermally isotropic plasma. In the presence of a temperature anisotropy, however, this may change as we have seen for the firehose mode. The above damping rate (\ref{chap2-eq-kineticgrowthrate}) does not account for thermal anisotropy which is, however, simple matter to include \citep[cf., e.g., ][]{Gary1993}. 

We note finally that $\gamma$ is a resonant damping/growth rate yielding resonant instability driven by small groups of resonant particles. The firehose mode is non-resonant since all particles contribute to it. Generally most non-resonant instabilities can only be found by solving the full dispersion relation numerically.
\begin{figure}[t!]
\hspace{0.0cm}\centerline{\includegraphics[width=0.85\textwidth,clip=]{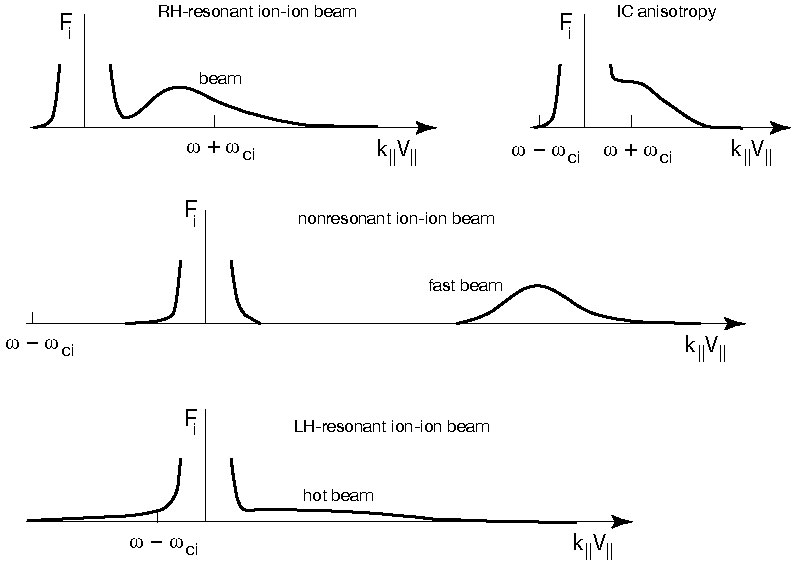} }
\caption[1]
{\footnotesize The three cases of ion beam - plasma interaction and the location of the unstable frequencies. Shown is the parallel (reduced) distribution function $F_i(k_\|v_\|)$, where for simplicity the (constant) parallel  wave-number $k_\|$ has been included into the argument. Right handed resonant modes (RH) are excited by a cool not too fast beam. When the beam is too fast the interaction becomes nonresonant. When the beam is hot, a resonant left hand mode (LH) i excited. In addition the effect of temperature anisotropy is shown when a plateau forms on the distribution function \citep[after][]{Gary1993}.}\label{chap2-fig-ioninst}
\end{figure}

At low frequencies  it suffices for our purposes of understanding shock physics to deal with a three-component plasma consisting of two ion species and one neutralising electron component which we assume to follow a Maxwellian (thermal) velocity distribution. Moreover, we assume that the drifting ion components are Maxwellians as well. In conformity with the above remarks on a resonant instability we assume that the dominant ion component has large density $N_i\gg N_b$, and the second component represents a weak fast beam of density $N_b$ propagating on the ion-electron background with velocity $V_b\gg V_i\approx 0$. Following \cite{Gary1993} it is convenient to distinguish the three regimes: cool beams ($0<v_b<V_b$), warm beams ($v_b\sim V_b$), and hot beams ($v_b\gg V_b$). Figure\,\ref{chap2-fig-ioninst} shows the beam configurations for these three cases and the location of the wave resonances respectively the position of the unstable frequencies.

\paragraph{Cool ion beam: Right-hand instability.} Assume that the ion beam is thermally isotropic and cool in the above sense, i.e. its velocity relative to the bulk plasma is faster than its thermal speed. In this case a right-handed resonant instability occurs. In the absence of a beam $V_b= 0$ the parallel propagating mode is a right-circularly polarised magnetosonic wave propagating on the lowest frequency whistler dispersion branch with $\omega\approx k_\|V_A$. In presence of a drift this wave becomes unstable, and the fastest growing frequency is at frequency $\omega\simeq k_\|V_b-\omega_{ci}$. This mode propagates parallel to the beam, because $\omega>0, k_\|>0$, and $V_b>0$. The numerical solution of this instability  for densities $0.01\lesssim N_b/N_i\lesssim 0.10$ at the wave-number $k_\|$ of fastest growth rate identifies a growth rate of the order of the wave frequency $\gamma\sim\omega$ and 
\begin{equation}\gamma_m\simeq \omega_{ci}(N_b/2N_e)^\frac{1}{3}\end{equation}
for the maximum growth rate $\gamma_m$, where $N_e=N_i+N_b$ is the total density from quasi-neutrality. This instability drives waves propagating together with the beam in the direction of the ion beam on the plasma background which has been assumed at rest. If applied, for instance, to shock reflected ions then for 2\% reflected ions the maximum growth rate is $\sim0.2\omega_{ci}$, and $V_b\sim 1.2\omega_{ci}/k_\|$, and $k_\|\sim0.2\omega-{ci}/ V_A$ which gives $V_b\sim 6V_A$. In the solar wind the Alfv\'en velocity is about $V_A\approx 30$\,km/s. Hence the velocity difference between shock reflected ions and solar wind along the magnetic field should be roughly $\sim180$\,km/s. The thermal velocity of the ion beam must thus be substantially less than this value, corresponding to a thermal beam energy less than $T_b\ll 100$ eV which  in the solar wind, for instance, is satisfied near the tangential field line. The solar wind travels at 300-1200 km/s. Complete reflection should produce difference speeds twice these values. The above value is thus not unreasonable for travelling shocks, but for bow shock reflected ions applies to the quasi-perpendicular portion of the bow shock only. We may thus conclude that this wave mode could be excited in the solar wind by shock reflected ion beams near quasi-perpendicular shocks. 

\paragraph{Warm ion beam: Left-hand instability.} \index{instability!warm ion beam}The above instability is present when the ion beam is rather cold. When the temperature of the ion beam increases and the background ions remain to be cold, then beam ions appear on the negative velocity side of the bulk ion distribution and go into resonance their with the left-hand polarised ion-Alfv\'en wave. Their maximum growth rate is a fraction of the growth rate of the right-hand low frequency whistler mode. Nevertheless it can excite the  Alfv\'en-ion cyclotron wave\index{waves!Alfv\'en-ion cyclotron} which also propagates parallel to the beam. For this instability the beam velocity must exceed the Alfv\'en speed $V_b>V_A$. 

At oblique propagation both the right and left hand instabilities have smaller growth rates. But interestingly, it has been shown \citep{Goldstein1985} that the fastest growing modes then appear for oblique ${\bf k}$ and harmonics\index{instability!ion-cyclotron harmonic} of the ion cyclotron frequency $\omega\sim n\omega_{ci}$, with $n=1,2,\dots$.
\begin{figure}[t!]
\hspace{0.0cm}\centerline{\includegraphics[width=0.77\textwidth,clip=]{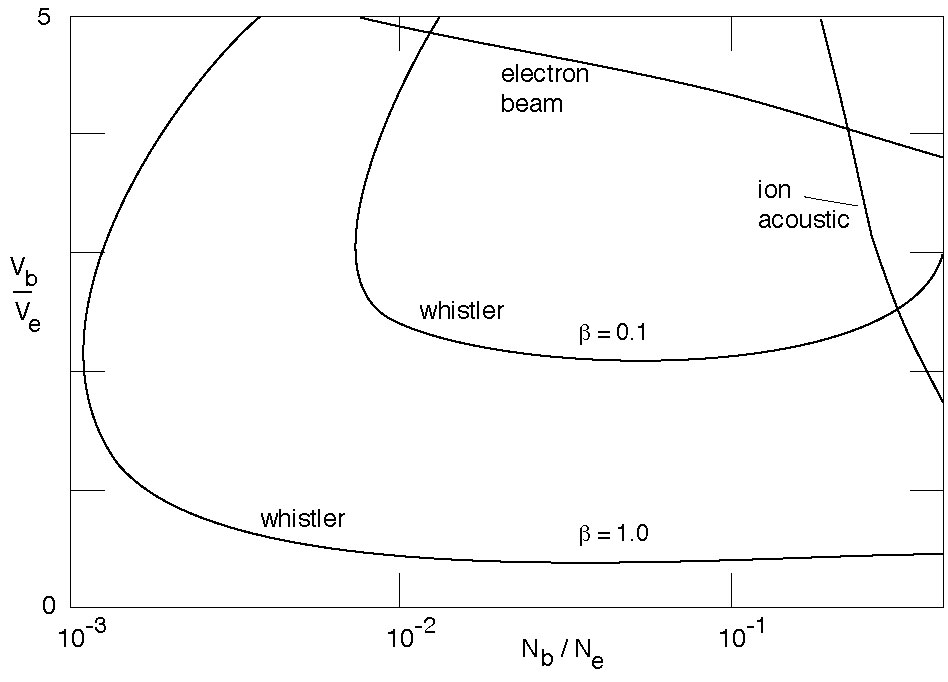} }
\caption[1]
{\footnotesize The regions of instability of the electron beam excited whistler mode in density and beam velocity space for two different $\beta$ compared to the ion acoustic and electron beam modes. Instability is above the curves. The whistler instability has the lowest threshold\index{threshold!whistler} in this parameter range \citep[after][]{Gary1993}.}\label{chap2-fig-unstab}
\end{figure}
\paragraph{Nonresonant ion instability: Firehose mode.} \index{instability!nonresonant}\index{waves!firehose}When the ion beam is fast and cold it does not go into resonance because its velocity is to high. In this case all ions participate in a nonresonant instability which in fact is a thermal firehose mode where the ion beam has sufficient energy to shake the field line. This mode propagates antiparallel to the ion beam, has small phase speed and negative helicity. This mode has large growth rate for large $N_b/N_e$ and $V_b/V_A$ simply because then there are many beam ions and the centrifugal force is large while the beam velocity lies outside any resonant wave speed. It is trivial that this instability becomes stronger when the ion beam is composed of heavier ions as the larger mass of these increases the centrifugal force effect.

\subsubsection{Electron instabilities and radiation -- {\bf $\omega \sim \omega_{pe}$}}
\noindent \index{instability!electron beam}Other than ion beam excited instabilities electron-beam instabilities are not involved in direct shock formation (unless the electron beams are highly relativistic which in the entire heliosphere is not the case; we do not investigate relativistic shocks in this text anyway). The reason is that the frequencies of electron instabilities are high. However, just because of this reason they are crucial in anomalous transport being responsible for anomalous collision frequencies and high frequency field fluctuations. The reason is that the high frequency waves lead to energy loss of the electrons retarding them while for the heavier ions they represent a fluctuating background scattering them.  In this way high frequency waves may contribute to the basic dissipation in shocks even though this dissipation for supercritical shocks will not be sufficient to maintain a collisionless shock or even to create a shock under collisionless conditions. This is also easy to understand intuitively, because the waves need time to be created and to reach a substantial amplitude. This time in a fast stream is much shorter than the time the stream needs to cross the shock. So waves will not be accumulated there; rather the fast stream will have convected them downstream long before they have reached substantial amplitudes for becoming important in scattering.

When we are going to discuss electromagnetic waves which can be excited by electrons we also must keep in mind that such waves can propagate only when there is an electromagnetic dispersion branch in the plasma under consideration. These electromagnetic branches in $\omega, {\bf k}$-space are located at frequencies below the electron cyclotron frequency $\omega_{ce}$. The corresponding branch is the whistler mode branch. Electrons will (under conditions prevailing at shocks) not be able to excite electromagnetic modes at higher frequencies than $\omega_{ce}$. We have seen before that ion beams have been able to excite whistlers at low frequencies but above the ion-cyclotron frequency. This was possible only because the whistler branch\index{waves!whistler} exists due to the presence of electron (because of quasi-neutrality) as a channel for wave propagation. In fact, ion cyclotron waves can of this reason also have higher electromagnetic harmonics. For electrons under conditions in the shock environment this is generally not possible. Electromagnetic waves excited by them propagate on the whistler branch or its low frequency Alfv\'enic extension\footnote{In fact there is one exception to this statement. There exist free space electromagnetic mode branches (radiation) above and even below the electron cyclotron frequency on which electron excited electromagnetic waves could in principle propagate. The mechanism to excite them is the \index{instability!cyclotron maser}\index{radiation!electron cyclotron maser}electron cyclotron maser instability \citep[for a recent review see, e.g., ][]{Treumann2006}, which is a very particular instability that becomes awakened under conditions which to our knowledge never are satisfied in the shock environment.}. Electron beams thus excite electromagnetic whistlers and right-handed Alfv\'en waves. They also excite a variety of electrostatic emissions which we will mention later as well.

\paragraph{Whistlers.} Whistlers can be driven in two ways, either by an electron temperature anisotropy \citep{Kennel1966}, or by electron beams (or heat fluxes) \citep{Gary1993}. In the former case the condition for instability are that the perpendicular electron temperature $T_{e\perp}>T_{e\|} $ exceeds the parallel electron temperature, and that the parallel energy of the resonant electrons ${\cal E}_\|=\frac{1}{2}m_eV_\|^2>B^2/2\mu_0N_e$ exceeds the magnetic energy per electron.  

\cite{Gary1993} has investigated the more relevant case of whistler excitation by an electron beam. He finds from numerical solution of the full dispersion relation including an electron beam in parallel motion that with increasing beam velocity $V_b$ the real frequency of the unstable whistler decreases, i.e. the unstably excited whistler shift to lower frequencies on the whistler branch while remaining in the whistler range $\omega_{ci}<\omega<\omega_{ce}$. Both, the background electrons and beam electrons contribute resonantly. The most important finding is that the whistler mode for sufficiently large $
\beta_i\sim 1$ (which means low magnetic field), $N_b/N_e$ and $T_b/T_e$ has a the lowest  beam velocity threshold when compared with the electrostatic electron beam instabilities as shown in Figure \ref{chap2-fig-unstab}. This finding implies that in a relatively high-$\beta$ plasma a moderately dense electron beam will first excite whistler waves. In the shock environment the conditions for excitation of whistlers should thus be favourable whenever an electron beam propagates across the plasma along the relatively weak magnetic field. The electrons in resonance satisfy $V_\|=(\omega-\omega_{ce})/k_\|$ and, because $\omega\ll \omega_{ce}$ the resonant electrons move in the direction opposite to the beam. Enhancing the beam temperature increases the number of resonant electrons thus feeding the instability. On the other hand, increasing the beam speed  shifts the particles out of resonance and decreases the instability.  Hence for a given beam temperature the whistler instability has a maximum growth rate a few times the ion cyclotron frequency.

\subsection{``Transport ratios"}\noindent
Measured wave spectra are complex and opaque, i.e. it is very difficult from an inspection of their shape to identify the wave modes that are present in the plasma volume under investigation. In some rare clean cases one can conclude from the observation of a particular maximum in the wave power or the observation of only one single field component which wave has been detected. In the general case of broad spectra or mixed spectral maxima and various field components lacking clear dominance of one field component it becomes nearly impossible to decide about the waves.Clearly, when the spectrum is shapeless power-law one in most cases is dealing with well developed turbulence in which case it makes no sense to distinguish and search for single modes. Then one must seek shelter among the well developed methods of turbulence. On the other hand, if the spectra indicate the presence of single waves, one would like to have some quantities at hand which help identifying which modes one is dealing with. It would be helpful if one could measure simultaneously both, the wave frequency and the wave number spectra. This is possible, however, only with sophisticated multi-spacecraft constellations. And even then only in the rarest cases the determination of the dispersion relation from experiment will be possible. In application of these theoretical arguments to real observations one therefore has defined some quantities, called ``transport ratios", which have turned out to be quite valuable in helping identity some of the wave modes. Such transport ratios for electromagnetic waves have been given by \cite{Gary1993}. Below we list the most interesting for our purposes.

\paragraph{Polarisation.} The polarisation of a wave magnetic field with respect to wave number ${\bf k}$ is given by
\begin{equation}
P=ib_{S}/b_A
\end{equation}
where $b_S,b_A$ are the components of the magnetic fluctuation field ${\bf b}$ in the directions ${\bf S, A}$ of magnetosonic and Alfv\'en waves, respectively, i.e. the vector ${\bf A = k\times B}_0$ is perpendicular to the wave vector and the ambient magnetic field, while the vector ${\bf S}$ is perpendicular to ${\bf k}$ (because of the vanishing divergence $\nabla\cdot{\bf b}=0$) in the plane (${\bf k,B}_0$). The waves are more magnetosonic or more Alfv\'enic whether $|P|> 1$ or $|P|<1$, rspectively. For Re\,$P>0$ ($<0$) the waves are right-hand (left-hand) polarised.

\paragraph{Compression.} The magnetic compression of the wave measures the relative variation in the parallel magnetic fluctuation field
\begin{equation}
C_B=\langle b_\|^2\rangle/\langle |{\bf b}|^2\rangle
\end{equation}
where the fluctuations are taken at a given pair ($\omega,{\bf k}$). The angular brackets $\langle ab\rangle$ mean taking the real part of the correlation function of the two bracketed quantities.

\paragraph{Parallel compressibility.} This ratio together with the compression ratio provides a tool for estimating how compressive a wave is. It is defined for species $s$ as
\begin{equation}
C_{\| s}=\frac{B_0^2}{\langle b_\|^2\rangle}\frac{\langle b_\|\Delta N_s\rangle}{N_sB_0}
\end{equation}

\paragraph{Non-coplanarity ratio.} This ratio measures the fluctuating field component out of the plane (${\bf k, B}_0$, and is given by
\begin{equation}
C_c=\langle |b_A|^2\rangle/\langle |{\bf b}|^2\rangle
\end{equation}

\paragraph{Alfv\'en ratio.} Defining $\Delta {\bf V}_A={\bf b}/\sqrt{\mu_0m_iN}$, where $N$ is the total plasma density, the Alfv\'en ratio is defined as
\begin{equation}
R_{As}=\langle |\Delta{\bf V}_s|^2\rangle/\langle|\Delta{\bf V}_A|^2\rangle
\end{equation}
Here $\Delta{\bf V}_s$ is the flow velocity of species $s$. An Alfv\'en wave has $\Delta{\bf V}_i=\pm{\bf V}_A{\bf b}/{\bf B}_0$, and its own Alfv\'en ratio is $R_{Ai}=1$. The Alfv\'en ratio thus measures the fraction of Alfv\'en waves contained in the near-zero frequency fluctuations. 

\paragraph{Cross-helicity.} Helicity of a wave is another identifier of the wave mode, it is in particular useful for determining the direction of propagation of the wave by considering its sign. One has therefore defined a ratio which provides a measure of it:
\begin{equation}
\frac{1}{2}H_{cs}=\langle{\bf b}\cdot\Delta{\bf V}_s\rangle/[\langle |\Delta{\bf V}_s|^2\rangle+\langle |{\bf b}|^2\rangle]
\end{equation}
Parallel propagating Alfv\'en waves have helicity $H_{ci}=-1$. Fast magnetosonic nearly parallel propagating modes have also $|H_{ci}|\sim 1$, and $H_{ci}=0$ for perpendicular propagation.

\section{Anomalous Transport}

\noindent It has been mentioned several time that the evolution of shocks requires the generation of some kind of dissipation. Under the conditions of non-collisionality the generation of dissipation must be intrinsic to the plasma. In fact, assume that a source of free energy is switched on in the plasma. It is then quite natural to imagine that this available free energy will act on the plasma in a way to dissipate itself, distribute itself all over the plasma and ultimately transform itself into heat, entropy and create a new thermal equilibrium. Seen from this point of view the occurrence of instability is the first step in this chain of processes directed toward thermal equilibrium. Putting an obstacle into a fast but otherwise thermalized plasma stream is clearly a way of providing free energy, because seen from the frame of the obstacle the plasma is not in equilibrium; there is a high velocity difference between plasma and obstacle and thus a large amount of free latent energy available in the system. Dissipation, however, requires small scale interactions between the plasma constituents. Macroscopic motions with their long scales contribute to large scale structures but do not directly act on the microscopic scales. In order to affect the particle motion and contribute to friction among the particles small scale processes have to be called for. 

These processes take mostly place on the electron scales. Moreover, since it is much easier to excite fluctuations in the electric field than in the magnetic field, these interactions are electrostatic. In the following we consider the electrostatic fluctuations which are expected to contribute to the generation of anomalous dissipation. These processes can be divided into those which are not affected by the presence of an external magnetic field called unmagnetized, and those where the external magnetic field must be taken into account in the particle motion, i.e. magnetized processes. The distinction is made by the relation between plasma and cyclotron frequencies. 
 
When  $V_A\ll c\sqrt{m_e/m_i}$, the electron cyclotron frequency is much less than the electron plasma frequency, $\omega_{pe}\gg\omega_{ce}$ and one is dealing with an \index{instability!unmagnetized}\index{dispersion relation!unmagnetized}unmagnetized case. Also, when the entire dynamics is restricted to the direction parallel to the magnetic field, the problem can be considered to be unmagnetized. The complete unmagnetized dispersion relation including all species $s$ and their drifts $V_s$ is
\begin{equation}\label{chap2-eq-unmagdisp}
1+\sum\limits_sK_s(\omega,{\bf k})=0, \qquad\qquad K_s(\omega,{\bf k})=\frac{\omega_s^2}{N_s}\frac{\partial}{\partial\omega}\int\frac{{\rm d}v^3F_{0s}({\bf v})}{\omega-{\bf k\cdot v}}
\end{equation}
The function $K_s$ is the susceptibility contribution of species $s$ with average distribution function $F_{0s}$.  In a Maxwellian component plasma  $K_s=-(1/2k^2\lambda_{Ds}^2)Z'(\zeta_s)$  can be expressed through the plasma dispersion function $Z(\zeta_s)$, with $\lambda_{Ds}$ the Debye-length, $\zeta_s=(\omega-{\bf k\cdot V}_{s})/\sqrt{2} kv_s$, and thermal speed $v_s$ of component $s$. Note that here the sign of charge is included in the cyclotron frequency, i.e. for electrons $-\omega_{ce}$, for ions $+\omega_{ci}$.\index{susceptibility!unmagnetized}

When the plasma is magnetized, which applies to all other cases, the susceptibility becomes more involved. For Maxwellian components it reads
\begin{equation}\label{chap2-eq-suscept}
K_s(\omega,{\bf k})=\frac{1}{k^2\lambda_{Ds}}\left[1+\frac{\omega{\rm e}^{-\eta_s}}{\sqrt{2}|k_\||v_s}\sum\limits_{l=-\infty}^\infty I_l(\eta_s)Z(\zeta^l_s)\right], \qquad \zeta_s^l\equiv \frac{\omega+l\omega_{cs}}{\sqrt{2}|k_\||v_s}
\end{equation}\index{susceptibility!magnetized}
Here $\eta_s=(k_\perp r_{cs})^2$ with $I_l(\eta_s)$ the order-$l$ Bessel function of imaginary argument. 

This last expression suggests that magnetized electrons will support \index{waves!electron cyclotron}electron cyclotron harmonics $l\omega_{ce}$, which, when purely perpendicular, are Bernstein modes. There are also ion Bernstein  modes $l\omega_{ci}$, but (with the exception of injection of a localized perpendicular ion beam) they do not play any susceptible role in transport. Exciting both electron and ion Bernstein modes requires beams perpendicular to the magnetic field, in which case when applied to shocks the lower hybrid instability will become the most important agent in generating anomalous dissipation.\index{waves!Bernstein modes} \index{waves!lower hybrid}

As with the electromagnetic waves discussed in the previous sections the electrostatic dispersion relations can be solved analytically in closed form only in particular simple cases, and one has to retreat to a numerical approach. In the following we present the few most important cases and their effect on the generation of anomalous dissipation keeping in mind, however, that most collisionless shocks in the heliosphere are supercritical, and anomalous dissipation does not contribute substantially to their evolution and maintenance. Dealing in Chapter 5 with supercritical shocks, particle reflection is the dominant dissipation process, and we will refer to the instabilities of the previous section.

\subsection{Electrostatic wave particle interactions}\noindent
The wave friction term Eq.\,(\ref{chap2-eq-qlcollisionterm}) requires the determination of the fluctuation amplitudes of the wave modes that are responsible for causing anomalous friction. Again, the first step is to identify the unstable wave modes. In the second step we will then either have to determine the saturation level amplitudes of these waves or to consider their further interaction with particles or other waves. 

Here we list instabilities of interest in the anomalous dissipation process only. When discussing application to observations we will later in passing also mention instabilities which are involved in radiation from shocks. However, radiation\index{radiation}\index{waves!electromagnetic} provides no substantial energy loss, and the dissipation caused by radiation under the conditions of the nonrelativistic shocks in the heliosphere is completely negligible and does neither affect shock formation nor shock structure. One possible exception are shocks in the solar atmosphere which sometimes are accompanied by x-ray emission which, however, is not a genuine unstable plasma process in this case. Mentioning of such processes in relation to solar coronal observations is deferred to Chapter 8. 

\subsubsection{Unmagnetized electron and ion instabilities}\noindent
The unmagnetized dispersion relation in a Maxwellian component plasma consisting of one electron and one ion component has two solution, electron plasma waves or\index{waves!Langmuir waves} Langmuir waves $\omega^2=\omega_{pe}^2+3k_\|^2v_e^2$ at $k_\|\lambda_{De}\ll 1$ and ion-acoustic waves. The Langmuir wave can be driven unstable by a parallel electron beam of velocity $V_e\gtrsim 3v_e$. 

\paragraph{Ion-acoustic waves.} 

More interesting are ion-acoustic waves\index{waves!ion acoustic}. In the absence of any difference velocity between electrons and ions these are strongly damped plasma waves propagating along the magnetic field with dispersion
\begin{equation}\label{chap2-eq-ionacousticdisp}
\frac{\omega^2}{k_\|^2}\simeq\frac{3T_i}{m_i}+\frac{T_e}{m_i}\frac{1}{1+k_\|^2\lambda_{De}^2}, \qquad\qquad\qquad k_\|\lambda_{Di} \ll 1
\end{equation}
These waves become dispersive at larger $k_\|$ and for large electron temperatures with concave dispersion curve. At small parallel wave numbers their dispersion is linear resembling sound waves with velocity $c_{ia}^2\simeq (T_e+3T_i)/m_i$. Obviously for dispersion to compensate nonlinearity large electron temperatures $T_e>3T_i$ are required in which case also the damping is small. Ion acoustic waves if managing to overcome damping are therefore a candidate for electrostatic unmagnetized shock formation: Their dispersion favours shock ramps with slow moving wave trails and, as we will see later, they also can contribute to dissipation thus satisfying all conditions as candidates for shock formation. However, such shocks are purely electrostatic and do not affect the magnetic field. They will therefore only be of interest in sub-structuring magnetized shocks, possibly contributing to the formation of subshocks of short wavelength of the order of several Debye lengths.

In fact going from the kinetic description to the  fluid description and writing down the continuity and momentum conservation equations for ion acoustic waves for parallel propagation and one-dimensionality and combining it with Poisson's equation
\begin{equation}
\frac{\partial N}{\partial t}+\nabla_\|(NV)=0, \qquad \frac{\partial V}{\partial t}+V\nabla_\|V=-\frac{e}{m_i}E_\|, \qquad E_\|=-\nabla_\|\phi
\end{equation}
under the well justified assumption that for the low ion-acoustic frequencies the electrons behave as  thermalized hot Boltzmannians with density $N_e=N\exp (e\phi/T_e)$ depending exponentially on the electrostatic potential $\phi$, assuming quasi-neutrality and localized stationary solutions, we manipulate all these equations into the Sagdeev pseudo-potential form
\begin{equation}
\frac{1}{2}\left(\nabla_\|\phi\right)^2=-S(\phi), \qquad S=-\frac{m_iN_1V_1^2}{\epsilon_0}\left[\left(1-\frac{2e\phi}{m_iV_1^2}\right)^{\!\!\frac{1}{2}}+\frac{T_e}{m_iV_1^2}\,{\exp}\,\frac{e\phi}{T_e}\right]
\end{equation}
As usual the subscript 1 refers to values far upstream of the localized solution. In the absence of dissipation this solution for $S<0$ yields solitons of maximum potential amplitude $\phi_m$ found from setting the bracket to zero. $\phi_m$ corresponds to a maximum compressive amplitude $N_m=N\exp(e\phi_m/T_e)$. With ion acoustic Mach number ${\cal M}_{ia}=V_1/c_{ia}$ it is found that solitons exist only in the supersonic regime ${\cal M}_{ia}>1$. The soliton speed can be expressed through the maximum potential (or density via Boltzmann's expression) as
\begin{equation*}
\frac{V_1}{c_{ia}}=\frac{1}{\sqrt{ 2N}}\frac{N_m-N}{\{N_m-N[1+\ln(N_m/N)]\}^\frac{1}{2}}, \qquad\qquad m_iV_1^2>2e\phi_m
\end{equation*}
The condition on the maximum compression amplitude $N_m$ in this expression simply requires that the potential energy must be less than the initial flow energy. This sets a limit on the possible Mach numbers ${\cal M}_{ia}^2>2e\phi_m/T_e$ or, when combined with the definition of the latter,  ${\cal M}_{ia}^2/2>\ln(N_m/N)$. \index{soliton!ion acoustic}

Since the denominator in the former equation must be real, this condition requires that 
\begin{equation*}
\ln(N_m/N)+1-N_m/N<0
\end{equation*}
Expressing herein the density ratio through the Mach number, one finds that for ion-acoustic solitons to exist the Mach number is limited to values below a surprisingly small critical Mach number
\begin{equation*}
{\cal M}_{ia}<{\cal M}_{ia}^{\rm crit}\simeq 1.6
\end{equation*}
Ion acoustic solitons cannot exists for Mach numbers exceeding ${\cal M}_{ia}^{\rm crit}$. For higher inflow velocities $V_1$ the ion acoustic soliton will either not evolve or break down. \index{Mach number!critical ion acoustic}

The range of possible Mach numbers is rather limited which is simply due to the fact that for higher speeds the dispersion is unable to sustain a stationary state. Any solution will be non-stationary, wave\index{soliton!nonstationary} like or unstable. Concerning the formation of shocks one in addition to dispersion requires that dissipation is produced. Since it is known that ion acoustic waves are Landau damped with damping rate $\gamma_{L,ia}$, one can argue that Landau damping\index{waves!Landau damping} will cause a shock profile on the ion acoustic soliton with downstream state different from the upstream state \citep{Ott1969,Tidman1971}. The downstream density is then found to be $N_2\simeq N_1\exp [2({\cal M}_{ia}-1)]$ and will exhibit trailing oscillations, as has been discussed above. However, Landau damping takes time, and therefore the general argument applies to this kind of shock formation that the damping will not have time to work for large Mach numbers. Thus the damping argument applies only to subcritical shocks\index{shocks!subcritical} of Mach numbers smaller than ${\cal M}_{ia}\simeq1.6$. Such weak (electrostatic) ion acoustic shocks can indeed evolve and may contribute to sub-structuring of stronger supercritical shocks in the region where the Mach number has already dropped to values below the critical. 

\cite{Sagdeev1966} has favoured reflection of inflowing particles from the leading edge of the soliton \citep{Moiseev1963} over Landau damping. This reflection affects ions with energy less than the soliton potential $\phi_m$ and causes oscillations of long wave length. More important is that the reflected ions form an ion-ion beam configuration and are thus subject to the ion instabilities discussed previously yielding waves which may generate dissipation but do also propagate upstream of the shock where they cause wave particle interactions and retard the inflow ahead of the shock.
\begin{figure}[t!]
\hspace{0.0cm}\centerline{\includegraphics[width=0.65\textwidth,clip=]{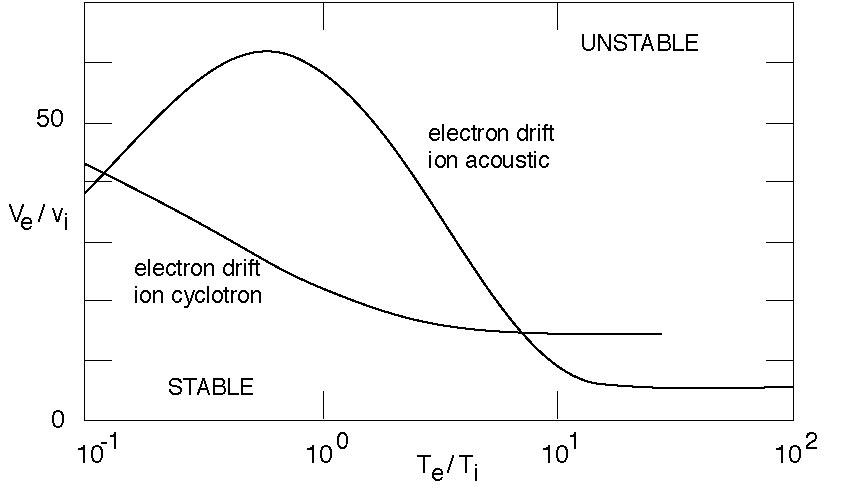} 
\includegraphics[height=0.375\textwidth,width=0.35\textwidth,clip=]{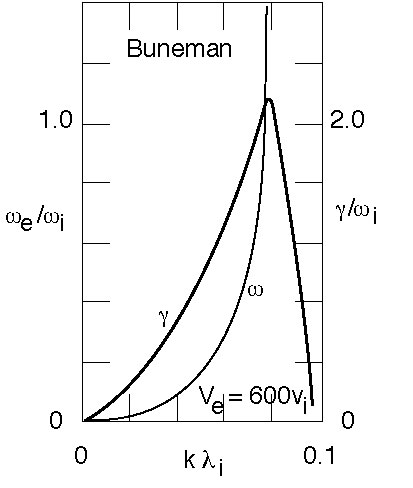} }
\caption[1]
{\footnotesize {\it Left}: The threshold drift speed for the electron current driven ion-acoustic instability as function of the electron to ion temperature ratio. For comparison the threshold for the parallel electron current driven ion cyclotron instability in a magnetized plasma is show for a dense plasma with $\omega_{pe}/\omega_{ce}=10$ \citep[after][]{Kindel1971}. {\it Right}: Buneman electron drift-ion two-stream instability, frequency and growth rate as function of wave number at large electron current drifts when the ion-acoustic instability has changed to the Buneman mode. }\label{chap2-fig-ionacthresh}
\end{figure}

\paragraph{Electron current driven ion acoustic instability.} 

So far we have not asked for the reason of an ion acoustic wave to grow. This can be achieved in the simplest way by letting one of the plasma components drift with respect to the other. If in a two-component plasma the electron drift with respect to the ions they effectively carry a current $j_\|=-eNV_{e\|}$ which is in most cases -- but not necessarily -- along the magnetic field. Here we assumed again quasi-neutrality $N_e=N_i=N$ which dispenses us from considering space charges and solving Poisson's equation. In this case assuming weak growth such that we can apply the general instability \index{instability!ion acoustic}theory with $\gamma\ll\omega$ the growth rate of the ion acoustic wave Eq.\,(\ref{chap2-eq-ionacousticdisp}) becomes
\begin{equation}\label{chap2-eq-ionacoustigrowth}
\frac{\gamma_{ia}}{\omega}=\sqrt{\frac{\pi}{2}}\frac{\omega^2}{k^3c_{ia}^2}\frac{{\bf k\cdot V}_e-\omega}{2v_e}\exp\,\left[-\frac{(\omega-k_\|V_e)^2}{2k_\|^2v_e^2}\right]
\end{equation}
Instability sets on for $V_e>\omega/k_\|$ when the electron velocity, which is the current drift velocity, exceeds the phase velocity of the ion acoustic wave, i.e. when -- approximately -- $V_e>c_ia$. These waves have relatively long wavelength $k_\|\lambda_{De}\ll 1$. The threshold for marginal stability of these electron current driven ion acoustic waves can be obtained from setting $\gamma_{ia}=0$, yielding
\begin{equation}
V_e\simeq \frac{\omega}{k_\|}\left[1+\left(\frac{m_i}{m_e}\right)^\frac{1}{2}\left(\frac{T_e}{T_i}\right)^\frac{3}{2}{\exp}\left(-\frac{3}{2}-\frac{T_e}{2T_i}\right)\right]
\end{equation}
The second term in the brackets results from the Landau damping of the ion acoustic waves. It is seen that this term disappears for hot electrons with $T_e\gg T_i$ thus lowering the threshold for instability to its marginally smallest value $V_e=c_{ia}$. The threshold is shown graphically as function of the temperature ratio in Figure\,\ref{chap2-fig-ionacthresh}. The threshold is measured in ion thermal speed $v_i$ and is quite high. Moreover, the electron temperature must be high implying that the electron distribution must be hotter than the ion distribution. In the solar wind this is usually satisfied. The physical reason for the electron temperature to be high is that the distribution must have a positive slope in $v$  in the region of overlap with the cold ion distribution for resonant instability, $\partial F_{0e}/\partial v_\| |_{\omega/k_\|}>0$. There must be more fast than slow electrons in the phase velocity frame of the wave in order to push the wave to higher momentum and energy, i.e. causing instability.

It is clear that this kind of interaction between the ion acoustic wave goes on the expense of the motional energy of the resonant electrons. Hence one expects that ion acoustic waves retard and scatter the current electrons thereby reducing the current flow. This resembles collisional friction which the resonant electrons experience and can thus be interpreted as the generation of an anomalous resistance in the plasma. An interpretation like this has been put forward by \cite{Sagdeev1966} and has been elaborated in depth afterwards. Below we will return to this theory.

\paragraph{Buneman electron current two stream instability.}

Returning to Figure\,\ref{chap2-fig-ionacthresh} we observe a change in the threshold\index{threshold!two-stream instability} curve for $V_e/v_i>\sqrt{m_i/m_e}T_e/T_i$. Here the electrons become thermally slow with respect to their drift velocity, $v_e<V_e$. The ion-acoustic instability under this condition changes into the Buneman instability which is an electron current fluid instability and is also known under the name electron current drift or electron-ion two stream instability. It has been discovered by \cite{Buneman1958} and favoured for application in shock physics already by \cite{Sagdeev1966}. It should be noted that the transition from ion-acoustic to two-stream instability\index{instability!two-stream} has been investigated in depth by \cite{Dum1979}.

Now, treating the electrons as cold the kinetic effects disappear, and the complex dispersion relation of the Buneman instability becomes
\begin{equation}\label{chap2-eq-buneman}
1-\frac{\omega_{pi}^2}{\omega^2}-\frac{\omega_{pe}^2}{(\omega-k_\|V_e)^2}=0
\end{equation}
Note that under these conditions the weak growth theory cannot be applied anymore. Instead one must find the growth rate from the complex solutions of this quartic expression. Fortunately, this equation can be solved since for resonant electrons the third term becomes dominant. The instability has real frequency $\omega\sim k_\|V_e$ and maximum growth rate of the order of the ion plasma frequency
\begin{equation}
\gamma_{\,\,\rm Bun, max}\simeq\frac{\sqrt{3}}{16^\frac{1}{3}}\left(\frac{m_i}{m_e}\right)^\frac{1}{6}\omega_{pi} \,\,\sim\omega_{pi}
\end{equation}
Figure\,\ref{chap2-fig-ionacthresh} on its right shows the Buneman frequency and growth rate for a case of very large electron current drift $V_e=600\, v_i$ in dependence on the wave number $k_\|\lambda_{Di}$. The maximum growth of the instability is close to $k_\|\lambda_{Di}\sim 0.9$ at short wavelengths slightly larger than the Debye length. The growth rate of this instability is very large. This implies that the instability is very strong and grows very fast thereby consuming a substantial fraction of the current streaming energy. The nonlinear treatment of this instability is of particular interest for shock physics. Since the instability grows so fast it makes little sense to treat it analytically for reasons which will become clear when dealing with the application of numerical simulations to shocks.\index{instability!Buneman}  
\begin{figure}[t!]
\hspace{0.0cm}\centerline{\includegraphics[width=0.9\textwidth,clip=]{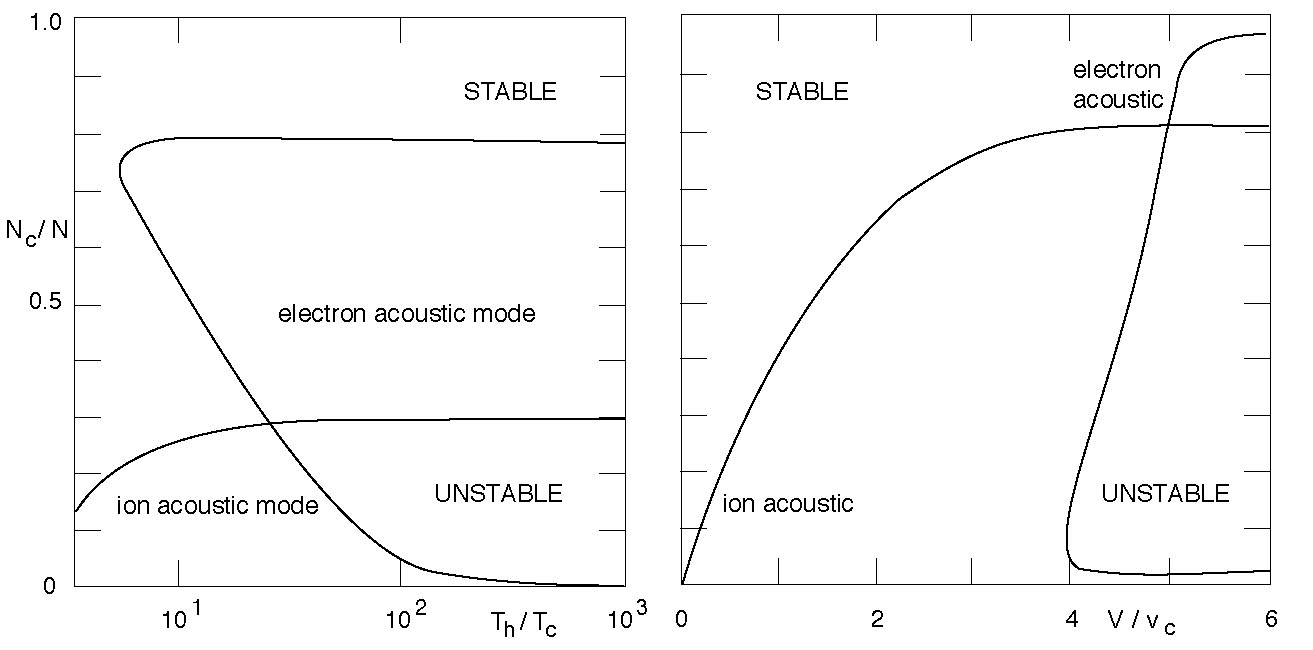} 
}
\caption[1]
{\footnotesize {\it Left}: Regions of existence of electron and ion acoustic modes in the density/temperature plane. Here $N=N_c+N_h=N_i$ is the total quasi-neutral density. The temperatures given as the hot to cold temperature ratio. The electron acoustic mode extends to larger cold density and higher hot temperatures than the ion acoustic mode. {\it Right}: Unstable versus stable domains for electron and ion acoustic modes in the density/drift velocity plane for the case when the electron acoustic mode is destabilised by a drifting hot plasma carrying a current. }\label{chap2-fig-elecacoust}
\end{figure}
The physics involved into the two stream instability can be described as follows. Both the electrons carrying the current and the ions are cold. The situation is thus two stream, and the instability is not resonant but reactive with all particles participating. This is the reason for its strength and rapidity. Because it consumes a fraction of the bulk flow energy of the electron current, the current becomes decelerated, and the energy is going mainly into the electrons which are heated by the instability until the instability stabilises when $V_e<v_e$. Then the ion-acoustic instability takes over. The Buneman two stream instability is thus accompanied by a burst in electron temperature and a rapid decrease in current. However, the final state of the instability is not a stationary state because the ion acoustic instability continues to grow with its own dynamics, possibly ending up in the formation of solitons when the Mach number has sufficiently decreased by current relaxation and heating, the former decreasing $V_e$, the latter increasing $c_{ia}$. But even during the blow-up phase of the two-stream instability structuring similar to soliton formation occurs. This can only be inferred from numerical simulation.

\paragraph{Modified two-stream instability.} \index{instability!modified-two-stream}This is a variant of the two-stream instability driven by a relative drift between electrons and ion across the ambient magnetic field ${\bf B}$ \citep{McBride1972}. The dispersion relation for the modified two-stream instability is
\begin{equation}
\left[1-\frac{\omega_{pi}^2}{(\omega-kV_i)^2}-\frac{\omega_{pe}^2}{\omega^2}\right]\left[1-\frac{\omega_{ce}^2\cos^2\thetabn}{\omega^2(1+k^{-2}\lambda_D^{-2})}\right]=\frac{\omega_{ce}^2\sin^2\thetabn}{\omega^2(1+k^{-2}\lambda_D^{-2})}\left[1-\frac{\omega_{pi}^2}{(\omega-kV_i)^2}\right]
\end{equation}
This expression is written here in the electron frame of reference and with angle $\thetabn$. One recognizes that the first term is the ordinary Buneman two-stream term. However, for oblique propagation $\thetabn\neq 90^\circ$ and $\thetabn\neq 0^\circ$ the two-stream mode couples to the whistler mode. It is this coupling which makes the modified two-stream instability interesting for shocks. Dispersion curves and growth rates are shown in Figure\,\ref{chap2-fig-mtsi} for $\thetabn=60^\circ$, and $V_i=V_A$, and an artificial mass ratio $m_i/m_e=80$ which has been taken in view of numerical simulations to be discussed later.
 
The modified two-stream instability (MTSI) operates also for relative drifts smaller than the electron thermal but larger than the ion-acoustic velocity and even for $T_i\sim T_e$ which makes it potentially important if only such perpendicular drifts can be generated. The unstable frequency is in the range of the lower-hybrid frequency. Hence the ions can be taken unmagnetized with strongly magnetized electrons. However, it requires oblique relative electron drifts since for perpendicular drift the instability disappears meaning that the unmagnetized ions propagate under an angle to the magnetic field while the magnetized electrons move only parallel to the magnetic field.
\begin{figure}[t!]
\hspace{0.0cm}\centerline{\includegraphics[width=0.7\textwidth,clip=]{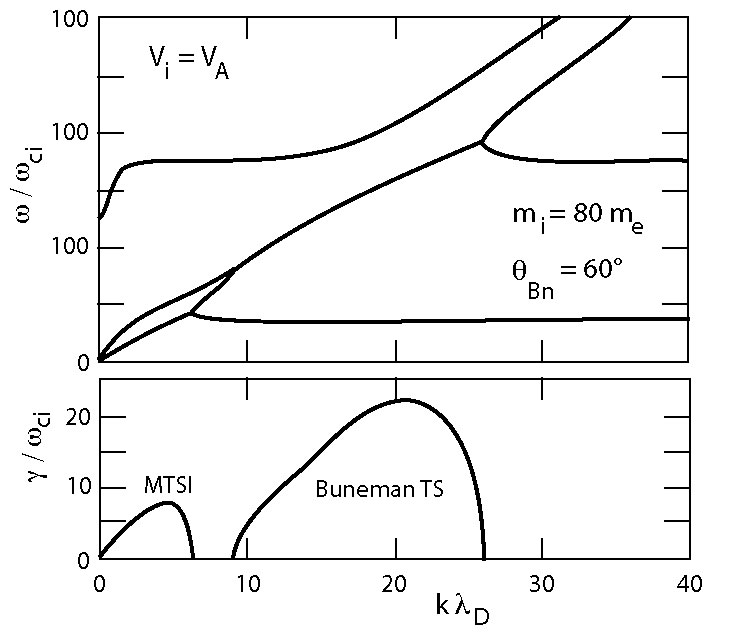} 
}
\caption[1]
{\footnotesize {\it Top}: Dispersion curves for the MTSI and Buneman TSI showing the coupling of the modes in dependence on wave number. {\it Bottom}: Growth rates for the Buneman TSI and MTSI. The Buneman TSI is at shorter wavelengths and higher frequencies but has larger growth rate while the MTSI has longer wavelengths, lower frequency. However the MTSI does not exist for 90$^\circ$, while it has much lower threshold than the Buneman TSI. }\label{chap2-fig-mtsi}
\end{figure}

\paragraph{Electron-beam electron-acoustic instability.} \index{instability!electron acoustic} 
A mode very similar to the ion-acoustic mode is the electron acoustic mode. Here the role of the ions is played by the cold (subscript $c$) electron background. In presence of another hot (subscript $h$) electron component the physics becomes very similar to the ion-acoustic wave, and a weakly damped resonant mode with real dispersion relation
\begin{equation}
\omega^2\simeq \omega_{pe,c}^2\frac{1+3k_\|^2\lambda_{De,c}^2}{1+1/k_\|^2\lambda_{De,h}^2},  \qquad\qquad k_\|\lambda_{DE,c}<1<k_\|\lambda_{DE,h}
\end{equation}
can propagate in the range of wave numbers indicated on the right. This weak damping can, like in the case of ion acoustic instability, overcome if the hot electron component drifts or if another electron beam is injected into the plasma. In the former case with ions and cold electrons at rest the drift implies current flow. The regions of existence and stability of the electron acoustic instability are shown in Figure \ref{chap2-fig-elecacoust} in comparison with the ion acoustic instability. Growth rates have been found numerically to be quite large, the order of the ion plasma frequency and thus similar to the Buneman mode. On the other hand, numerical experiments have not detected strong electron acoustic waves. 

Inspection of the electron acoustic dispersion relation shows that the mode is dispersive as well. For very low cold electron temperatures this dispersion is similar to ion acoustic waves suggesting that electron acoustic solitons could be formed in a similar way. In fact, such solitons have been calculated analytically for different parameter regimes \citep{Dubouloz1993}. However, observations do not seem to support their existence.  Also numerical simulations \citep{Matsukiyo2004} have not shown the formation of such solitons.  Clearly, electron acoustic waves can have a different dynamics because of the fast reaction times of the electrons, leading to rapid thermalization of the plasma. These questions  still remain to be open. In any case, if electron acoustic waves exist, the electron acoustic mode is quite well suited for plasma heating in shocks where plasmas of different temperatures mix. Its advantage is also that it proceeds on a very fast time scale close to the inverse of the plasma frequency. It is hence well suited for fast production of anomalous dissipation of energy. Moreover, since it very obviously damps rapidly it may act as an agent of about immediate transformation of excess energy in the electrons into heating electrons. None of these question has been understood nor answered properly at the time of writing.

\paragraph{Ion-beam ion-acoustic instability.} \index{instability!ion beam} 

An instability very similar to electron beam driven electron acoustic instability is its ion pendant when a cold ion core and hot ion beam in presence of a neutralising electron background become unstable \citep{Gary1987}. The mode excited in this case is again the ion acoustic mode, this time propagating at very low phase velocities $\omega/k\simeq c_{ia}(N_{i,c}-N-{i,b})/N_e$ less than $c_{ia}$. In this range the waves have no dispersion. This wave is, however, very easy to destabilise because of its low beam velocity threshold\index{threshold!ion-acoustic} which can lie even below the thermal speed of the ion core component. On the other hand the growth rates of this instability are very low. Measured as before for the electron-acoustic, ion-acoustic and Buneman instabilities in terms of the ion plasma frequency, its maximum growth rate is just of the order of a few per cent of $\gamma_m\sim 0.3\,\omega_{pi}$, very small compared to the growth rate of the electron-current driven ion acoustic instability which was of the order of $\gamma_m\sim \omega_{pi}$. It should thus be less important in anomalous processes.

\paragraph{Electrostatic ion-cyclotron harmonic instabilities.}\index{instability!ion-cyclotron harmonic} 
So far we have considered only unmagnetized instabilities. We now turn to listing the most important -- in view of the anomalous processes we have in mind when considering shock formation -- {\it magnetized electrostatic} instabilities. These instabilities occur when a magnetic field is present in the plasma and the electrons must be considered to be magnetized. 

The frequencies of the waves in question will therefore be well below the electron cyclotron frequency $\omega\ll\omega_{ce}$ falling into the range of and above the ion cyclotron frequency $\omega\gtrsim\omega_{ci}$. According to linear theory the magnetized modes in this range occur at harmonics $\omega_l\sim l\omega_{ci}$ of the ion cyclotron frequency. We are thus dealing here with with electrostatic ion-cyclotron harmonics.

Following a suggestion by \cite{Drummond1962} ion cyclotron instabilities have been proposed as generating anomalous collisions by \cite{Kindel1971} who advocated their importance because of their apparently lower instability threshold than the ion acoustic instability for electron current shown in Figure\,\ref{chap2-fig-ionacthresh}. In fact ion-cyclotron harmonic waves oblique to the magnetic field can become unstable in the presence of an electron current flowing along the magnetic field. The reason is that --  in contrast to the earlier mentioned strictly perpendicular Bernstein mode\index{waves!Bernstein modes}  resonances -- such oblique waves posses a field-aligned projection component of their electrostatic field which can resonant with the parallel current electrons via the Doppler-shifted resonance condition $k_\|V_{\|e}=\omega\pm l\omega_{ci}$ of which only Landau resonance $l=0$ is effective for $|\omega-k_\|V_{e\|}|<\sqrt{2}|k_\| |v_e$. The ions contribute only a weak resonant damping of the waves for $l=-1$.
Note that the obliqueness of propagation of these modes in contrast to bernstein modes implies that the resonance does not occur precisely at the harmonics but lies somewhere in between in the harmonic bands where the dispersion curves have particular geometrical forms \citep[cf., any book on basic plasma wave physics, e.g.,][]{Baumjohann1996}. Its precise location depends heavily on the exact prescribed conditions of the plasma and current velocity $V_{e\|}$, and no analytic expression can be provided.  

Strongest weak instability occurs in the harmonic range $1<l<2$ with growth rate $\gamma\ll\omega$ and $k_\|r_{ci}\sim 1$, i.e. wavelengths comparable to the thermal ion gyro-radius in the frame of the ions and propagation angles $\theta<85^\circ$. The velocity threshold\index{threshold!ion-cyclotron}  decreases with increasing electron temperature $T_e/T_i$ simply because more electrons go into resonance then. But for large ratios it is taken over by the ion acoustic instability as seen from Figure\,\ref{chap2-fig-ionacthresh}. (Note that in the solar wind/bow shock system, for instance, the ratio is about $T_e/T_i\approx 10$ changing across the bow shock to become $T_eT_i<1$; hence one may expect the first electrostatic ion-cyclotron harmonic to be present wherever parallel electron currents flow in the shock on the upstream side, while downstream neither current driven ion cyclotron nor ion acoustic instabilities should exist.)

We have already noted that ion beams can also excite ion cyclotron harmonic waves. Parallel beams excite similar waves with similar properties like parallel electron currents for propagation angles $0^\circ\ll\theta<90^\circ$ rather close to perpendicular. Perpendicular ion beams, on the other hand excite a broad spectrum of ion cyclotron harmonics on the background ion component depending on beam energy and background temperature. These excitations are restricted to a distance of the order of the ion gyro-radius $r<r_{ci,b}$ of the beam ions only, because at larger perpendicular scales the beam ions are themselves magnetized. Thus this kind of excitation is of importance merely when heavy ion beams penetrate the plasma, for instance when a heavy ion beam component of large perpendicular energy is reflected from a supercritical shock wave back upstream, or when ions become highly accelerated in interaction with the shock and penetrate across the shock onto the downstream side where they have much larger gyro-radii than background ions.

\paragraph{Electrostatic electron-cyclotron instability.} This instability \index{instability!electron-cyclotron harmonic} is the pendant to the former ion cyclotron instability at the much higher frequency perpendicular (or oblique) electron cyclotron harmonics (for purely perpendicular propagation these are the Bernstein modes). \index{waves!Bernstein modes}It is driven by the coupling between a sufficiently fast ion beam and the electron cyclotron harmonics at perpendicular wave numbers $k_\perp> 0$ and can also be driven unstable when reflected ions are present for instance in the foot of the supercritical perpendicular shock. 

\subsubsection{Electrostatic drift instabilities in inhomogeneous plasma.}\noindent\index{instability!drift} 
The last group of instabilities we will refer to in the context of shock physics are instabilities that are caused in presence of plasma inhomogeneity. Three basic kinds of plasma inhomogeneities can be identified: density $N({\bf x})$, magnetic ${\bf B(x)}$, and temperature $T({\bf x})$ real space inhomogeneities. The dependencies of these (average) quantities on space may in most cases not be independent. In the following, for the purposes of investigation of their effects on wave excitation, we will take them as being in fact mutually independent. For simplicity we will take into account only density inhomogeneities perpendicular to ${\bf B}_0$ on scales much larger than the wave length. We also assume quasi-linearity which is justified because under the assumption of weak gradients the effects of the inhomogeneity will be weak as well and thus cause only slow  wave growth. In this case one  can expand the density with respect to the perpendicular direction $x$ up to first order and write \index{instability!absolute} 
\begin{equation*}
N_s(x)=N_{s0}(1+\epsilon_N x), \qquad\qquad {\rm where}\qquad\qquad \epsilon_N\equiv [\nabla_x N_s(x)]_{x=0}
\end{equation*}
The effect of the inhomogeneity is that the magnetized particle component behaves adiabatically and starts performing a diamagnetic drift motion $V_{Ns}\,\hat{\bf y}= (\epsilon_Nv_s^2/\pm\omega_{cs})\,\hat{\bf y}$ perpendicular to the magnetic field and density gradient into $\pm y$ direction, depending on the sign of the particle charge. The $\pm$-sign in the denominator indicates that the cyclotron frequency is taken here including the sign of the charge. Drift motions of this kind cause a perpendicular drift current $j_{dy}$ to flow in the plasma because particles of different charge sign move in opposite directions. The waves excited under such conditions satisfy the modified dispersion relation
\begin{equation}\label{chap2-eq-inhomdisp}
1-\epsilon_N+\sum_sK_s({\bf k}, \omega, \epsilon_N)=0
\end{equation}
The new susceptibility $K_s(\epsilon_N)$ is still local, i.e. it changes on scales larger than the gradient scale, but locally depends only on $\epsilon_N$. It is of the same for as Eq.\,(\ref{chap2-eq-suscept}) with two differences: because of the occurrence of the finite perpendicular drift $V_{dy}\hat{\bf y}$ the frequency  in the factor in front of the sum in (\ref{chap2-eq-suscept}) is exchanged with the Doppler-shifted frequency $\omega\to\omega-k_\perp V_{Ns}$. In addition a new drift term
\begin{equation*}
+\frac{\omega\,V_{Ns}}{k_\perp v_s^2}{\rm e}^{-\eta_s}\sum\limits_{l=-\infty}^\infty\frac{l\omega_s\,I_l(\eta_s)}{\omega+l\omega_{cs}}
\end{equation*}
must be added inside the brackets. Instabilities resulting from this dispersion relation in the weak instability limit bear the general name of {\it drift} or {\it universal} instabilities. They resonant with the drift motion and have frequencies $\omega\simeq k_\perp V_{Ni}$ and long wave lengths satisfying $k_\perp r_{ci}\lesssim 1$. for smaller wavelengths these waves are highly dispersive and thus can form nonlinear structures.

This mode becomes particularly interesting and important in shock physics when the drift speed is so large that the ions can be considered as unmagnetized. This happens because the frequency of the drift mode increases with $V_N$ and quickly exceeds the ion cyclotron frequency.  In this case the frequency of maximum growth is close to the lower-hybrid frequency $\omega_{lh}$ and the drift mode becomes a lower-hybrid drift wave.  

\paragraph{Lower-hybrid drift instability.}\index{instability!lower-hybrid drift}  For unmagnetized ions, when the frequency of the drift wave is $\omega\gg\omega_{ci}$, the ion cyclotron frequency can be neglected and the susceptibility simplifies
\begin{equation}
K_s(\omega,{\bf k})=\frac{1}{k^2\lambda_{Ds}^2}\left[1+\frac{\omega-{\bf k_\perp\cdot V}_{Ns}}{\sqrt{2}kv_i}Z\left(\frac{\omega}{\sqrt{2}kv_i}\right)\right]
\end{equation}
The positive slope on the distribution function which is responsible for instability is in this case on the perpendicular part which depends on the drift velocity. There the maximum of the distribution is shifted out of the origin to the location of the drift velocity. The real frequency and growth rates are given by \citep{Gary1979}
\begin{eqnarray}
\frac{\omega}{k_\perp V_{Ne}}&\simeq& -\left[\left(1+\frac{T_e}{T_i}\right)\frac{{\rm e}^{\eta_e}}{I(\eta_e)}-1\right]^{-1} \\ 
\frac{\gamma}{\omega}&\simeq&\sqrt{\frac{\pi}{2}}\frac{T_e}{T_i}\frac{V_{Ne}}{v_i}\frac{{\rm e}^{\eta_e}}{I_0(\eta_e)}\left[\left(1+\frac{T_e}{T_i}\right)\frac{{\rm e}^{\eta_e}}{I(\eta_e)}-1\right]^{-2}
\end{eqnarray}
The unstable wave propagates antiparallel to the direction of the electron gradient drift. i.e. in the direction of the electric drift current. Maximum growth of this lower hybrid wave has been found at long wavelengths $k_{\perp, m}r_{ci}\sim \sqrt{m_i/m_e}$ over a relatively broad frequency range close to the lower hybrid frequency $\omega_{\,lh}\simeq \omega_{\,ce}\sqrt{m_e/m_i} $. 

The reason for the lower-hybrid drift instability to just excite the lower-hybrid frequency is that it is a perpendicular two-stream instability\index{instability!two-stream} \index{instability!Buneman}  similar to the Buneman mode that is, however, driven by the bulk velocity difference between electron and ion gradient drifts. It thus propagates on the background ion component. In fact in the ($\omega,k_\perp$)-plane the drift-beam mode $\omega_d\simeq k_\perp V_{Ne}$ couples to many ion cyclotron harmonics thereby exciting almost all of them. However, largest growth occurs in the harmonic dispersion band that contains the lower hybrid frequency which is a strong plasma resonance. 

The lower-hybrid drift instability is  the strongest in the family of the electrostatic ion cyclotron instabilities. As a two stream instability its maximum growth rate remains still modestly large being of the order of a few ion cyclotron frequencies, $\gamma_{\,lh,m}\simeq (1-3)\omega_{\,ci}$. In a plasma leaving sufficient time $\tau_{nl, sat}\gg\omega_{ci}^{-1}$ for growth and saturation it may well play a substantial role in generating dissipation. We will see later that this instability indeed provides the highest so far inferred from instabilities anomalous collision frequency which turns out to be of the order of the lower-hybrid frequency itself. Its relevance in application to collisionless shocks is however questionable, because of the above argument.

In addition, the lower-hybrid drift instability\index{instability!stabilization} appears to stabilise under $\beta> 1$ conditions  \citep{Davidson1977}. In application to shocks this restriction, if it translates into the nonlinear regime,  presents a severe barrier to the use of the lower hybrid drift instability as generator of anomalous resistance, dissipation and entropy generation.

\subsection{Anomalous resistivity}\noindent\index{resistivity!anomalous}
Resistivity is defined via the Drude formula $\eta=\nu/\epsilon_0\omega_{pe}^2$, with $\nu=\sigma_cNv_e$ the collision frequency. The latter, under collisionless conditions, becomes the anomalous collision frequency $\nu_a$ and is the quantity containing the interaction between electrons and the nonlinear wave fluctuations. 

This becomes obvious when realising that the Spitzer-Coulomb \index{collisions!frequency, Spitzer}collision frequency $\nu_{\rm C}\sim\omega_{pe}/N\lambda_{De}^3$ is proportional to the ratio of the  {\it plasma wave fluctuation level in thermal equilibrium} $W_{th}=\frac{1}{2}\epsilon_0 \langle {\bf e}^2_{th}\rangle$ to thermal energy, $\nu_{\rm C}\sim\omega_{pe}W_{th}/NT_e$. Under saturated instability conditions it is then reasonable to assume that the actual fluctuations $\langle{\bf e}^2\rangle$ replace the thermal fluctuations in this expression which yields the Sagdeev formula
\begin{equation}\label{chap2-eq-sagdeevformula}
\nu_a\simeq\frac{W_{sat}}{NT_e}\omega_{pe} 
\end{equation}

The problem is thus reduced to the determination of the {\it nonlinear saturation level of the unstable wave spectrum}.  Its determination requires knowledge of the electric current ${\bf j}\simeq-e\langle N{\bf V}_e\rangle$ as a functional of the electric wave fluctuation field ${\bf e}$. (We are speaking here only of electrons since in collisionless plasma electrons -- because of their much faster mobility than ions -- are the particles which carry the electric current. The electrons feel the friction of the waves and become retarded by anomalous collisions thereby dissipating the kinetic energy of the current and contributing to collisionless Joule heating of the plasma.) 
The evolution of the electron current is -- in principle -- given by the electronic part of Eq.\,(\ref{chap2-eq-momentequ}), respectively Ohm's law (\ref{chap2-eq-ohmslaw}), if on the right-hand sides the average anomalous electronic friction terms (\ref{chap2-eq-qlcollisionterm}) are added, since these are the crucial terms containing the wave-particle interactions. \index{Ohm's law}

We are interested only in the parallel collision frequency here. In the nonlinear stationary state the time derivatives can be neglected. In order to obtain a first expression for the parallel anomalous collision frequency $\nu_a$ we assume that the last term in Eq.\,(\ref{chap2-eq-momentequ}) is of the form 
\begin{equation*}
-\frac{\nu_a}{m_e} NV_\|= \int {\rm d}v^3  v_\| {\cal C}_e
\end{equation*}
When inserting from Eq.\,(\ref{chap2-eq-qlcollisionterm}), keeping only the parallel component and the electric part which in the microscopic interactions dominates over the electromagnetic fluctuations since these affect only frequencies below $\omega_{ci}$. One obtains for the anomalous collision frequency 
\begin{equation}\label{chap2-eq-nuanomalous}
\nu_a \simeq \frac{1}{NV_\|} \nabla_\| \langle W_{sat}\rangle
\end{equation}
which is general but still preliminary. It requires knowledge of the average power density of the electric wave field $\langle W_{sat}\rangle=\frac{1}{2}\epsilon_0\langle |{\bf e}^2|\rangle$ which here can be of arbitrarily large amplitude and arbitrary spatial structure. Remember that the only condition implied in deriving (\ref{chap2-eq-qlcollisionterm}) was that the fluctuations were fast  both in space and time compared to the slow changes in the plasma background quantities. In this sense the parallel gradient operator $\nabla_\|$ acts on the slow variability of the wave power. 

In order to proceed, another equation is required which determines the evolution of the wave power. This lacking equation can only be formulated in Fourier space $(\omega, {\bf k})$ and should contain all the nonlinear interactions and thus cannot be of general nature. A simplifying assumption is that it describes the nonlinear evolution according to kind of wave-kinetic equation\index{equation!wave kinetic}
\begin{equation}\label{chap2-eq-kinwaveeq}
\frac{{\rm d}}{{\rm d}t}W_k \simeq \frac{\partial W_k}{\partial t}+\left({\bf V}-\frac{\partial \omega}{\partial{\bf k}}\right)\cdot\nabla W_k= 2\gamma\,(\omega, {\bf k}, W_k) W_k + \dots
\end{equation}
We assume the system has reached stationarity such that the wave spectrum $W_k$ does not  evolve with time anymore. In this case the left-hand side simplifies, and we can express the spatial derivative of the stationary wave spectrum as 
\begin{equation}
\nabla_\| W_k \simeq \frac{2\gamma\,(\omega,{\bf k}, W_k)}{|V_\|-\partial\omega/\partial k_\| |} W_k
\end{equation}
which after transforming from Fourier into real space and averaging over the fast fluctuations can be used in the above expression (\ref{chap2-eq-nuanomalous}) to express the anomalous collision rate through the average wave power. Such expressions will be given below. Usually the current-drift speed $|V_\||\gg|\partial\omega/\partial k_\||$ is much larger than the wave group velocity, and the latter can be neglected. This yields the inverse square dependence of $\nu_a\propto |V_\||^{-2}$ on the current drift velocity.\index{collisions!frequency, anomalous}

The derivation of an anomalous collision rate contains a number of crucial assumptions of which the most severe concern the simplifications in the terms in the kinetic wave equation (\ref{chap2-eq-kinwaveeq}) where we suppressed a refraction term $(\nabla\omega)\cdot(\partial W_k/\partial_{\bf k})$ -- which becomes important in strongly inhomogeneous plasmas like in a shock ramp -- and neglected all terms on the right with the exception of the growth rate term. Since the dependence of $\gamma$ on the wave amplitude has been retained, some generality in the nonlinearity is still retained. Stronger simplifications are made when restricting to pure quasilinear theory. In this theory the nonlinear dependence of the growth rate on the wave amplitude is dropped. This confirms to the  conventional approach to anomalous dissipation. The most elaborated {\it quasilinear} theory (valid for any direction of propagation including electromagnetic contributions) can be found in \cite{Yoon2007} where also the application to one particular mode (the lower-hybrid drift mode discussed below) is given and a rudimentary contribution of the neglected Coulomb collision term is retained.\index{resistivity!quasilinear}\index{dissipation!anomalous}

In any case the mechanism of saturation of the nonlinear wave field must be known in order to obtain a useful practical expression for $\nu_a$. In the following we review only the three wave modes that contribute strongest to anomalous resistance, ion-acoustic, Buneman two-stream, and lower-hybrid drift modes.

\subsubsection{Nonlinear evolution of waves}\noindent\index{waves!nonlinear evolution}
Quasilinear theory \citep{Sagdeev1979,Yoon2007} is so far the simplest and most effective approach to the calculation of anomalous collision frequencies. This approach uses the linear growth rate $\gamma({\bf k})$ of the instability yielding the simplified formula
\begin{equation}
\nu_a\simeq\frac{1}{m_eNV_\|^2}\int\frac{{\rm d^3}k}{8\pi^3}\frac{{\bf k\cdot V}_e}{\omega({\bf k})}\gamma({\bf k})W_{sat}({\bf k}), \qquad W_{sat}({\bf k}) \equiv \frac{\epsilon_0}{2}\frac{\partial \omega\epsilon(\omega,{\bf k})}{\partial\omega}\langle|{\bf e}^2|\rangle 
\end{equation}
It requires in addition knowledge of $\gamma({\bf k})$ and the drift velocity ${\bf V}_e= V_\|\hat{\bf z}$.

\paragraph{Anomalous ion acoustic collision frequency.} For instance, from this expression, assuming $V_\|>c_{ia}$ and $k_{max}\lambda_{De} \sim 1$, and $\gamma_{ia}\sim\omega V_\|/v_e$ which holds for the ion acoustic instability in the large drift limit, one just obtains the above Sagdeev expression (\ref{chap2-eq-sagdeevformula}) for the anomalous collision rate which is good for application when the saturation level of the instability is measured. \index{collisions!frequency, ion-acoustic}

Ion acoustic waves saturate by scattering off thermal plasma ions according to the resonant scattering process $\omega-\omega'=({\bf k-k'})\cdot{\bf v}_i$ where the prime $' $ indicates the frequency and wave numbers of the scattered wave. This process has been used by \cite{Sagdeev1966} to explicitly calculate the ion acoustic anomalous collision frequency
\begin{equation}
\nu_{a,ia}\simeq 0.01\omega_{pi}\frac{V_\|}{c_{ia}}\frac{T_e}{T_i}\theta^{-2}
\end{equation}
which holds for large electron current drifts $V_\|\gg c_{ia}$ and for narrow angles $\theta>0$. Actually, experiments have shown that this expression overestimates the anomalous resistance suggesting that anomalous collisions are less effective than theory predicts. More precise theories than the above estimate have been developed by \cite{Vekshtein1970}, and have been reviewed by \cite{Sagdeev1979}. 

\paragraph{Anomalous two-stream collision frequency.} As noted earlier, the two-stream instability is a very strong instability causing large current momentum losses and rapid plasma heating. It switches itself of during evolution and will therefore not be a stationary instability. It causes, in fact, different effects which probably destroy its direct importance in collision generation. \index{collisions!frequency, two-stream}

It has phase speed $\omega/k\sim v_e\sqrt{m_e/m_i}$ substantially below electron thermal speed implying that it stays relatively long in the volume of excitation which supports its effect on the local plasma. Its wave energy density, from simple consideration is less than ion energy $W_{ts}\leq T_i$. It yields a large theoretical collision frequency 
\begin{equation}\label{chap2-eq-tscollrate}
\nu_{a,ts}\simeq \omega_{pi} \gg \nu_{a,ic} \qquad {\rm with}\qquad \nu_{a,ic}\sim 0.3\,k_\|v_e<\omega_{ci} 
\end{equation}
of the order of the ion plasma frequency and several orders of magnitude larger than Spitzer-Coulomb collision frequency. As such it is a serious candidate for generating anomalous dissipation and heating in shocks whenever a two-stream situation is encountered. This is indeed frequently the case as we will see in the supercritical shock Chapters 4 and 5. 

If in the above Eq.\,(\ref{chap2-eq-tscollrate}) we compare the two-stream collision rate, for instance, with the ion-cyclotron collision rate $\nu_{a,ic}$ that had been favoured by \cite{Kindel1971}. We find that it has far larger growth rate than the ion cyclotron wave and will thus always dominate when the current is strong, $V_\|>v_e$. At weaker currents the ion-acoustic will be faster (because of the slowness of the ion-cyclotron instability) as long as the magnetic field is weak and $V_\|\gg c_{ia}$.

\paragraph{Anomalous lower-hybrid drift collision rate.} As noted earlier, the lower-hybrid drift instability is particularly important as it is the exceptional representant of a fast growing (universal) drift wave instability which in the presence of gradients in plasma can always be expected to grow. Clearly shocks are a particularly good candidate for such an instability because of the steep density and magnetic field gradients occurring in compressive shocks. Moreover, in application to shocks, other than at pressure equilibrium boundaries like the magnetopause \citep[see, e.g.,][]{Treumann1991,Winske1995}, the magnetic gradient adds positively to the growth rate.\index{collisions!frequency, lower-hybrid drift} 

In calculating the quasilinear saturation level of this instability makes use of the wave number at maximum growth $k_{max}^2\lambda_{Di}^2=2/(1+\omega_{pe}^2/ \omega_{ce}^2)$ and $\partial(\omega\epsilon)/\partial \omega=1+\omega_{pe}^2/\omega_{ce}^2\gg 1$ in dense plasma like the vicinity of the bow shock, for instance. The saturation wave level \citep{Davidson1978} then follows after solving the quasilinear diffusion equation to be fraction of the drift energy
\begin{equation}
W_{sat,lh}\simeq \frac{m_eNV_{de}^2}{8(1+\omega_{pe}^2/ \omega_{ce}^2)}
\end{equation}
From here the anomalous collision rate follows as
\begin{equation}
\nu_{a,lh}\simeq \sqrt{\frac{\pi}{2}}\left(\frac{\omega_{pe}^2}{\omega_{lh}}\right)\frac{W_{sat,lh}}{NT_i}\approx \sqrt{\frac{\pi}{2}}\left(\frac{r_{ci}}{4L_N}\right)^2\omega_{lh}
\end{equation}
proportional to the ratio of ion-gyroradius to density gradient scale length $L_N=|\nabla \ln N|^{-1}$. This growth rate is large for steep gradient scales close to the ion gyro-radius, a condition that holds in the shock ramp. The propagation of electrons in the lower-hybrid drift case is  perpendicular to the magnetic field since the electrons perform a diamagnetic drift which constitutes the electric current. In the above one-dimensional theory everything has been reduced to the coordinate parallel to the current. In a shock wave this current will flow along the shock surface, perpendicular to the magnetic field and shock ramp density/field gradient.\index{scales!density gradient scale} 

The anomalous lower-hybrid drift collision frequency is large and renders the lower-hybrid drift instability a viable candidate for generation of anomalous dissipation in shocks if only the condition that maximum growth is found for $\beta < 1$ can be circumvented. Shocks in the heliosphere in most cases satisfy the inverse condition $\beta \gtrsim 1$. Whether this indeed provides a serious restriction has not yet been clearly verified, however, neither theoretically nor in numerical simulation or observation.

Recently \cite{Yoon2007} reviewed the theory of anomalous resistivity for the lower-hybrid drift and two-stream instabilities and derived some (slightly) more precise (but considerably more involved) expressions than the formulae given above. They also included electromagnetic effects and arbitrary directions of propagation to find that the lower-hybrid drift anomalous resistance can be very large indeed, up to a factor of $10^{10}$ or more larger than Spitzer-Coulomb resistance.

\paragraph{Runaway effects.} Since the collision frequency is an inverse function of the particle current drift velocity it allows for the interesting effect that particles of sufficiently high speed cannot be captured by the electric field. They instead get another push and escape as so-called runaway electrons. This effect is known since long and applies to some of the anomalous collision processes as well as to Spitzer collisions. The reason is that the wave level saturates yielding a constant collision frequency for every mode in question. Hence, fast particles do simply not interact but escape like in free ballistic flight. Hence there will always be some particles which behave like freely streaming particles. These, when flowing along the magnetic field, constitute a moderately energetic particle beam and may provide a seed particle population for further acceleration even in the presence of anomalous collisions. \index{acceleration!runaway}

\paragraph{Other effects.} Several other effects have not been mentioned here in relation to anomalous effects. These are wave decay \index{process!wave decay}processes, generation of radiation\index{process!radiation generation}, and resonant wave absorption processes\index{process!resonant absorption} in inhomogeneities. It is interesting to note that, historically, \cite{Sagdeev1966} in following calculations of \cite{Moiseev1963} proposed that such wave decay processes would contribute substantially to anomalous dissipation in subcritical shocks by enhancing the number of waves present and thus enhancing the probability of particle scattering by waves. 
\subparagraph{Wave decay processes.} The first of these belong to the class of wave-wave interaction which are the pendant of collisions between particles in the wave picture without invoking particles. Such interactions must satisfy the wave momentum and wave energy conservation\index{waves!wave-wave interaction} laws which can be written simply as
\begin{equation}
\sum\limits_\alpha {\bf k}_\alpha=0, \qquad\qquad \sum\limits_\alpha \omega_\alpha({\bf k}) =0
\end{equation}
where the index $\alpha$ marks the particular wave mode, and each pair $[\omega_\alpha({\bf k}), {\bf k}_\alpha]$ satisfies its own linear (or nonlinear) dispersion relation.  According to the number of waves involved, the smallest possible number is 3, there are three-wave interactions, four-wave interactions, $\dots$ with for weak interactions with decaying probability. For the three-wave case one important case is that a strong wave (one that has become very strong due to linear wave growth) such a three wave process can lead to wave decay into two other weaker modes of different dispersion due to the relations
\begin{equation}\label{chap2-eq-decay}
\omega_{\,\,0}\to \omega_1+\omega_{\,2}, \qquad {\bf k}_0 +{\bf k}_1 +{\bf k}_3=0, \qquad \omega_{\,\,0}\gg (\omega_1,\omega_{\,2})
\end{equation}
We do not go into the details of these process; they are rather involved because of the complexity of the three dispersion relations which have to be taken into account and which can be different when waves from one branch jump over into another branch feeding a wave mode there. These processes have, however, a number of consequences of which four are important: 
\begin{itemize}
\item They contribute to the excitation of wave modes in a plasma which are not directly driven by an instability but result from the decay of an instability-driven large amplitude intense wave, in this way contributing to the production of a broad spectrum of turbulence \index{waves!turbulence spectrum}that consists of many different and even possibly weakly damped modes\index{waves!weakly damped} in the plasma which otherwise would not exist., when only the decay is stronger than the natural damping of the wave. 
\item Decay processes limit the intensity of an instability-driven mode and reduce it substantially to the advantage of other modes. In this way they weaken the contribution of the particular mode to anomalous collisions while they might enhance collisionality by producing a broad turbulent  background fluctuation spectrum. 
\item By generating other waves in the plasma they may provide a background from which other instabilities may grow which are driven by sources which otherwise would not overcome the instability threshold.\index{threshold!for instability} 
\item The broad spectrum produced in plasma wave decay processes may move upstream of the shock and modify the upstream conditions in a way not foreseen in the Rankine-Hugoniot relations.  Hence such processes cannot be handled in simple plasma modelling of shock wave generation. They can only be investigated by particularly tailored numerical simulation techniques.
\end{itemize}

\subparagraph{Radiation.} \index{radiation}Only radiation generated from plasma waves is of interest in heliospheric shocks because the densities are generally far too low for reaching a substantial emission measure in synchrotron or x-ray radiation. Radiation can then be generated only by mode conversion\index{waves!mode conversion}\index{waves!mode coupling} or mode coupling. The difference between the two is that in mode conversion an intense high frequency plasma wave  propagating up a density gradient gradually transforms into a free space mode. \index{radiation!harmonic}\index{radiation!fundamental}

The more interesting case is a special case of the wave decay process, it is in fact its inversion, when two plasma waves join in interaction to inject their energies into a high frequency radio wave that is able to escape from the plasma. Radiation production is thus always a process of energy loss that leads to cooling of the plasma. However, in the heliosphere this cooling is weak and can be safely neglected. this kind of radiation is in fact degraded to an energetically completely unimportant process that has only indicative power.   

The free space modes can be either an ordinary or an extraordinary wave\index{waves!ordinary (LO)}\index{waves!extra-ordinary (RX)}, both propagating above  their upper cut-off frequencies which for the left-hand polarised ordinary (LO) mode with dispersion $\omega^2=\omega_{pe}^2+k^2c^2$ is the plasma frequency, and for the right-hand extraordinary (RX) mode is a cut-off frequency\index{waves!cut-off} slightly higher than the  upper-hybrid frequency $\omega_{uh}^2=\omega_{pe}^2+\omega_{ce}^2$. 

The following wave-wave processes are of interest in generating radiation: 
\begin{itemize}
\item the interaction of two counter-streaming electron plasma (Langmuir)\index{waves!Langmuir waves} waves into the transverse (T) electromagnetic wave, following the symbolic process L + L $\to$ T, where the symbols L, T mean the tuples $(\omega_{\,\rm L}, k_{\rm L})$ and $(\omega_{\,\rm T}, k_{\rm T})$, respectively. The energy and momentum conservation equations have been given above in Eq.\,(\ref{chap2-eq-decay}). This process produces a transverse wave with nearly perpendicular propagation $k_{\rm T}\ll k_{\rm L}$ in the RX-mode and of frequency $\omega_{\,\rm T}\simeq 2\omega_{pe}$, which can clearly propagate above the cut-off in weakly magnetized plasma,
\item the process L + L$' \to$ T of interaction of a Langmuir wave with another Langmuir wave that has been scattered off thermal ions (i) according to the process L + i $\to$ L$'$ + i$^*$, where the prime indicates the scattered Langmuir wave, and the star on the ion the excited ion. The wave interaction in this case causes a lower frequency transverse wave still above the plasma frequency but closer to the cut-off of the RX-mode,
\item a process similar to the one under the first item in which harmonic Langmuir waves have been generated in L-L interaction. This yields weak higher plasma harmonic radiation at frequencies  $\omega\lesssim l\omega_{pe}$ with harmonic numbers $l=3,\dots$ in the RS-T mode with intensity that decreases steeply with increasing harmonic number $l$,
\item merging of a Langmuir and an ion acoustic wave (S) according to L + S $\to$ T, which produces a transverse wave with frequency above but very close to $\omega_{pe}$. Whether this wave can escape from the plasma depends on whether its frequency is above the cut-off of the RX-mode, which is possible only for high frequency S waves, or whether it can tour into the OL mode for which generally the conditions of being selected in the merging process are worse than for the RX mode,
\item merging of another electrostatic plasma wave like, for instance, an electron acoustic mode (EA) with a Langmuir mode according to L + EA $\to$ T. Since the EA wave is near $\frac{1}{2}\omega_{pe}$, this produces a T wave at frequency $\omega\sim \frac{3}{2}\omega_{pe}$ well above  the cut-off still even though still counted as  fundamental mode radiation, 
\item then, the lower-hybrid mode can also merge with the Langmuir mode. The product of this process can, however, not escape in the RX mode as it usually does not exceed the cut-off. On the other hand, it could excite the LO mode, 
\item finally several nonlinear processes exciting transverse free space electromagnetic waves have been proposed like Langmuir wave collapse\index{waves!collapse}. Collapse is very attractive because Langmuir waves become trapped in this process inside deep density depressions where they bounce back and forth between the walls. Thus in quite a natural way collapse generates counter-streaming Langmuir waves in highly localized places. these waves are particularly well configurated for merging and escaping in the RX mode. During collapse moreover the internal plasma frequency decreases rapidly, causing a decrease of the radiation frequency as well. This allows only the higher frequency part of the radiation with frequency just above the plasma frequency of the environment to escape. Radiation occurs as intense fundamental but highly bursty radiation at the plasma frequency because at the end of the collapse the intensity of the waves explodes and therefore the emitted power also explodes. However, experimentally this process could never been proven. It has been replaced by another mechanism known as `stochastic growth' which itself is doubtful as well but very popular. It takes into account the stochastic modulation of Langmuir growth in a medium of spatially fluctuating density, i.e. containing a broad spectrum of weak ion acoustic turbulence. This causes the growth rate to experience spatial modulations leading to exponential modulation of the Langmuir amplitude. Contribution to the intensity comes only from localized places, and thus the volume contributing to radiation is a fraction only of the total volume while locally radiation may be rather intense. 
\end{itemize}

\subsection{Shock particle reflection}\index{shocks!particle reflection}\noindent
The process of particle reflection from a shock wave is one of the most important processes in the entire physics of collisionless shocks, as we have noted already in several places. The mechanism of particle reflection has not yet been fully illuminated, however. 

Particle reflection is required in supercritical shocks as, to our knowledge, it is the only process that can compensate for the incapability of dissipative processes inside the shock ramp to digest the fast inflow of momentum and energy into the shock. Shock particle reflection is not dissipative by itself even though in a fluid picture which deals with moments of the distribution function it can be interpreted as kind of an ion viscosity \cite{Macmahon1965}, i.e. it generates an anomalous viscosity coefficient \index{viscosity!ion viscosity}\index{particles!shock reflection}\index{collisions!ion viscosity}\index{process!ion viscosity}\index{process!ion reflection}which appears as a factor in front of the second derivative of the ion velocity in the ionic equation of motion. As such it also appears in the ion heat-transport equation. The kinematic ion viscosity can be expressed as 
\begin{equation}
\mu_{\rm vis}=m_iNv_i\lambda_{\rm mfp}\simeq P_i/2\omega_{ci}
\end{equation} 
through the ion pressure $P_i$ and the ion-cyclotron frequency $\omega_{ci}$ when replacing the mean free path through the ion gyro-radius.. 

In this sense shock particle reflection constitutes by itself a very efficient non-dissipative dissipation mechanism. However, its direct dissipative action is to produce real dissipation as far as possible {\it upstream of the shock} in order to dissipate as much energy of motion as remains to be in excess after formation of a shock ramp, dissipation inside the ramp, and reflection of ion back upstream.  The shock does this by inhibiting a substantial fraction of inflow ions to pass across the shock from upstream into the downstream region. It is sending these ions back into the upstream region where they cause a violently unstable upstream ion beam-plasma configuration which excites a large amplitude turbulent wave spectrum that scatters the uninformed  plasma inflow, heats it and retards it down to the Mach number range that can be digested by the shock. In this way the collisionless shock generates a shock transition region that extends far upstream with the shock ramp degrading to the role of playing a  subshock at the location where the ultimate decrease of the Mach number from upstream to downstream takes place.

Shock reflection has another important effect on the shock as the momentum transfer from the reflected particle component to the shock retards the shock in the region of reflection thereby decreasing the effective Mach number of the shock.\index{process!momentum transfer}

The outcome of the previous paragraphs is that shock particle reflection is of incomparable importance in shock formation and in the understanding of collisionless shock physics. On the other hand it is barely understood and can, in principle, be treated only by numerical simulations. Before, in the next chapters, coming to discuss those problems in greater depth we will present below a few attempts to understand shock reflection.
\begin{figure}[t!]
\hspace{0.0cm}\centerline{\includegraphics[width=1.0\textwidth,clip=]{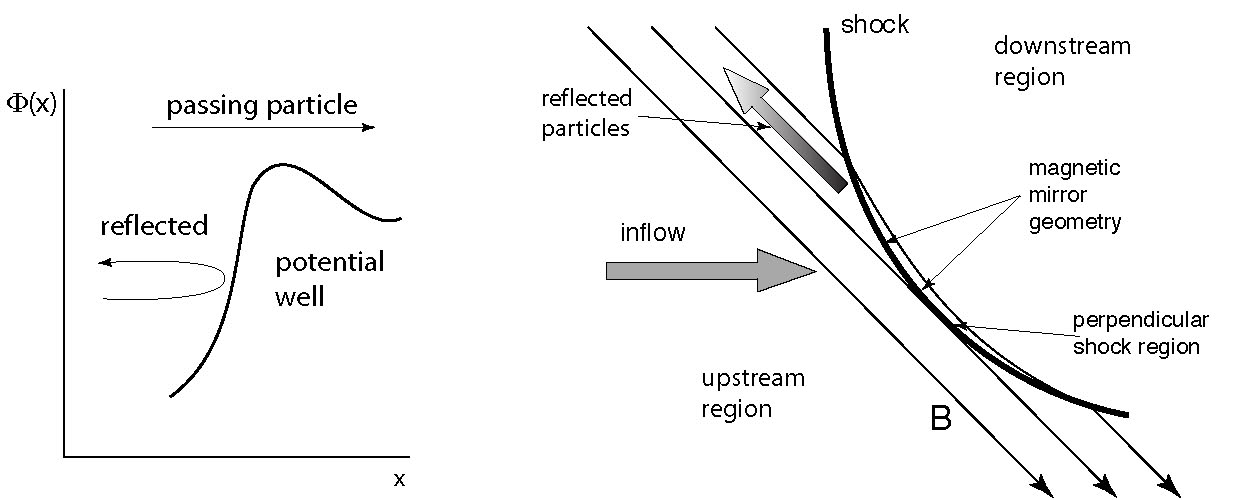} 
}
\caption[1]
{\footnotesize The two cases of shock reflection. {\it Left}: Reflection from a potential well $\Phi(x)$. Particles of energy higher than the potential energy $e\Phi$ can pass while lower energy particles become reflected. {\it Right}: Reflection from the perpendicular shock region at a curved shock wave as the result of magnetic field compression. Particles move toward the shock like in a magnetic mirror bottle, experience the repelling mirror force and for large initial pitch angles are reflected back upstream. }\label{chap2-fig-reflection}
\end{figure}
\subsubsection{Specular reflection}\noindent\index{process!specular reflection}
\noindent Specular reflection of ions from a shock front is the simplest case to be imagined. It requires that the ions experience the shock ramp as an impenetrable wall.  This can be the case only when the shock itself contains a positive reflecting electric potential which builds up in front of the approaching ion. Generation of this electric potential is not clarified yet. In the soliton picture the shock potential is related to the density gradient, however, dissipation processes are also involved. Understanding its formation requires understanding the entire collisionless shock physics. In a very naive approach one assumes that in flowing magnetized plasma a potential wall is created as the consequence of charge separation between electrons and ions in penetrating the shock ramp. It occurs over a scale typically of the spatial difference between an ion and an electron gyro-radius, because in the ideal case the electrons, when running into the shock ramp, are held temporarily back in the steep magnetic field gradient over this distance while the ions feel the magnetic gradient only over a scale longer than their gyro-radius and thus penetrate deeper into the shock transition. \index{scales!gyroradius}

\paragraph{Reflection from shock potential.} \index{shocks!shock potential, reflection from}Due to this simplistic picture the shock ramp should contain a steep increase in the electric potential $\Delta\phi$ which will reflect any ion which has less kinetic energy $m_iV_N^2/2<e\Delta\phi$. This condition contains the perpendicular ion velocity component along the shock normal. Since the ion gyrates it depends on the instantaneous angle the ion velocity has with respect to the magnetic field at the location of the ramp. In this reflection the ion velocity component $-V_N\to +V_N$ simply changes sign. For the gyrating particles this component adds up of the normal components of the bulk flow velocity $V_N^f$ and the microscopic particle speed ${\bf v}=(v_\perp\cos\alpha,v_\perp\sin\alpha,v_\|)$, with $\alpha$ the gyration angle, projected on the direction of the shock normal ${\bf n}$. This yields
\begin{equation*}
-V_N=-V_N^f+v_\|\cos\thetabn +v_\perp\cos\alpha\sin\thetabn 
\end{equation*}
and the condition for specular reflection can be written as
\begin{equation*}
(-V_N^f+v_\|\cos\thetabn +v_\perp\cos\alpha\sin\thetabn)^2<2e\Delta\phi/m_i 
\end{equation*}
This is a condition on the gyration angle $\alpha$ restricting the gyration phases of the reflected particles. For a gyrotropic upstream distribution one can average over all gyration angles from 0 to $\pi/2$ since only upstream directed velocity components reduce the inflow velocity to values below the reflection threshold,\index{threshold!for reflection} obtaining
\begin{equation*}
(v_\|\cos\thetabn-V_N^f)^2 +\frac{4}{\pi}(v_\|\cos\thetabn-V_N^f)v_\perp\sin\thetabn +\frac{1}{\pi}v_\perp^2\sin^2\thetabn<\frac{2e\Delta\phi}{m_i}
\end{equation*}
This condition  must be used on one of the velocity components $v_\|, v_\perp$ when determining the number of specularly reflected particles from the upstream ion distribution function $F_i^{up}(v_\|,v_\perp)$. 

\begin{figure}[t!]
\hspace{1.0cm}{\includegraphics[width=0.8\textwidth,clip=]{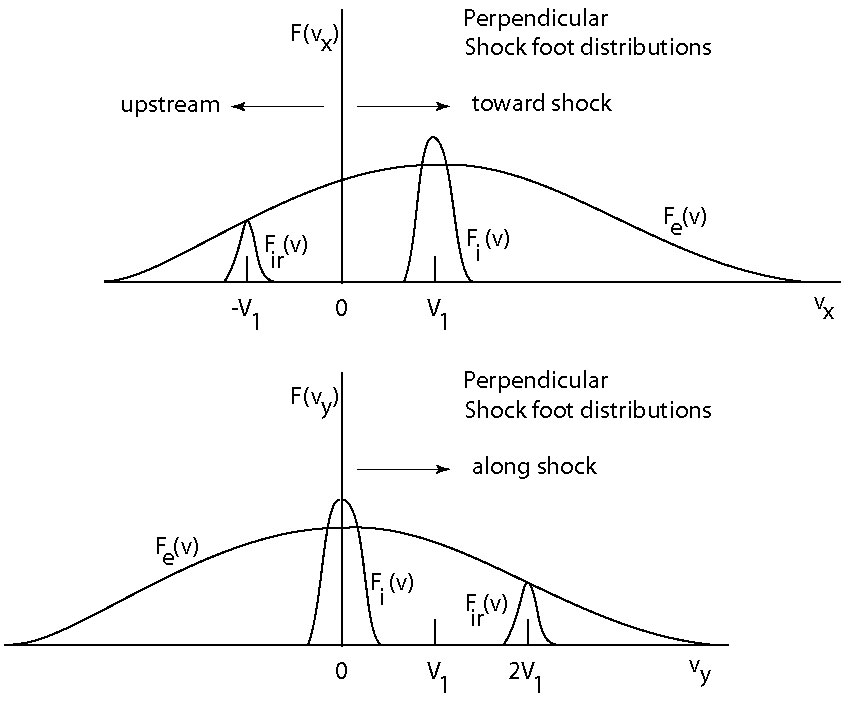} }
\caption[1]
{\footnotesize Expected particle distributions in the shock foot of a supercritical perpendicular shock which reflects ions back upstream. {\it Top}: Distribution functions in flow direction. The ion and electron distributions flow with velocity $V_1$ in shock direction. For the electrons with their high thermal speed the flow velocity is practically negligible. Reflected ions in the foot occur at velocity $-V_1$ on the scale of about one ion gyro-radius. This yields a two-ion beam configuration which is electrostatically unstable. Moreover, ion acoustic waves are excited to the right of the reflected ion distribution by the ion-acoustic instability in the positive slope of the hot electron distribution. {\it Bottom}: Same in the direction along the shock front. Inflowing ions and electrons have only thermal velocities in this direction. The reflected ions are accelerated to about twice the  inflow velocity in the tangential inflow electric field $|{\bf E}|=|{\bf V}_1\times{\bf B}_1|$. This causes an unstable ion beam-ion configuration and a two-stream configuration between accelerated reflected ions and electrons.}\label{chap2-fig-footdistr}
\end{figure}
\paragraph{Mirror reflection.} \index{process!mirror reflection}Another simple possibility for particle reflection from a shock ramp in magnetized plasma is mirror reflection. An ion approaching the shock has components $v_{i\|}$ and $V_\|=V_1\cos({\bf V}_1\cdot{\bf B}_1)$ along the magnetic field. Assume a curved shock like Earth's bow shock. Close to its perpendicular part where the upstream magnetic field becomes tangential to the shock the particles approaching the shock with the stream and moving along the magnetic field with their parallel velocities experience a mirror magnetic field configuration that results from the converging magnetic field lines near the perpendicular point. Conservation of the magnetic moment $\mu=T_{i\perp}/B$ implies that the particles become heated adiabatically in the increasing field; they also experience a reflecting \index{force!mirror}mirror force $-\mu\nabla_\|B$ which tries to keep them away from entering the shock along the magnetic field. Particles will mirror at the perpendicular shock point and return upstream when their pitch angle becomes $90^\circ$ at this location \citep[this theory has been developed in detail by][]{Leroy1984,Wu1984}. We will return to this mechanism in Chapter 6 when describing shock particle acceleration.

A rough estimate of the marginal upstream pitch angle for mirror reflection can be given from conservation of $\mu$. Since the increase in field strength is according to the magnetic gradient across the shock ramp one has roughly at the perpendicular shock point $B=B_1+(\nabla_n B)\Delta$ where $\Delta$ is the shock width and $\nabla_n$ the field gradient across the shock. Hence, to lowest order, $B= B_1+\delta B$ with $\delta B$ the magnetic compression. This yields for the upstream pitch angle at reflection $\sin^2\alpha_1>(1+\delta B/B_1)^{-1}$. With compression factor $\delta B/B_1\sim 3$ particles of upstream pitch angles $\alpha_1>60^\circ$ will become mirror reflected from the perpendicular shock area due to the action of the mirror force, a condition which has to be used upstream in the inflowing distribution if one wants estimating the fraction of reflected particles. This requires knowledge of the compression factor, however. The compression factor and the number of reflected particles are not independent. Hence, a selfconsistent determination requires developing the full shock theory. This can be done only by numerical simulations.

Of course, the above estimate is very crude. It demonstrates, however, that a fraction of upstream particles can, in principle, become reflected from a curved shock surface by mirroring in the converging magnetic field geometry around the perpendicular area of the shock. For fast flows reflection will always be located on the nose inflow side of the shock. This holds for ions as well as for electrons. Reflection of both sorts of particles has continuously been observed at the bow shock as is shown schematically, for instance, in Figure\,\ref{Tsurutani-f1} which has been drawn for condition in front of a perpendicular shock at a distance inside the foot, roughly within $<1\,r_{ci}$ from the shock ramp. 

\paragraph{Consequences of shock reflection.} How far the reflected ions return upstream depends on the direction of the magnetic field with respect to the shock, i.e. on the shock normal angle $\thetabn$. For perpendicular shocks the reflected ions only pass just one gyro-radius back upstream. Seeing the convection electric field $|E_y|=|V^fB|$ they become accelerated along the shock forming a current, the velocity of which in any case exceeds the inflow velocity (which is zero in the perpendicular direction) and for sufficiently cold ions also the ion acoustic velocity $c_{ia}$ in which case the ion-beam plasma instability will be excited in the shock foot \index{shocks!foot}region where the ion current flows. This may generate and anomalous collision frequency in the shock foot region. Moreover, since the excited waves accelerate electrons along the magnetic field other secondary instabilities can arise as well.

In quasi-perpendicular and oblique shocks the ions can escape along the magnetic field. In this case an ion-beam/ion beam situation arises between the upstream beam and the plasma inflow with the consequence of excitation of a variety of instabilities, electromagnetic and electrostatic. In addition, however, an ion beam-electron beam two stream situation is caused between the upstream ins and the inflow electrons which because of the large upstream electron temperatures probably excites mainly ion-acoustic modes but can also lead to Buneman two-stream\index{instability!two-stream} mode excitation. These modes contribute to turbulence in the upstream foreshock region creating a weakly dissipative state in the foreshock\index{shocks!foreshock} where the plasma inflow becomes informed about the presence of the shock. The electromagnetic low frequency instabilities on the other hand, which are excited in this region, will grow to large amplitude, form localized structures and after being convected by the main flow towards the shock ramp interact with the shock and modify the shock profile or even contribute to shock formation and shock regeneration.

Specular reflection from shocks is the extreme case of shock particle reflection. Other mechanisms like turbulent reflection are, however, not well elaborated and must in any case be investigated with the help of numerical simulations. In the following subsection we will in passing encounter one such mechanism in phase space hole formation in a two-stream unstable collisionless plasma.

Figure\,\ref{chap2-fig-footdistr} shows the expected particle distributions for shock reflection of ions in a perpendicular shock right in the foot region. The upper part of this figure is perpendicular to the shock in flow direction. The lower part is along (tangential to) the shock surface. The different configurations of the distributions in this region can lead to the excitation of ion acoustic and two-stream instabilities. Ion beam-ion beam interaction is expected in the direction perpendicular to the shock in addition to ion-acoustic instability between the reflected ions and the main electron component in the region of positive gradient on the electron distribution. In the direction parallel to the shock surface, on the other hand, one expects and ion beam-ion interaction and a two stream instability of the main electron component with the accelerated along the shock surface reflected ion beam. 
\begin{figure}[t!]
\hspace{0.0cm}\centerline{\includegraphics[width=1.0\textwidth,clip=]{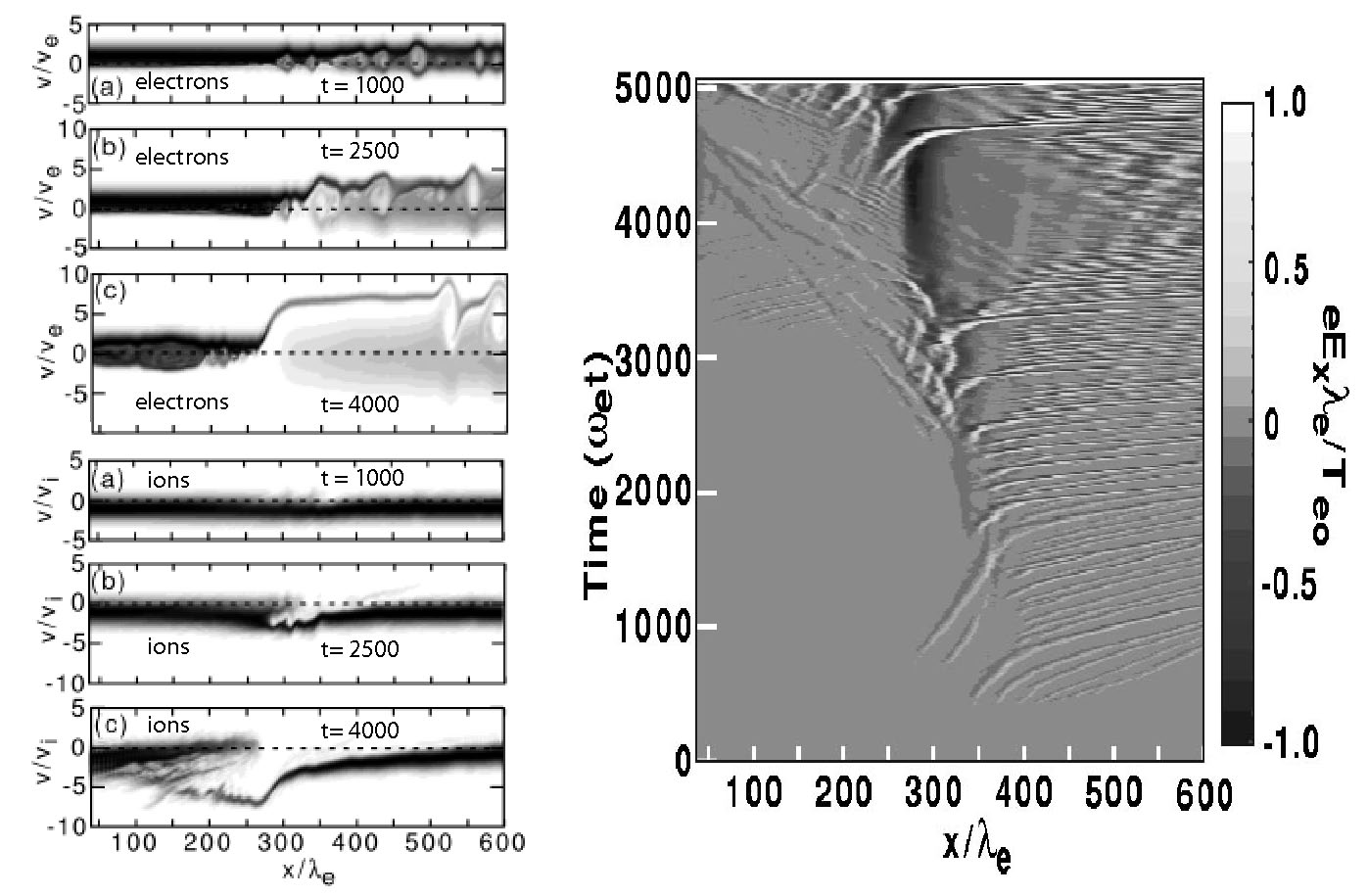} 
}
\caption[1]
{\footnotesize {\it Left}: Electron and ion phase space plots at three simulation times of the interaction of a marginally two-stream unstable plasma with a shock ramp. The ramp has been modelled as a density dip (potential wall for electrons) in the centre of the simulation box. The box has one space and one velocity coordinate. The fast electron beam current is injected into the quiet ion plasma and  \citep[after][]{Newman2001}. {\it Right}: Time history of the elecric field during the interaction with the shock ramp showing the evolution of electron and ion phase space holes and their interaction \citep[from][]{Newman2002}.  }\label{chap2-fig-newman}
\end{figure}
Inside the ramp conditions are more involved and will be describe in more detail in the context with observations in the respective chapters for both quasi-perpendicular and quasi-parallel shocks. 

We note in addition that there is a peculiarity concerning quasi-parallel supercritical shocks. Due to the presence of an intense reflected and transformed ion component in the foreshock of a quasi-parallel shock there is a broad spectrum of large amplitude low frequency electromagnetic waves which are convected by the inflow stream towards the shock ramp, steepen and interact with the shock. These waves are predominantly transverse having components tangential to the shock on scales of the order and shorter than the ion gyro-radius. Consequently, quasi-parallel shocks remain to be quasi-parallel for ions, in particular for the more energetic accelerated ions. however, for the electrons all quasi-parallel supercritical shocks become quasi-perpendicular in the vicinity\index{shocks!electron reflection}\index{acceleration!quasi-parallel shock, electrons} of the shock ramp transition such that for electrons no quasi-paallel supercritical shocks exist. This has the consequence that electrons will become reflected and accelerated all-over the shock front independent on its quasi-parallel character. We will later prove this statement by referring to observations and simulations.

\subsubsection{Hole formation}\index{instability!nonlinear}\noindent
As the last item in this section we consider the stability of an high velocity $V_b>v_e$ electron beam or electron current. In the fluid picture we have found that such high speed current flowing through the plasma along the magnetic field or (in the presence of steep density gradients) perpendicular to both, the gradient and the magnetic field, will excite the fast growing Buneman two-stream instability. It has, however, been predicted early \citep{Schamel1972,Dupree1975} that currents of this strength will undergo a kinetic instability which structures the electron and ion phase space into so-called phase space holes which are regions of lacking particles localized in phase space while in real space represent localized electric fields, trapped particles and particle acceleration. Such holes have meanwhile been found to exist all-over in collisionless space plasmas in relation to spatially localised strong current flow as in reconnection, auroral phenomena, and also in shocks \citep{Bale2002}. Since strong currents are expected in shocks as well in the ramp as in the foot, as we have discussed above, it is not unreasonable to assume that phase space holes might form under shock conditions as well. \index{instability!phase-space holes} 

Hole formation follows a nonlinear interaction known as Bernstein-Green-Kruskal (BGK) mode\index{waves!BGK modes} formation \citep[cf., e.g.,][]{Davidson1972} and is based on the splitting of the phase space distribution function into two components, particles that are energetic enough to surpass the electric potential of a localized electric field inside a soliton,  for instance, and particles of lesser energy that become either trapped or reflected from these potentials\index{waves!particle trapping} depending on the sign of particle and potential. 
Even if the potential is repulsive and ejects, say, electrons from the region, some trapped electrons will remain there performing oscillations and become heated up to a certain energy that leaves them still trapped. These electrons are in disordered motion and are assuming a high temperature, while the rejection and expulsion of other particles from the potential site causes their acceleration. This mechanism is quite complicated and has been treated analytically only up to a certain approximate degree in the above cited papers. In order to investigate it one better performs numerical simulations. 

Figure\,\ref{chap2-fig-newman} shows the example of a simulation of hole formation in interaction of a marginally two-stream stable current (of electron and ion of opposite bulk velocity and same initial temperature  $T_e=T_i$ with a localized inhomogeneity \citep[from][]{Newman2001,Newman2002}. The localized plasma inhomogeneity has been modelled as a simple density dip $\propto -\cos^2(x-x_0)$ at the centre. This is not a shock, it is, however, a potential wall which should reflect one sort of particles, in this case electrons. 

The figure shows on the left phase space plots at three different simulation times for electrons and ions when due to electron reflection at the potential ramp a strong two-stream instability evolves. The holes formed on the electron distribution appear early in the simulation as egg-like distortions of the distribution in the reflected electron component. Widening of the distribution indicates the strong heating of the electrons. The holes contain dilute trapped electron, and some part of the beam becomes accelerated. At later times the heating becomes violent with a strong broadening of the electron distribution when the ion hole starts forming in the lower panels.  Strong acceleration of a narrow and thus very cool  electron beam is also observed in the final state. In addition the holes move along the beam, while the ion hole moves in the opposite direction.

Most interesting is the time history shown on the right in the figure. It shows the initial evolution of many small amplitude electron holes moving at fast speed to the right away from the potential ramp. At later times the ramp steepens, and the electron holes start interacting with the ion hole which moves slowly to the left. The holes intensify and finally can break through the potential ramp to escape to the right where a whole fabric of interfering holes evolves. 

The importance of this observation is that two-stream instability can form as a cause of reflection at a potential ramp. This is expected for shocks as well. Moreover, the instability causes electron and ion phase space holes to evolve and leads to completely collisionless heating due to electron trapping inside the holes, i.e. it causes irreversible heating and entropy which is needed for shocks, and it also generates a very cool electron beam to escape from the holes by continuous acceleration and collimation of cool but fast electrons. This is a very interesting and important mechanism which is capable of injecting a fast seed\index{acceleration!seed population} particle population into shock acceleration.

\section{Briefing on Numerical Simulation Techniques}
\noindent The modern age of physics is to a large degree determined by the availability of high speed and high capacity computer systems. The use of these computing facilities for performing numerical experiments on collisionless plasmas covers now almost half a century of experience. It started with the introduction of Fermi's  Monte-Carlo method and blossomed after the formulation of the Fermi-Ulam 1961 numerical model approach to cosmic ray acceleration \citep[cf., e.g.,][]{Lichtenberg1991}  which was based on nonlinear particle motions in electromagnetic fields. The modern state of the art in application to plasma physics has been formulated in several textbooks \citep[e.g.,][]{Birdsall1985} and review articles \citep[e.g.,][]{Dawson1983,Dawson1995}. Many problems in plasma and in particular space plasma physics with their enormous complexity could not have been solved or even attacked without computers and numerical simulations. Also, most of the discussion on shock in the following chapters will be based on such numerical simulations which must accompany observation and experiment in order to understand what is going on in the shock environment. A brief discussion about numerical methods is therefore not only unavoidable but even necessary. \index{simulations!discretization}
\subsection{Basic Equations}\noindent
Computers are capable to deal simultaneously with the dynamics of many particles as we have already described in Chapter 1. The most fundamental approach in numerical simulations is hence based on the full Newton-Maxwell set of microscopic equations of which we write down here only the Newtonian subset 
\begin{equation}
\frac{{\rm d}{\bf v}_{is}}{{\rm d}t}=\frac{e_s}{m_s}({\bf E}+{\bf v}_{is}\times{\bf B}), \qquad\qquad \frac{{\rm d}{\bf x}}{{\rm d}t}={\bf v}
\end{equation}
The connection to Maxwell's equations is done by the microscopic definition equations of the space charge $\rho({\bf x}, t)$ and conduction current ${\bf j}({\bf x}, t)$ according to
\begin{equation}
\rho({\bf x}, t)=\sum\limits_{is} e_s\delta({\bf x-x}_{is}), \qquad\qquad{\bf j}({\bf x}, t)=\sum\limits_{is} e_s{\bf v}_s\delta({\bf x-x}_{is})
\end{equation}
In fact, these are the most general equations of a classical plasma consisting of $i$ point-like space charges $e_s$ of species $s$
with mass $m_s$ and momentum $m_s{\bf v}_{is}$ located at time $t$ at location ${\bf x}_{is}(t)$. The point-like character is taken care of by the $\delta$-functions. This whole set is the set of Liouville's equations [which could also formally be combined into one single {\it exact} equation in phase space (${\bf x, v}$) by introducing an {\it exact phase space distribution} function $F({\bf x, v})=\sum_{is}\delta({\bf x-x}_{is})\delta({\bf v-v}_{is})$. Such an equation is know as the Liouville equation].\index{equation!Liouville}

These equations can be simplified depending on the nature of the problem. For instance, when electromagnetic effects are not of interest, the magnetic field will drop out and one uses only Poisson's equation for the electrostatic potential and the electric field in Amp\`ere's law. When the plasma is collisionless one can use another simplification, i.e. replace the Liouville distribution with the one-particle distribution and switch to the Vlasov equation. This then produces Vlasov-codes. When one is interested only in low frequency responses of the plasma, the electrons can be treated as Boltzmannian electrons, and the electron equations are replaced by the Boltzmann dependence of the electron density on the electric potential with the ions being treated as single particles. The corresponding codes are the hybrid codes. An even stronger simplification is the assumption of quasi-neutrality when the Poisson equation is replaced by the condition $N_e=N_i$ and the electrons are merely an instantaneously reacting neutralising background.

\subsection{General methods}\noindent
Either of the resulting set of equations must be represented in digitalized form in order to be prepared for treatment on a computer. One represents the spatial coordinates as a discrete grid and advances the equations over discrete time intervals. The choice of space and time steps is prescribed by the necessary accuracy of the result and by the stability of the code. The particles in the code are, however, no more point particles but of finite size, i.e. the Delta-functions assume bell-function shapes. This has the consequence that particle experience only small angle collisions well suited for studying collisionless plasmas. The integration of the equations is then performed in a series on a large number of such discrete steps of the finite sized particles the Lagrangean positions of which are used to deposit the charges and currents onto the fixed discrete grid points, followed by solving the field equations on these grid points of the many spatial cells. This produces self-consistent fields which are used in the next time step and which are smeared out over the volume by interpolation in order to construct a field continuum in which the particle orbits are advanced further. One is thus working with space and time differences and an interpolation from charge to grid and  subsequently of the fields from grid to particle. 
In view of application to the heliospheric shocks in the following we briefly discuss the discretization only for the electromagnetic case.\index{equation!normalization}

\subsubsection{Units}\noindent 
Numerical methods require that all quantities including coordinates space and time are pure numbers. One thus needs to normalize them by introducing constant reference values, a density $N_0$, charge-to-mass ratio $e/m_0$, velocity of light $c$ (for instance, other choices are Alfv\'en velocity and so on). Time is then normalized, for instance, to plasma frequency expressed in these reference units $\omega_p^{-1}=(e^2N_0/\epsilon_0m_0)^{-\frac{1}{2}}$, space in inertial lengths $c/\omega_p$, the electric potential for instance in $m_0c^2/e$, the magnetic vector potential in $m_0c/e$. This choice of units is one of many possibilities only.

\subsubsection{Discretization}\noindent 
Both the field and particle equations must be discretized. The idea of discretization is quite simple. One returns in history to the time just one step before Newton and Leibniz. Differential quotients become quotients of differences, higher order differential quotients become the corresponding powers of quotients of differences, mixed differential quotients become products of quotients of differences. The only trick is to assign the results to some location inside the difference interval, not necessarily the centre (!), and to do this properly. Also, time runs only in one direction: forward. Applying such a scheme one arrives at recursive equations which can be solved on a sufficiently powerful computer.

 Let us assume we reduce the electromagnetic set of equations\index{equation!electromagnetic set} to the electromagnetic wave equation for the vector potential component ${\bf A}$. This must be written in difference form
\begin{equation*}
\hspace{-0.5cm}\frac{1}{4}\nabla^2({\bf A}^{(\frac{3}{2})}+2{\bf A}^{(\frac{1}{2})}+{\bf A}^{(-\frac{1}{2})})-\frac{1}{c^2\Delta t^2}(1+\beta D^2\nabla^2)({\bf A}^{(\frac{3}{2})}-2{\bf A}^{(\frac{1}{2})}+{\bf A}^{(-\frac{1}{2})})=-{\bf j}_T^{(\frac{1}{2})}
\end{equation*}
The superscripts indicate the time levels of the various terms, the ad hoc parameter $\beta$ is introduced to modify the dispersion at short wavelengths, and $\nabla^2$ is defined as
\begin{equation*}
\hspace{-0.5cm}\nabla^2 A=\frac{A_{j+1,m}-2A_{j,m}+A_{j-1,m}}{\Delta x^2}+\frac{A_{j,m+1}-2A_{j,m}+A_{j,m-1}}{\Delta y^2}, \qquad \frac{1}{D^2}=\frac{1}{\Delta x^2}+\frac{1}{\Delta y^2}
\end{equation*}
The transverse current density in the Coulomb gauge is ${\bf j}_T={\bf j} -\nabla\partial \phi/\partial t$. since $\nabla\cdot{\bf j}_T=0$, the additional equation $\nabla^2(\partial\phi/\partial t)=\nabla\cdot{\bf j}$ must be solved. The electrostatic potential $\phi^{(1)}$ is to be taken a full time step (1), while ${\bf j^{(\frac{1}{2})}}$ is at half time step $(\frac{1}{2})$. The former is obtained from $\nabla^2\phi^{(1)}=-\rho^{(1)}$. Finally, the fields follow from
\begin{equation*}
{\bf E}^{(1)}=-\nabla\phi^{(1)}-\frac{{\bf A}^{(\frac{3}{2})}-{\bf A}^{(\frac{1}{2})}}{\Delta t}, \qquad {\bf B}^{(1)}=\frac{1}{2}\nabla\times({\bf A}^{(\frac{3}{2})}+{\bf A}^{(\frac{1}{2})})
\end{equation*}
Fields, charge and currents are defined at the centre of the cells. These equations must be solved with a so-called Poisson solver. In addition one needs to specify appropriate boundary conditions at the boundaries of the simulation box.\index{equation!discretized}

In a similar way one discretizes the equation of motion of the particles. For this one defines $h=e\Delta t/m$ and obtains a centred form of the equation of motion as a recursion equation with unknown ${\bf v}^{(1)}$
\begin{equation*}
{\bf v}^{(\frac{3}{2})}={\bf v}^{(\frac{1}{2})}+h({\bf E}^{(1)}+{\bf v}^{(1)}\times{\bf B}^{(1)})
\end{equation*}
In order to determine ${\bf B}= \frac{1}{2}({\bf v}^{(\frac{3}{2})}+{\bf v}^{(\frac{1}{2})})$ the former equation is solved implicitly taking the scalar and vector products of the former equation with ${\bf B}^{(1)})$. This yields up to terms second order in $\Delta t$ the expression
\begin{equation*}
{\bf v}^{(\frac{3}{2})}={\bf v}^{(\frac{1}{2})}\left(1-\frac{h^2}{2}B^2\right)+h({\bf E}+{\bf v}^{(\frac{1}{2})}\times{\bf B})+\frac{h^2}{2}\left({\bf E\times B}+{\bf B}{\bf B}\cdot{\bf v}^{(\frac{1}{2})}\right)
\end{equation*}
for use in the expression for ${\bf v}^{(1)}$ only. 
The set of equation obtained is then ready for computing. \index{simulations!explicit}\index{simulations!implicit}

However, there are two ways of computing, so-called explicit and implicit techniques. In the {\it explicit} technique one solves the equations as they are obtained after discretization. In such an approach the internal errors will necessarily grow and at some stage become unstable such that the calculation must be truncated. One can artificially introduce some kind of damping for these growing error modes in order to suppress them. Justification for this is discussed in the literature. The {\it implicit} technique solves the problem by calculating {\it backward} in time \citep{Friedman1981} such that the error modes decay away when time runs negative. We do not describe this technique here. It suffices to note that in this approach the dangerous unstable short-wavelength error modes disappear by definition and become partially eliminated from the system. The most efficient ways of calculation are the combinations of both methods.

All these methods work on a fixed grid. In application to shock ramp research the grid mash has to be chosen refined enough for resolving shock structure. Recently \cite{Karimabadi2006} developed a self-adaptive technique  which takes care of the steepening and narrowing of a shock front in order to resolve its substructure. This is a significant progress in shock simulation technique.

We do not go into detail of the various methods and refinements of simulation techniques. Those readers who are interested and prepare by themselves for doing simulation research in collisionless shocks we rather refer to the mentioned basic literature on the techniques of numerical simulation. 

}

\end{document}